\documentclass{article}
\usepackage{amsmath}
\usepackage{amssymb}
\usepackage{bm}
\begin{document}
\title{$N$ Identical Particles
Under Quantum Confinement: A Many-Body Dimensional Perturbation
Theory Approach II,\\ The Lowest-Order Wave Function II}
\author{M.\ Dunn, D.K.\ Watson\\
University of Oklahoma \\ Department of Physics and Astronomy \\
Norman, OK 73019 \and J.G.\ Loeser \\
Oregon State University \\ Department of Chemistry \\ Corvallis,
OR 97331}
\date{\today}
\maketitle

\begin{abstract}
In this paper, the second in a series of two, we complete the
derivation of the lowest-order wave function of a dimensional
perturbation theory (DPT) treatment for the $N$-body
quantum-confined system.
Taking advantage of the symmetry of the zeroth-order configuration, we
use group theoretic techniques and the FG matrix method from quantum
chemistry to obtain analytic results for frequencies and normal modes.
This method directly accounts for each two-body
interaction, rather than an average interaction so that even
lowest-order results include beyond-mean-field effects.
It is thus appropriate for the study of both weakly and strongly
interacting systems and the transition between them. While
previous work has focused on energies, lowest-order wave functions
yield important information such as the nature of excitations and
expectation values of physical observables at low orders including
density profiles. Higher orders in DPT also require as input the
zeroth-order wave functions. In the earlier paper we presented a
program for calculating the analytic normal-mode coordinates of the
large-$D$ system and
illustrated the procedure by deriving the two simplest normal
modes. In this paper we complete this analysis by deriving the
remaining, and more complex, normal coordinates of the system.
\end{abstract}

\section{Introduction}
Quantum confined systems involving $N$ identical particles are
common in many areas of physics, including chemical, condensed
matter and atomic physics. In earlier papers\cite{loeser,FGpaper,
energy} we have developed the method of dimensional perturbation
theory\cite{copen92} at low orders to examine the energies of
quantum-confined systems such as atoms, quantum dots and
Bose-Einstein condensates, both the ground and excited states.
Dimensional perturbation theory (DPT) has many advantages. These
include the fact that the number of particles, $N$\,, enters into
the theory as a parameter, and so it is easy to calculate results
for an arbitrary number of particles. Also the theory is a
beyond-mean-field method, directly accounting for each
interaction, rather than some average representation of the
interactions. It is therefore appropriate for confined
systems of both weakly interacting and strongly interacting
particles. In the case of a trapped gaseous atomic Bose-Einstein
condensate (BEC) the system is typically a weakly interacting
system for which a mean-field approximation is valid. However, if
$N$ is increased, or Feshbach resonances are exploited to increase
the effective scattering length, or the system is squeezed in one
or more directions, it will transition into a strongly-interacting
regime where the mean-field approach breaks down. Such systems
have been created in the laboratory.\cite{wieman} Dimensional
perturbation theory is equally applicable to both
regimes and may be used to study the transition from weak to
strongly interacting regimes. Even at lowest order DPT includes
beyond-mean-field effects.

In an earlier paper\cite{FGpaper}  we presented a detailed
discussion of the dimensional continuation of the $N$-particle
Schr\"odinger equation, the $D\to\infty$ equilibrium ($D^0$)
structure, and the energy through order $D^{-1}$\,.  We used the
FG matrix method to derive general, analytical expressions for the
many-body normal-mode vibrational frequencies, and we gave
specific analytical results for three confined $N$-body quantum
systems: the $N$-electron atom, $N$-electron quantum dot, and
$N$-atom inhomogeneous Bose-Einstein condensate with a repulsive
hardcore potential.

In a subsequent paper\cite{energy} we studied the $N$-atom
inhomogeneous Bose-Einstein condensate (BEC) with a repulsive
hardcore potential, optimizing our low-order DPT energy by fitting
to low-$N$ diffusion Monte Carlo
data\cite{blume} and then calculated results
out to large $N$. Results were obtained for weak, intermediate and
strong interatomic interactions.

Low-order many-body dimensional perturbation methods have been
applied to the $N$-electron atom in Ref.~\cite{loeser} where
analytical expressions are obtained for the ground-state energy of
neutral atoms. For $Z=1$ to $127$, the energies compare
well to Hartree-Fock energies with a correlation correction.

In a subsequent paper\cite{paperI} (Paper~I) we have begun a
detailed derivation of
the zeroth-order DPT wave function for quantum-confined systems

With the wave function many more properties of the system become
accessible. The normal-mode coordinates of the zeroth-order wave
function reveal the nature of the excitations of the system, and
the lowest-order wave function provides expectation values and
transition matrix elements to zeroth order in DPT. For macroscopic
quantum-confined systems, such as the BEC, the wave function is
manifested in an explicit way as the density profile may be viewed
directly in the laboratory. This is readily calculable at zeroth order in
DPT from the zeroth-order wave function.

Calculating energies and wave functions to higher orders in $1/D$
also requires as input the lowest-order wave function. Indeed, as
we have discussed in an earlier paper\cite{energy} which considers energies
and excitation frequencies of the BEC, low-order DPT calculations
of confined interacting particles, including strongly
interacting systems, are accurate out to $N$ equals a few thousand
particles. To obtain yet more accurate energies, and to extend the
calculation beyond a few thousand particles we need to go beyond
these low-order calculations of the energies and excitation
frequencies. These calculations are facilitated by Paper~I and
the present paper since, as we have already noted above, it
requires the zeroth-order DPT wave function. This will be pursued
in later papers.

In Paper~I we set up the general theory for determining the
zeroth-order DPT wave function of a quantum confined system. For
quantum-confined systems in large dimensions, the
Jacobian-weighted\cite{avery} wave function is harmonic, the
system oscillating about a configuration termed the Lewis
structure with every particle equidistant and equiangular from
every other particle. Notwithstanding its relatively simple form,
the large-dimension, zeroth-order wave function includes
beyond-mean-field effects. The Lewis structure is a completely
symmetric configuration invariant under a point group which is
isomorphic to the symmetric group $S_N$. Using group theory
associated with this $S_N$ symmetry\cite{hamermesh} and the
$FG$ matrix method familiar from quantum chemistry\cite{WDC},
there is a remarkable reduction from $N(N+1)/2$ possible distinct
frequencies to only five distinct frequencies. The $S_N$ symmetry
also greatly simplifies the determination of the normal
coordinates. There are five sets of normal coordinates, where the
normal modes of a set have the same frequency and transform
under the same irreducible representation of $S_N$\,. Two sets of
normal coordinates transform under the $[N]$ irreducible
representation, two sets of normal coordinates transform under the
$[N-1, \hspace{1ex} 1]$ irreducible representation, and one set of
normal coordinates transforms under the $[N-2, \hspace{1ex} 2]$
irreducible representation.

In Paper~I we illustrated this general theory, for the zeroth-order
DPT wave function of a quantum confined system, by determining the
breathing and center-of-mass modes of the $[N]$ species
(coordinates transforming under the same irreducible
representation are said to belong to the same species). In this
paper, we complete this analysis by determining the modes of the
more complex $[N-1, \hspace{1ex} 1]$ and $[N-2, \hspace{1ex} 2]$
species. The zeroth-order density profile of the ground-state
condensate
is relatively simple to
calculate once the zeroth-order wave function is known.

Paper~I and this paper are restricted to consideration of quantum
confined systems with a confining potential of spherical symmetry.
The extension of this method to systems with cylindrical symmetry
is relatively straightforward and will be discussed in subsequent
papers. This is particularly important for the BEC as few current
laboratory traps have spherical symmetry.

In Sections~\ref{sec:SE} through \ref{sec:NormCalc} we briefly review the
general theory of the zeroth-order DPT wave function for a quantum
confined system from Paper I. In contrast to Paper I where we first discussed
the symmetry coordinates which motivated the introduction of
the primitive irreducible coordinates, in this paper we
lay out the steps involved sequentially in the order in which they
are applied to a particular problem. In Section~\ref{sec:SNm1pNm2}
we begin the extension to the $[N-1,\hspace{1ex} 1]$ and
$[N-2, \hspace{1ex} 2]$ species by
discussing the primitive irreducible coordinates for both the
angular and radial sectors of these species. Angular and radial symmetry
coordinates for the $[N-1, \hspace{1ex} 1]$ and $[N-2,
\hspace{1ex} 2]$ species are then derived by forming appropriate
linear combinations of the primitive irreducible coordinates. In
Section~\ref{sec:SysFreqNorm} the frequencies and normal-mode
coordinates of the system are derived by
transforming the $\bm{G}$ and $\bm{FG}$ matrices in the internal
displacement coordinate basis to the symmetry coordinate basis.
The problem is simplified by an extraordinary
degree to two $2 \times 2$ and one $1 \times 1$ matrix eigenvalue
equations. This reduction for $[N]$ species has already been discussed
in Paper~I. Here we focus on the remaining, more complicated
$[N-1, \hspace{1ex} 1]$ and $[N-2, \hspace{1ex} 2]$ species. From
the solution of these reduced eigenvalue equations, the $[N-1,
\hspace{1ex} 1]$ and $[N-2, \hspace{1ex} 2]$ species normal
coordinates are derived in Section~\ref{subsec:FNM}\,. In
Section~\ref{subsec:MotNormM} the normal mode motions of the
particles in the original internal coordinates are derived.
Section~\ref{sec:Sum} and Section~\ref{sec:Conc} summarize and
conclude this paper.

\section{The ${\bm{D}}$-dimensional ${\bm{N}}$-body Schr\"odinger
Equation}\label{sec:SE}  In Paper~I we considered an $N$-body system
of particles confined by a spherically symmetric potential and
interacting via a common two-body potential $g_{ij}$. The
Schr\"odinger equation for this system in $D$-dimensional
Cartesian coordinates is
\begin{equation}
\label{generalH} H \Psi = \left[ \sum\limits_{i=1}^{N} h_{i} +
\sum_{i=1}^{N-1}\sum\limits_{j=i+1}^{N} g_{ij} \right] \Psi = E
\Psi \,,
\end{equation}
where
\begin{equation} \label{eq:hi}
h_{i}=-\frac{\hbar^2}{2
m_{i}}\sum\limits_{\nu=1}^{D}\frac{\partial^2}{\partial
x_{i\nu}^2} +
V_{\mathtt{conf}}\left(\sqrt{\sum\nolimits_{\nu=1}^{D}x_{i\nu}^2}\right)
\,,
\end{equation}
\begin{equation}
\mbox{and} \;\;\;
g_{ij}=V_{\mathtt{int}}\left(\sqrt{\sum\nolimits_{\nu=1}^{D}\left(x_{i\nu}-x_{j\nu}
\right)^2}\right)
\end{equation}
are the single-particle Hamiltonian and the two-body interaction
potential, respectively. The operator $H$ is the $D$-dimensional
Hamiltonian, and $x_{i\nu}$ is the $\nu^{th}$ Cartesian component
of the $i^{th}$ particle. $V_{\mathtt{conf}}$ is the confining
potential.

\subsection{The Effective $\bm{S}$-Wave Schr\"{o}dinger Equation.}
Restricting our attention to spherically symmetric ($L=0$) states
of the many-body system, we transform from Cartesian to internal
coordinates. A convenient internal coordinate system for confined
systems is
\begin{equation}\label{eq:int_coords}
r_i=\sqrt{\sum_{\nu=1}^{D} x_{i\nu}^2} \;\;\; (1 \le i \le N)
\;\;\; \mbox{and} \;\;\;
\gamma_{ij}=cos(\theta_{ij})=\left(\sum_{\nu=1}^{D}
x_{i\nu}x_{j\nu}\right) / r_i r_j \;\;\; (1 \le i < j \le N),
\end{equation}
which are the $D$-dimensional scalar radii $r_i$ of the $N$
particles from the center of the confining potential and the
cosines $\gamma_{ij}$ of the $N(N-1)/2$ angles between the radial
vectors. Under this coordinate change the effective $S$-wave
Schr\"{o}dinger equation in these internal coordinates becomes
\begin{equation} \label{eq:laplacian} \begin{array}{r@{}l} {\displaystyle \left[ \sum\limits_{i} \left\{ -\frac{\hbar^2}{2
m_{i}} \left( \vphantom{\frac{D-1}{r_i^2}\sum\limits_{j\not=
i}\gamma_{ij}\frac{\partial} {\partial \gamma_{ij}}} \right.
\right. \right. } & {\displaystyle \frac{\partial^2}{{\partial
r_i}^2} + \frac{D-1}{r_i}\frac{\partial}{\partial r_i} +
\sum\limits_{j\not= i}\sum\limits_{k\not=
i}\frac{\gamma_{jk}-\gamma_{ij}\gamma_{ik}}
{r_i^2}\frac{\partial^2}{\partial \gamma_{ij} \partial
\gamma_{ik}} } \\
&  {\displaystyle \left. \left. \left. -
\frac{D-1}{r_i^2}\sum\limits_{j\not= i}\gamma_{ij}\frac{\partial}
{\partial \gamma_{ij}} \right) + V_{\mathtt{conf}} (r_i) \right\}
+ \sum_{i=1}^{N-1}\sum\limits_{j=i+1}^{N} V_{\mathtt{int}}(r_{ij})
\right] \Psi = E \Psi \,,}
\end{array}
\end{equation}
where $(r_{ij})^2 = (r_i)^2 + (r_j)^2 - \,\, 2 \, r_i \, r_j \,
\gamma_{ij}$\,.

\subsection{The Jacobian-Weighted Schr\"{o}dinger Equation}
\label{sub:simtransf} Dimensional perturbation theory utilizes a
similarity transformation so that the kinetic energy operator is
transformed into a sum of two types of terms, namely, derivative
terms {\em and} a repulsive centrifugal-like term. The latter
repulsive centrifugal-like term stabilizes the system against
collapse in the large-$D$ limit when attractive interparticle
potentials are present. The zeroth and first orders of the
dimensional ($1/D$) expansion of the similarity-transformed
Schr\"{o}dinger equation are then exactly soluble for any value of
$N$. In the $D\to\infty$ limit, the derivative terms drop out,
resulting in a static problem at zeroth order, while first order
corrections correspond to simple-harmonic normal-mode oscillations
about the infinite-dimensional structure.

In Paper~I the weight function was chosen to be the square root
of the inverse of the Jacobian, $J$, where\cite{avery}
\begin{equation} \label{eq:Jac}
J = (r_1 r_2 \ldots r_N)^{D-1} \Gamma^{(D-N-1)/2}
\end{equation}
and $\Gamma$ is the Gramian determinant, the determinant of the
matrix whose elements are $\gamma_{ij}$ (see Appendix~B of
Paper~I), so that the similarity-transformed wave function
$_{(i)}\Phi$ and operators $_{(i)}\widetilde{O}$ are
\begin{equation}
\label{eq:simtransf} _{(i)}\Phi = J^{\frac{1}{2}} \, \Psi, \;\;
\mbox{and} \;\; _{(i)}\widetilde{O}= J^{\frac{1}{2}} \,
\widehat{O} \, J^{-\frac{1}{2}} \,.
\end{equation}
Under this Jacobian transformation, a first derivative of an
internal coordinate is the conjugate momentum to that coordinate.
The matrix elements of coordinates and their derivatives
between the zeroth-order normal-mode functions, which are involved
in the development of higher-order DPT expansions, are much easier
to calculate since the weight function in the integrals is now
unity.

Carrying out the transformation of the Schr\"{o}dinger equation of
Eq.~(\ref{generalH}) via Eqs.~(\ref{eq:Jac}) and
(\ref{eq:simtransf}), we obtain\cite{avery}:
\begin{equation}
 (_{(i)}T+V)\, {_{(i)}}\Phi = E \,\, {_{(i)}}\Phi \,,
\label{eq:SE}
\end{equation}
where
\begin{eqnarray}
_{(i)}T&=& {\displaystyle \hbar^2
\sum\limits_{i=1}^{N}\Biggl[-\frac{1}{2
m_i}\frac{\partial^2}{{\partial r_i}^2}- \frac{1}{2 m_i r_i^2}
\Bigg(
\sum\limits_{j\not=i}\sum\limits_{k\not=i}(\gamma_{jk}-\gamma_{ij}
\gamma_{ik})
\frac{\partial^2}{\partial\gamma_{ij}\partial\gamma_{ik}} -
N\sum\limits_{j\not=i} \gamma_{ij}
\frac{\partial}{\partial\gamma_{ij}} \Bigg)} \nonumber \\
&& {\displaystyle +\frac{N(N-2)+(D-N-1)^2 \left(
\frac{\Gamma^{(i)}}{\Gamma} \right) }{8 m_i r_i^2} \Biggr] }
\nonumber \\
&=& {\displaystyle \hbar^2 \sum\limits_{i=1}^{N}\Biggl[-\frac{1}{2
m_i}\frac{\partial^2}{{\partial r_i}^2}- \frac{1}{2 m_i r_i^2}
\sum\limits_{j\not=i}\sum\limits_{k\not=i}
\frac{\partial}{\partial\gamma_{ij}}(\gamma_{jk}-\gamma_{ij}
\gamma_{ik})
\frac{\partial}{\partial\gamma_{ik}}} \nonumber \\
&& {\displaystyle +\frac{N(N-2)+(D-N-1)^2 \left(
\frac{\Gamma^{(i)}}{\Gamma} \right) }{8 m_i r_i^2} \Biggr]\,, }
 \label{eq:SE_T}
\end{eqnarray}
and
\begin{equation}
V=\sum\limits_{i=1}^{N}V_{\mathtt{conf}}(r_i)+
\sum\limits_{i=1}^{N-1}\sum\limits_{j=i+1}^{N}
V_{\mathtt{int}}(r_{ij}) \,.
\end{equation}
The latter expression for $_{(i)}T$ is explicitly self-adjoint
since the weight function, $W$, for the matrix elements is equal
to unity. The similarity-transformed Hamiltonian for the energy
eigenstate $_{(i)}\Phi$ is $ _{(i)}H=( _{(i)}T+V)$.

\section{Infinite-${\bm{D}}$ analysis: Leading order
energy}\label{sec:infD}

As in Paper I, we begin the perturbation analysis by regularizing the
large-dimension limit of the Schr\"odinger equation by defining
dimensionally-scaled variables:
\begin{equation} \label{eq:kappascale}
\bar{r}_i = r_i/\kappa(D) \;\;\; , \;\;\; \bar{E} = \kappa(D) E
\;\;\; \mbox{and} \;\;\; \bar{H} = \kappa(D) \,\, {_{(i)}H}
\end{equation}
where $\kappa(D)$ is a dimension-dependent scale factor.  From
Eq.~(\ref{eq:SE_T}) the kinetic energy T scales
in the same way as $1/r^2$, so the scaled version of
Eq.~(\ref{eq:SE}) becomes
\begin{equation} \label{eq:scale1}
\bar{H} \Phi =
\left(\frac{1}{\kappa(D)}\bar{T}+\bar{V}_{\mathtt{eff}}
\right)\Phi = \bar{E} \Phi,
\end{equation}
where barred quantities indicate that the variables are now
in scaled units.
The centrifugal-like term in T of Eq.~(\ref{eq:SE_T}) has
 quadratic $D$ dependence so the scale factor $\kappa(D)$
 must also be quadratic in $D$,
otherwise the $D\to\infty$ limit of the Hamiltonian would not be
finite. The precise form of $\kappa(D)$ depends on the particular system
and is chosen so that the
result of the scaling is as simple as possible.
In previous work\cite{FGpaper} we have chosen
$\kappa(D)=(D-1)(D-2N-1)/(4Z)$ for the $S$-wave, $N$-electron
atom; $\Omega \, l_{\mathtt{ho}}$ for the $N$-electron quantum dot
where $\Omega=(D-1)(D-2N-1)/4$ and the dimensionally-scaled
harmonic oscillator length and trap frequency respectively are
${l}_{\mathtt{ho}}=\sqrt{\frac{\hbar}{m^*\bar{\omega}_{\mathtt{ho}}}}$
and $\bar{\omega}^2_{\mathtt{ho}}=\Omega^3
{\omega}^2_{\mathtt{ho}}$\,; and $D^2 \bar{a}_{\mathtt{ho}}$ for
the BEC where $\bar{a}_{\mathtt{ho}}=\sqrt{\frac{\hbar}{m
\bar{\omega}_{\mathtt{ho}}}}$ and
${\bar{\omega}_{\mathtt{ho}}}=D^3{\omega_{\mathtt{ho}}}$\,. The
factor of $\kappa(D)$ in the denominator of Eq.~(\ref{eq:scale1})
suppresses the derivative terms as $D$ increases leaving behind a
centrifugal-like term in an effective potential,

\begin{equation}
\label{veff}
\bar{V}_{\mathtt{eff}}(\bar{r},\gamma;\delta=0)=\sum\limits_{i=1}^{N}\left(\frac{\hbar^2}{8
m_i
\bar{r}_i^2}\frac{\Gamma^{(i)}}{\Gamma}+\bar{V}_{\mathtt{conf}}(\bar{r},\gamma;\delta=0)\right)
+\sum\limits_{i=1}^{N-1}\sum\limits_{j=i+1}^{N}
\bar{V}_{\mathtt{int}}(\bar{r},\gamma;\delta=0)\,,
\end{equation}
with $\delta=1/D$, in which the particles become frozen at large
$D$. In the $D\to\infty$ ($\delta \to 0$) limit, the excited
states collapse onto the ground state at the
minimum of $V_{\mathtt{eff}}$.

We assume a totally symmetric minimum characterized by the
equality of all radii and angle cosines of the particles when
$D\to\infty$, i.e.\
\begin{equation}
\bar{r}_{i}=\bar{r}_{\infty} \;\; (1 \le i \le N), \;\;\;\;
\gamma_{ij}=\overline{\gamma}_{\infty} \;\; (1 \le i < j \le N).
\end{equation}
In scaled units the zeroth-order ($D\to\infty$) approximation for
the energy becomes
\begin{equation}
\label{zeroth}
\bar{E}_{\infty}=\bar{V}_{\mathtt{eff}}(\bar{r}_{\infty},\overline{\gamma}_{\infty};
\hspace{1ex} \delta=0).
\end{equation}
In this leading order approximation, the centrifugal-like term
that appears in $\bar{V}_{\mathtt{eff}}$, even for the ground
state, is a zero-point energy contribution satisfying the minimum
uncertainty principle\cite{chat}.

\section{Normal-mode analysis and the ${\bm{1/D}}$ first-order
quantum energy correction}\label{sec:firstorder}

At zeroth-order, the particles are frozen in a
completely symmetric, high-$D$ configuration which is
somewhat analogous to the Lewis structure in atomic physics
terminology. Similarly, the first-order $1/D$ correction can be
viewed as small oscillations of this structure, analogous to
Langmuir oscillations. To obtain the $1/D$ quantum correction to
the energy for large but finite values of $D$, we expand about the
minimum of the $D\to\infty$ effective potential. We first define a
position vector, consisting of all $N(N+1)/2$ internal
coordinates:
\begin{equation}\label{eq:ytranspose}
{\bar{\bm{y}}} = \left( \begin{array}{c} \bar{\bm{r}} \\
\bm{\gamma} \end{array} \right) \,,
\end{equation}
where
\begin{equation}
\bar{\bm{r}} = \left(
\begin{array}{c}
\bar{r}_1 \\
\bar{r}_2 \\
\vdots \\
\bar{r}_N
\end{array}
\right)
\end{equation}
and
\begin{equation}
\bm{\gamma} = \left(
\begin{array}{c}
\gamma_{12} \\ \cline{1-1}
\gamma_{13} \\
\gamma_{23} \\ \cline{1-1}
\gamma_{14} \\
\gamma_{24} \\
\gamma_{34} \\ \cline{1-1}
\gamma_{15} \\
\gamma_{25} \\
\vdots \\
\gamma_{N-2,N} \\
\gamma_{N-1,N} \end{array} \right) \,.
\end{equation}
Making the following substitutions for all radii and angle
cosines:
\begin{eqnarray}
\label{eq:taylor1}
&&\bar{r}_{i} = \bar{r}_{\infty}+\delta^{1/2}\bar{r}'_{i}\\
&&\gamma_{ij} =
\overline{\gamma}_{\infty}+\delta^{1/2}\overline{\gamma}'_{ij}
\label{eq:taylor2},
\end{eqnarray}
where $\delta=1/D$ is the expansion parameter,
we define a displacement vector consisting of the internal
displacement coordinates [primed in Eqs. (\ref{eq:taylor1}) and
(\ref{eq:taylor2})]
\begin{equation}\label{eq:ytransposeP}
{\bar{\bm{y}}'} = \left( \begin{array}{c} \bar{\bm{r}}' \\
\overline{\bm{\gamma}}' \end{array} \right) \,,
\end{equation}
where
\begin{equation}\label{eq:bfrp}
\bar{\bm{r}}' = \left(
\begin{array}{c}
\bar{r}'_1 \\
\bar{r}'_2 \\
\vdots \\
\bar{r}'_N
\end{array}
\right)\,,
\end{equation}
and
\begin{equation}\label{eq:bfgammap}
\overline{\bm{\gamma}}' = \left(
\begin{array}{c}
\overline{\gamma}'_{12} \\ \cline{1-1}
\overline{\gamma}'_{13} \\
\overline{\gamma}'_{23} \\ \cline{1-1}
\overline{\gamma}'_{14} \\
\overline{\gamma}'_{24} \\
\overline{\gamma}'_{34} \\ \cline{1-1}
\overline{\gamma}'_{15} \\
\overline{\gamma}'_{25} \\
\vdots \\
\overline{\gamma}'_{N-2,N} \\
\overline{\gamma}'_{N-1,N} \end{array} \right) \,.
\end{equation}
We find
\begin{equation}
\label{Taylor} \bar{V}_{\mathtt{eff}}({\bar{\bm{y}}'};
\hspace{1ex} \delta)=\left[ \bar{V}_{\mathtt{eff}}
\right]_{\delta^{1/2}=0} + \frac{1}{2} \delta \left\{
\sum\limits_{\mu=1}^{P} \sum\limits_{\nu=1}^{P} \bar{y}'_{\mu}
\left[\frac{\partial^2 \bar{V}_{\mathtt{eff}}}{\partial
\bar{y}_{\mu}
\partial \bar{y}_{\nu}}\right]_{\delta^{1/2}=0} \bar{y}'_{\nu} + v_o \right\} +
O\left(\delta^{3/2}\right) \,,
\end{equation}
where
\begin{equation}
P \equiv N(N+1)/2
\end{equation}
is the number of internal coordinates and $v_o$ is a constant which
comes from the explicit $\delta$ dependence in the centrifugal term.
The first term of the
$O((\delta^{1/2})^2)$ term defines the elements of the Hessian
matrix\cite{strang} $\bm{F}$ of Eq.~(\ref{Gham}) below. The
derivative terms in the kinetic energy are taken into account by a
similar series expansion, beginning with a first-order term that
is bilinear in ${\partial/\partial \bar{y}'}$, i.e.\,
\begin{equation}\label{eq:T}
{\mathcal T}=-\frac{1}{2} \delta \sum\limits_{\mu=1}^{P}
\sum\limits_{\nu=1}^{P} {G}_{\mu\nu}
\partial_{\bar{y}'_{\mu}}
\partial_{\bar{y}'_{\nu}} + O\left(\delta^{3/2}\right),
\end{equation}
where ${\mathcal T}$ is the derivative portion of the kinetic
energy $T$ [see Eq.~(\ref{eq:SE_T})]. It follows from Eqs.
(\ref{Taylor}) and (\ref{eq:T}) that $\bm{G}$ and $\bm{F}$, both
constant matrices, are defined in the first-order $\delta=1/D$
Hamiltonian as follows:
\begin{equation}\label{Gham}
\widehat{H}_1=-\frac{1}{2} {\partial_{\bar{y}'}}^{T} \bm{G}
{\partial_{\bar{y}'}} + \frac{1}{2} \bar{\bm{y}}^{\prime T} {\bm
F} {{\bar{\bm{y}}'}} + v_o \,.
\end{equation}
Thus, obtaining the first-order energy correction is reduced to a
harmonic problem, which is solved by obtaining the normal modes of
the system.

We use the FG matrix method\cite{dcw} to obtain the
normal-mode vibrations and, thereby, the first-order energy
correction. A derivation of the FG matrix method may be found in Appendix A of
Paper~I, but the main results may be stated as follows. The $b^{\rm th}$
normal mode coordinate may be written as:
\begin{equation} \label{eq:qyt}
[{\bm q'}]_b = {\bm{b}}^T {\bar{\bm{y}}'} \,,
\end{equation}
where the coefficient vector ${\bm{b}}$ satisfies the eigenvalue
equation,
\begin{equation} \label{eq:FGit}
\bm{F} \, \bm{G} \, {\bm{b}} = \lambda_b \, {\bm{b}}\,,
\end{equation}
with the resultant secular equation
\begin{equation} \label{eq:character}
\det(\bm{F}\bm{G}-\lambda\bm{I})=0.
\end{equation}
The coefficient vector also satisfies the normalization condition
\begin{equation} \label{eq:normit}
{\bm{b}}^T \bm{G} \, {\bm{b}} = 1.
\end{equation}
The frequencies $\bar{\omega}_b^2$ are related to $\lambda$ by:
\begin{equation}\label{eq:omega_b}
\lambda_b=\bar{\omega}_b^2,
\end{equation}
while the wave function is a product of $P = N(N+1)/2$ harmonic
oscillator wave functions
\begin{equation}
\Phi_0({\bar{\bm{y}}'}) = \prod_{b=1}^{P} h_{n_b}\left(
\bar{\omega}^{1/2}_b [{\bm q'}]_b \right) \,, \label{eq:Phi_0}
\end{equation}
where $h_{n_b}\left( \bar{\omega}^{1/2}_b [{\bm q'}]_b \right)$ is
a one-dimensional harmonic-oscillator wave function of frequency
$\bar{\omega}_b$, and $n_{b}$ is the oscillator quantum number, $0
\leq n_{b} < \infty$, which counts the number of quanta in each
normal mode.

In an earlier paper\cite{FGpaper} we solve
Eqs.~(\ref{eq:character}) and (\ref{eq:omega_b}) for the
frequencies. The number of roots $\lambda$ of
Eq.~(\ref{eq:character}) -- there are $P \equiv N(N+1)/2$ roots
-- is potentially huge. However, due to the $S_N$ symmetry of the
problem discussed in Sect.~5 of Paper~I, there is a stunning
simplification. Equation~(\ref{eq:character}) has only five
distinct roots, $\lambda_{\mu}$, where $\mu$ runs over ${\bf
0}^-$, ${\bf 0}^+$, ${\bf 1}^-$, ${\bf 1}^+$, and ${\bf 2}$,
regardless of the number of particles in the system (see
Refs.~\cite{FGpaper} and Sect.~\ref{subsec:eigreduct}). Thus the
energy through first-order (see Eq.~(\ref{eq:E1})) can be written
in terms of the five distinct normal-mode vibrational frequencies
which are related to the roots $\lambda_{\mu}$ of $FG$ by
\begin{equation}\label{eq:omega_p}
\lambda_{\mu}=\bar{\omega}_{\mu}^2 \,.
\end{equation}
The energy through first-order in $\delta = 1/D$ is then\cite{FGpaper}
\begin{equation}
\overline{E} = \overline{E}_{\infty} + \delta \Biggl[
\sum_{\renewcommand{\arraystretch}{0}
\begin{array}[t]{r@{}l@{}c@{}l@{}l} \scriptstyle \mu = \{
  & \scriptstyle \bm{0}^\pm,\hspace{0.5ex}
  & \scriptstyle \bm{1}^\pm & , & \\
  & & \scriptstyle \bm{2} & & \scriptstyle  \}
            \end{array}
            \renewcommand{\arraystretch}{1} }
\hspace{-0.50em} \sum_{\mathsf{n}_{\mu}=0}^\infty
({\mathsf{n}}_{\mu}+\frac{1}{2}) d_{\mu,\mathsf{n}_{\mu}}
\bar{\omega}_{\mu} \, + \, v_o \Biggr] \,, \label{eq:E1}
\end{equation}
where the $\mathsf{n}_{\mu}$ are the vibrational quantum numbers
of the normal modes with the same frequency $\bar{\omega}_{\mu}$.
 The quantity $d_{\mu,\mathsf{n}_{\mu}}$ is the
occupancy of the manifold of normal modes with vibrational quantum
number $\mathsf{n}_{\mu}$ and normal mode frequency
$\bar{\omega}_{\mu}$.  The total occupancy of the normal
modes with frequency $\bar{\omega}_{\mu}$ is equal to the
multiplicity of the root $\lambda_{\mu}$, i.e.\
\begin{equation}
d_{\mu} = \sum_{\mathsf{n}_{\mu}=0}^\infty
            d_{\mu,\mathsf{n}_{\mu}} \,,
\end{equation}
where $d_{\mu}$ is the multiplicity of the $\mu^{th}$ root. The
multiplicities of the five roots are\cite{FGpaper}
\begin{eqnarray}
d_{{\bf 0}^+} &=& 1 \,,\nonumber\\
d_{{\bf 0}^-} &=& 1 \,,\nonumber\\
d_{{\bf 1}^+} &=& N-1 \,,  \label{eq:dpm} \\
d_{{\bf 1}^-} &=& N-1 \,,\nonumber\\
d_{{\bf 2}} &=& N(N-3)/2 \,.\nonumber
\end{eqnarray}
Note that although the equation in Ref.~\cite{loeser} for the
energy through Langmuir order is the same as Eq.~(\ref{eq:E1}), it
is expressed a little differently (See Ref. \cite{different}).

\section{The Symmetry of the Large-${\bm{D}}$, ${\bm{N}}$-body Quantum
Confinement Problem} \label{sec:symm}

\subsection{$\bm{Q}$ matrices in terms of simple invariant submatrices}
\label{subsec:Qsubm} Such a high degree of degeneracy of the
frequencies in large-$D$, $N$-body quantum confinement problem
indicates a high degree of symmetry. The $\bm{F}$, $\bm{G}$, and
$\bm{FG}$ matrices, which we generically denote by $\bm{Q}$, are
$P \times P$ matrices. The $S_N$ symmetry of the $\bm{Q}$ matrices
($\bm{F}$, $\bm{G}$, and $\bm{FG}$) allows us to write these
matrices in terms of six simple submatrices which are invariant
under $S_N$. We first define the number of $\gamma_{ij}$
coordinates to be
\begin{equation}\label{eq:M}
M \equiv N(N-1)/2,
\end{equation}
and let $\bm{I}_N$ be an $N \times N$ identity matrix, $\bm{I}_M$
an $M \times M$ identity matrix, $\bm{J}_N$ an $N \times N$ matrix
of ones and $\bm{J}_M$ an $M \times M$ matrix of ones. Further, we
let $\bm{R}$ be an $N \times M$ matrix such that
${R}_{i,jk}=\delta_{ij}+\delta_{ik}$, $\bm{J}_{NM}$ be an $N
\times M$ matrix of ones, and $\bm{J}^T_{NM}=\bm{J}_{MN}$. These
matrices are invariant under interchange of the particles, the
$S_N$ symmetry, and form a closed algebra (see Appendix B of
Ref.~\cite{FGpaper}).

We can then write the $\bm{Q}$ matrices as
\begin{equation}\label{eq:Q}
\bm{Q}=\left(\begin{array}{cc} \bm{Q}_{\bar{\bm{r}}'
\bar{\bm{r}}'} & \bm{Q}_{\bar{\bm{r}}' \overline{\bm{\gamma}}'}
\\ \bm{Q}_{\overline{\bm{\gamma}}' \bar{\bm{r}}'} & \bm{Q}_{\overline{\bm{\gamma}}' \overline{\bm{\gamma}}'}
\end{array}\right),
\end{equation}
where the block $\bm{Q}_{\bar{\bm{r}}' \bar{\bm{r}}'}$ has
dimension $(N \times N)$, block $\bm{Q}_{\bar{\bm{r}}'
\overline{\bm{\gamma}}'}$ has dimension $(N \times M)$, block
$\bm{Q}_{\overline{\bm{\gamma}}' \bar{\bm{r}}'}$ has dimension $(M
\times N)$, and block $\bm{Q}_{\overline{\bm{\gamma}}'
\overline{\bm{\gamma}}'}$ has dimension $(M \times M)$.  Now, as
we show in Appendix~B of Ref.~\cite{FGpaper}, we can write the
$\bm{Q}$ matrices as follows:
\begin{eqnarray}
\bm{Q}_{\bar{\bm{r}}'
\bar{\bm{r}}'} & = & (Q_a-Q_b) \bm{I}_N + Q_b \bm{J}_N \label{eq:Qrr}\\
\bm{Q}_{\bar{\bm{r}}'
\overline{\bm{\gamma}}'} & = & (Q_e-Q_f) \bm{R} + Q_f \bm{J}_{NM} \label{eq:Qrg} \\
\bm{Q}_{\overline{\bm{\gamma}}' \bar{\bm{r}}'}
& = & (Q_c-Q_d) \bm{R}^T + Q_d \bm{J}_{NM}^T \label{eq:Qgr} \\
\bm{Q}_{\overline{\bm{\gamma}}' \overline{\bm{\gamma}}'} & = &
(Q_g-2Q_h+Q_{\iota}) \bm{I}_M + (Q_h-Q_{\iota}) \bm{R}^T \bm{R} +
Q_{\iota} \bm{J}_M \,. \label{eq:Qgg}
\end{eqnarray}
It is this structure that results in the remarkable reduction from
a possible $P=N(N+1)/2$ distinct frequencies to just five distinct
frequencies.

For $\bm{Q}=\bm{FG}$, the matrix that must be
diagonalized in Eq.~(\ref{eq:FGit}), Eq.~(\ref{eq:Q}) becomes
\begin{equation} \label{GFsub}
\bm{FG}= \left(
\begin{array}{cc}
\tilde{a} \bm{I}_N + \tilde{b} \bm{J}_N & \tilde{e} \bm{R} + \tilde{f} \bm{J}_{NM} \\
\tilde{c} \bm{R}^T + \tilde{d} \bm{J}_{MN} & \tilde{g} \bm{I}_M +
\tilde{h} \bm{R}^T \bm{R} + \tilde{\iota} \bm{J}_M
\end{array}\right) \,,
\end{equation}
where we have used the following abbreviations:
\begin{eqnarray}  \label{GFsym}
\tilde{a} & \equiv & ({FG})_{a} - ({FG})_{b} = ({F}_{a} - {F}_{b}) {G}_{a}   \nonumber \\
\tilde{b} & \equiv & ({FG})_{b} = {F}_{b}{G}_{a}   \nonumber\\
\tilde{c} & \equiv & ({FG})_{c} - ({FG})_{d} = ({F}_{e} - {F}_{f}) {G}_{a}   \nonumber \\
\tilde{d} & \equiv & ({FG})_{d} = {F}_{f}{G}_{a}   \nonumber\\
\tilde{e} & \equiv & ({FG})_{e} - ({FG})_{f} = ({F}_{e} - {F}_{f}) ({G}_{g}+(N-4){G}_{h}) \\
\tilde{f} & \equiv & ({FG})_{f} = 2{F}_{e}{G}_{h} +  {F}_{f}({G}_{g} + 2(N-3){G}_{h})    \nonumber \\
\tilde{g} & \equiv & ({FG})_{g} - 2({FG})_{h} + ({FG})_{\iota}
= ({F}_{g} - 2{F}_{h} + {F}_{\iota}) ({G}_{g} - 2{G}_{h}) \nonumber \\
\tilde{h} & \equiv & ({FG})_{h} - ({FG})_{\iota} =
{F}_{g}{G}_{h}+{F}_{h}({G}_{g}+(N-6){G}_{h})
- {F}_{\iota}({G}_{g} + (N-5){G}_{h}) \nonumber\\
\tilde{\iota} & \equiv & ({FG})_{\iota} =
4{F}_{h}{G}_{h}+{F}_{\iota}({G}_{g}+2(N-4){G}_{h}) \,. \nonumber
\end{eqnarray}
The $\bm{F}$ and $\bm{G}$ matrix elements on the right-hand sides
of Eq.~(\ref{GFsym})
may be derived using the graph-theoretic techniques discussed in
Appendix~B of Ref.~\cite{FGpaper}.

We also require the $\bm{G}$ matrix for the normalization
condition (Eq.~(\ref{eq:normit})). It has a simpler structure than
the $\bm{FG}$ matrix,
\begin{equation} \label{eq:Gsub}
\bm{G} = \left( \begin{array}{cc}
\tilde{a}' \bm{I}_N & \bm{0} \\
\bm{0} & \tilde{g}' \bm{I}_M + \tilde{h}' \bm{R}^T \bm{R}
\end{array} \right) \,,
\end{equation}
where
\begin{eqnarray}  \label{eq:Gsym}
\tilde{a}' & \equiv & ({G})_{a} \nonumber \\
\tilde{g}' & \equiv & ({G})_{g} - 2({G})_{h} \\
\tilde{h}' & \equiv & ({G})_{h} \nonumber
\end{eqnarray}
and
\begin{eqnarray} \label{eq:Goneorzero}
({G})_{b} & = & 0 \nonumber \\
({G})_{c} & = & 0 \nonumber \\
({G})_{d} & = & 0  \\
({G})_{e} & = & 0 \nonumber \\
({G})_{f} & = & 0 \nonumber \\
({G})_{\iota} & = & 0 \,. \nonumber
\end{eqnarray}
The quantities $({G})_{a}$, $({G})_{g}$ and $({G})_{h}$ depend on
the choice of $\kappa(D)$ (See Ref.~\cite{FGpaper}).

\section{Symmetry and Normal Coordinates} \label{sec:symnorm}
As discussed in Paper I the
$\bm{FG}$ matrix is a $N(N+1)/2 \times N(N+1)/2$ dimensional
matrix (there being $N(N+1)/2$ internal coordinates), and so
Eqs.~(\ref{eq:FGit}) and (\ref{eq:character}) could have up to
$N(N+1)/2$ distinct frequencies. However, as noted above, there
are only five distinct frequencies. The $S_N$ symmetry is
responsible for the remarkable reduction from $N(N+1)/2$ possible
distinct frequencies to five actual distinct frequencies. As we
shall also see, the $S_N$ symmetry greatly simplifies the
determination of the normal coordinates and hence the solution of
the large-$D$ problem.

\subsection{Symmetrized coordinates} \label{subsec:symCoor}
The $\bm{Q}$ matrices, and in particular the
$\bm{FG}$ matrix, are invariants under $S_N$, so they do not connect
subspaces belonging to different irreducible representations of
$S_N$\cite{WDC}. Thus from Eqs.~(\ref{eq:qyt}) and (\ref{eq:FGit})
the normal coordinates must transform under irreducible
representations of $S_N$\,. Since the normal coordinates will
be linear combinations of the elements of the internal coordinate
displacement vectors $\bar{\bm{r}}'$ and
$\overline{\bm{\gamma}}'$\,, we first look at the $S_N$
transformation properties of the internal coordinates.

The internal coordinate displacement vectors
$\bar{\bm{r}}'$ and $\overline{\bm{\gamma}}'$ of
Eqs.~(\ref{eq:bfrp}) and (\ref{eq:bfgammap}) are basis functions
which transform under matrix representations of $S_N$, and each
span the corresponding carrier spaces, however these representations of
$S_N$  are not irreducible representations  of $S_N$\,.

In Sec.~6.1 of Paper~I we have shown that the reducible
representation under which $\bar{\bm{r}}'$ transforms is reducible
to one $1$-dimensional irreducible representation labelled by the
partition $[N]$ (the partition denotes a corresponding Young
diagram ( = Young pattern = Young shape) of an irreducible
representation (see Appendix~C of Paper~I) and one
$(N-1)$-dimensional irreducible representation labelled by the
partition $[N-1, \hspace{1ex} 1]$. We also showed in Sec 6.1 of Paper I
that the
reducible representation under which $\overline{\bm{\gamma}}'$
transforms is reducible to one $1$-dimensional irreducible
representation labelled by the partition $[N]$, one
$(N-1)$-dimensional irreducible representation labelled by the
partition $[N-1, \hspace{1ex} 1]$ and one $N(N-3)/2$-dimensional
irreducible representation labelled by the partition $[N-2,
\hspace{1ex} 2]$. Thus if $d_\alpha$ is the dimensionality of the
irreducible representation of $S_N$ denoted by the partition
$\alpha$ then
\begin{equation}
\renewcommand{\arraystretch}{1.5}
\begin{array}{rcl}
d_{[N]} & = & 1 \\
d_{[N-1, \hspace{1ex} 1]} & = & N-1 \\
d_{[N-2, \hspace{1ex} 2]} & = & \displaystyle \frac{N(N-3)}{2}
\end{array}
\renewcommand{\arraystretch}{1}
\end{equation}
and so
\begin{equation}
d_{[N]} + d_{[N-1, \hspace{1ex} 1]} = N \,,
\end{equation}
giving correctly the dimension of the $\bar{\bm{r}}'$ vector and
that
\begin{equation}
d_{[N]} + d_{[N-1, \hspace{1ex} 1]} + d_{[N-2, \hspace{1ex} 2]} =
\frac{N(N-1)}{2} \,,
\end{equation}
giving correctly the dimension of the $\overline{\bm{\gamma}}'$
vector.

Since the normal modes transform under irreducible representations
of $S_N$ and are composed of linear combinations of the elements
of the internal coordinate displacement vectors $\bar{\bm{r}}'$
and $\overline{\bm{\gamma}}'$\,, there will be two $1$-dimensional
irreducible representations labelled by the partition $[N]$, two
$(N-1)$-dimensional irreducible representations labelled by the
partition $[N-1, \hspace{1ex} 1]$ and one entirely angular
$N(N-3)/2$-dimensional irreducible representation labelled by the
partition $[N-2, \hspace{1ex} 2]$. Thus if we look at
Eq.~(\ref{eq:dpm}) we see that the ${\bf 0}^\pm$ normal modes
transform under two $[N]$ irreducible representations, the ${\bf
1}^\pm$ normal modes transform under two $[N-1, \hspace{1ex} 1]$
irreducible representations, while the ${\bf 2}$ normal modes
transform under the $[N-2, \hspace{1ex} 2]$ irreducible
representation.

\section{Calculating the Normal Modes.} \label{sec:NormCalc}
\subsection{An Outline of the Program to Calculate the Normal Modes of the System.}
\label{subsec:ProgOutline} Summarizing the procedure in Paper I,
we determine the normal coordinates in a three-step process:
\newcounter{twostep}
\newcounter{twostepseca}
\newcounter{twostepsecb}
\newcounter{twostepsecc}
\begin{list}{\alph{twostep}).}{\usecounter{twostep}\setlength{\rightmargin}{\leftmargin}}
\item \setcounter{twostepseca}{\value{twostep}} We determine the
  primitive irreducible coordinates by the following procedure. First,
  we determine two sets
of linear combinations of the elements of coordinate vector
$\bar{\bm{r}}'$ which transform under particular non-orthogonal
$[N]$ and $[N-1, \hspace{1ex} 1]$ irreducible representations of
$S_N$\,. Using these two sets of coordinates we then determine two
sets of linear combinations of the elements of coordinate vector
$\overline{\bm{\gamma}}'$ which transform under exactly the same
non-orthogonal $[N]$ and $[N-1, \hspace{1ex} 1]$ irreducible
representations of $S_N$ as the coordinate sets in the
$\bar{\bm{r}}'$ sector. Finally, another set of linear combinations of
the elements of coordinate vector $\overline{\bm{\gamma}}'$ which
transforms under a particular non-orthogonal $[N-2, \hspace{1ex}
2]$ irreducible representation of $S_N$ is determined.
In particular we set out
to find sets of coordinates transforming irreducibly under $S_N$
which have the simplest functional form possible subject to the
requirement that they transform irreducibly under $S_N$. We call
these coordinates primitive irreducible coordinates.
\item \setcounter{twostepsecb}{\value{twostep}} Appropriate linear
combinations within each coordinate set from
item~\alph{twostepseca}).\ above are taken so that the results
transform under orthogonal irreducible representations of $S_N$\,.
These are then the symmetry coordinates of the problem\cite{dcw}.
Care is taken to ensure that the transformation from the
coordinates which transform under the non-orthogonal irreducible
representations of $S_N$ of item~\alph{twostepseca}).\ above, to
the symmetry coordinates which transform under orthogonal
irreducible representations of $S_N$ preserve the identity of
equivalent representations in the $\bar{\bm{r}}'$ and
$\overline{\bm{\gamma}}'$ sectors. Furthermore, we choose one of
the symmetry coordinates to be just a single primitive irreducible
coordinate so that it has the simplest functional form possible
under the requirement that it transforms irreducibly under $S_N$.
The succeeding symmetry coordinate is then chosen to be composed
of two primitive irreducible coordinates and so on. In this way
the complexity of the functional form of the symmetry coordinates
is kept to a minimum and only builds up slowly as more symmetry
coordinates of a given species are considered.
\item  \setcounter{twostepsecc}{\value{twostep}} The $\bm{FG}$
matrix is expressed in the $\bar{\bm{r}}'$,
$\overline{\bm{\gamma}}'$ basis. However, if we change the basis in
which the $\bm{FG}$ matrix is expressed to the symmetry
coordinates an enormous simplification occurs. The $N(N+1)/2
\times N(N+1)/2$ eigenvalue equation of Eq.~(\ref{eq:FGit}) is
reduced to one $2 \times 2$ eigenvalue equation for the $[N]$
sector, $N-1$ identical $2 \times 2$ eigenvalue equations for the
$[N-1, \hspace{1ex} 1]$ sector and $N(N-3)/2$ identical $1 \times
1$ eigenvalue equations for the $[N-2, \hspace{1ex} 2]$ sector. In
the case of the $2 \times 2$ eigenvalue equations for the $[N]$
and $[N-1, \hspace{1ex} 1]$ sectors, the $2
\times 2$ structure allows for the mixing in the normal
coordinates of the symmetry coordinates in the $\bar{\bm{r}}'$ and
$\overline{\bm{\gamma}}'$ sectors. The $1 \times 1$ structure of
the eigenvalue equations in the $[N-2, \hspace{1ex} 2]$ sector
reflects the fact that there are no $[N-2, \hspace{1ex} 2]$
symmetry coordinates in the $\bar{\bm{r}}'$ sector for the $[N-2,
\hspace{1ex} 2]$ symmetry coordinates in the
$\overline{\bm{\gamma}}'$ sector to couple with. The $[N-2, \hspace{1ex}
2]$ normal modes are entirely angular.
\end{list}

\subsection{The Primitive Irreducible Coordinate Vectors ${\bm{\overline{\bm{S}}_{\bar{\bm{r}}'}}}$
and ${\bm{\overline{\bm{S}}_{\overline{\bm{\gamma}}'}}}$\,.} The
primitive irreducible coordinates,
$\overline{\bm{S}}_{\bar{\bm{r}}'}$, of the $\bar{\bm{r}}'$ sector
which transform under irreducible, though non-orthogonal,
representations of the group $S_N$ are given by
\begin{equation}\label{eq:sbwbr}
\overline{\bm{S}}_{\bar{\bm{r}}'} = \overline{W}_{\bar{\bm{r}}'}
\bar{\bm{r}}' \,,
\end{equation}
where
\begin{equation}\label{eq:wbr}
\overline{W}_{\bar{\bm{r}}'} = \left( \begin{array}{l} \overline{W}_{\bar{\bm{r}}'}^{[N]} \\
\overline{W}_{\bar{\bm{r}}'}^{[N-1, \hspace{1ex} 1]}
\end{array} \right) \,,
\end{equation}
$\overline{W}_{\bar{\bm{r}}'}^{[N]}$ is a $1 \times N$ dimensional
matrix and $\overline{W}_{\bar{\bm{r}}'}^{[N-1, \hspace{1ex} 1]}$
is an $(N-1) \times N$ dimensional matrix. Hence we can write
\begin{equation}\label{eq:sbr}
\overline{\bm{S}}_{\bar{\bm{r}}'} = \left( \begin{array}{l} \overline{\bm{S}}_{\bar{\bm{r}}'}^{[N]} \\
\overline{\bm{S}}_{\bar{\bm{r}}'}^{[N-1, \hspace{1ex} 1]} \end{array} \right) = \left( \begin{array}{l} \overline{W}_{\bar{\bm{r}}'}^{[N]} \bar{\bm{r}}' \\
\overline{W}_{\bar{\bm{r}}'}^{[N-1, \hspace{1ex} 1]} \bar{\bm{r}}'
\end{array} \right) \,.
\end{equation}

Likewise the primitive irreducible coordinates,
$\overline{\bm{S}}_{\overline{\bm{\gamma}}'}$, of the
$\overline{\bm{\gamma}}'$ sector which transform under
irreducible, though non-orthogonal, representations of the group
$S_N$ are given by
\begin{equation}\label{eq:sbwbgamma}
\overline{\bm{S}}_{\overline{\bm{\gamma}}'} =
\overline{W}_{\overline{\bm{\gamma}}'} \, \overline{\bm{\gamma}}'
\end{equation}
and so writing
\begin{equation}\label{eq:wbgamma}
\overline{W}_{\overline{\bm{\gamma}}'} = \left( \begin{array}{l} \overline{W}_{\overline{\bm{\gamma}}'}^{[N]} \\
\overline{W}_{\overline{\bm{\gamma}}'}^{[N-1, \hspace{1ex} 1]} \\
\overline{W}_{\overline{\bm{\gamma}}'}^{[N-2, \hspace{1ex} 2]}
\end{array} \right) \,,
\end{equation}
where $\overline{W}_{\overline{\bm{\gamma}}'}^{[N]}$ is a $1
\times N(N-1)/2$ dimensional matrix,
$\overline{W}_{\overline{\bm{\gamma}}'}^{[N-1, \hspace{1ex} 1]}$
is an $(N-1) \times N(N-1)/2$ dimensional matrix and
$\overline{W}_{\overline{\bm{\gamma}}'}^{[N-2, \hspace{1ex} 2]}$
is an $N(N-3)/2 \times N(N-1)/2$ dimensional matrix, then
\begin{equation}\label{eq:sbgamma}
\overline{\bm{S}}_{\overline{\bm{\gamma}}'} = \left( \begin{array}{l} \overline{\bm{S}}_{\overline{\bm{\gamma}}'}^{[N]} \\
\overline{\bm{S}}_{\overline{\bm{\gamma}}'}^{[N-1, \hspace{1ex} 1]} \\
\overline{\bm{S}}_{\overline{\bm{\gamma}}'}^{[N-2, \hspace{1ex}
2]}
\end{array} \right) =
\left( \begin{array}{l} \overline{W}_{\overline{\bm{\gamma}}'}^{[N]} \, \overline{\bm{\gamma}}' \\
\overline{W}_{\overline{\bm{\gamma}}'}^{[N-1, \hspace{1ex} 1]} \,
\overline{\bm{\gamma}}' \\
\overline{W}_{\overline{\bm{\gamma}}'}^{[N-2, \hspace{1ex} 2]} \,
\overline{\bm{\gamma}}' \end{array} \right) \,,
\end{equation}
where
\begin{equation} \label{eq:Wbsum2}
\overline{W}_{\overline{\bm{\gamma}}'}^\alpha \,
\overline{\bm{\gamma}}' = \sum_{j=1}^N \sum_{i < j} \,\,
[\overline{W}_{\overline{\bm{\gamma}}'}^\alpha]_{ij} \,
\overline{\gamma}'_{ij} \,.
\end{equation}

\subsection{The Full Primitive Irreducible Coordinate Vector,
${\bm{\overline{\bm{S}}}}$\,.} It is useful to form a full
primitive irreducible coordinate vector that groups primitive
irreducible coordinates of the same species together as follows:
\begin{equation}\label{eq:FSbCV}
\overline{\bm{S}} = P \left( \begin{array}{l} \overline{\bm{S}}_{\bar{\bm{r}}'} \\
\overline{\bm{S}}_{\overline{\bm{\gamma}}'} \end{array} \right) = \left( \begin{array}{l} \overline{\bm{S}}_{\bar{\bm{r}}'}^{[N]} \\
\overline{\bm{S}}_{\overline{\bm{\gamma}}'}^{[N]}  \\ \hline \overline{\bm{S}}_{\bar{\bm{r}}'}^{[N-1, \hspace{1ex} 1]} \\
\overline{\bm{S}}_{\overline{\bm{\gamma}}'}^{[N-1, \hspace{1ex}
1]} \\ \hline
\overline{\bm{S}}_{\overline{\bm{\gamma}}'}^{[N-2, \hspace{1ex} 2]} \end{array} \right) = \left( \begin{array}{l} \overline{\bm{S}}^{[N]} \\
\overline{\bm{S}}^{[N-1, \hspace{1ex} 1]} \\
\overline{\bm{S}}^{[N-2, \hspace{1ex} 2]} \end{array} \right) \,,
\end{equation}
where the orthogonal matrix $P$ is given by

\begin{equation}\label{eq:p}
P = \left(
\begin{array}{ccccc}
  1 & 0 & 0 & 0 & 0 \\
  0 & 0 & 1 & 0 & 0 \\
  0 & 1 & 0 & 0 & 0 \\
  0 & 0 & 0 & 1 & 0 \\
  0 & 0 & 0 & 0 & 1
\end{array} \right) \,,
\end{equation}
and
\begin{equation}\label{eq:sbrep}
\begin{array}{cc@{\mbox{\hspace{2ex}and\hspace{2ex}}}c}
\overline{\bm{S}}^{[N]} = \left( \begin{array}{l} \overline{\bm{S}}_{\bar{\bm{r}}'}^{[N]} \\
\overline{\bm{S}}_{\overline{\bm{\gamma}}'}^{[N]} \end{array}
\right) \,, &
\overline{\bm{S}}^{[N-1, \hspace{1ex} 1]} = \left( \begin{array}{l} \overline{\bm{S}}_{\bar{\bm{r}}'}^{[N-1, \hspace{1ex} 1]} \\
\overline{\bm{S}}_{\overline{\bm{\gamma}}'}^{[N-1, \hspace{1ex}
1]}
\end{array} \right) &
\end{array}
\overline{\bm{S}}^{[N-2, \hspace{1ex} 2]} =
\overline{\bm{S}}_{\overline{\bm{\gamma}}'}^{[N-2, \hspace{1ex}
2]} \,.
\end{equation}

Now consider applying a transformation $\overline{W}$ to
Eqs.~(\ref{eq:qyt}), (\ref{eq:FGit}), (\ref{eq:character}) and
(\ref{eq:normit}), where
\begin{equation}\label{eq:Wb}
\overline{W}= P \left( \begin{array}{c|c} \overline{W}_{\bar{\bm{r}}'} & \bm{0} \\
\hline \bm{0} & \overline{W}_{\overline{\bm{\gamma}}'}
\end{array} \right) =
\left( \begin{array}{cc} \overline{W}_{\bar{\bm{r}}'}^{[N]} & \bm{0} \\
\bm{0} & \overline{W}_{\overline{\bm{\gamma}}'}^{[N]} \\
\hline
\overline{W}_{\bar{\bm{r}}'}^{[N-1, \hspace{1ex} 1]} & \bm{0} \\
\bm{0} & \overline{W}_{\overline{\bm{\gamma}}'}^{[N-1, \hspace{1ex} 1]} \\
\hline \bm{0} & \overline{W}_{\overline{\bm{\gamma}}'}^{[N-2,
\hspace{1ex} 2]}
\end{array} \right) \,,
\end{equation}
where
\begin{equation}\label{eq:SbWby}
{\overline{\bm{S}}} = \overline{W} \, {\bar{\bm{y}}'} \,.
\end{equation}

We can now relate the primitive irreducible coordinate vector
${\overline{\bm{S}}}$ of Eq.~(\ref{eq:FSbCV}) and
item~\alph{twostepseca}).\ above, to the symmetry coordinate
vector ${\bm{S}}$ of item~\alph{twostepsecb}).\ above.

\subsection{The Transformation, ${\bm{U}}$\,, from Primitive Irreducible
Coordinates to Symmetry Coordinates.} The symmetry coordinate
vector ${\bm{S}}$ will be related to the primitive irreducible
coordinate vector ${\overline{\bm{S}}}$ by a non-orthogonal linear
transformation $U$, i.e.\
\begin{equation} \label{eq:SUSb}
{\bm{S}} = U \, {\overline{\bm{S}}} \,,
\end{equation}
where
\begin{equation}\label{eq:FSCV}
{\bm{S}} = \left( \begin{array}{l} {\bm{S}}_{\bar{\bm{r}}'}^{[N]} \\
{\bm{S}}_{\overline{\bm{\gamma}}'}^{[N]}  \\ \hline {\bm{S}}_{\bar{\bm{r}}'}^{[N-1, \hspace{1ex} 1]} \\
{\bm{S}}_{\overline{\bm{\gamma}}'}^{[N-1, \hspace{1ex} 1]} \\
\hline
{\bm{S}}_{\overline{\bm{\gamma}}'}^{[N-2, \hspace{1ex} 2]} \end{array} \right) = \left( \begin{array}{l} {\bm{S}}^{[N]} \\
{\bm{S}}^{[N-1, \hspace{1ex} 1]} \\
{\bm{S}}^{[N-2, \hspace{1ex} 2]} \end{array} \right)
\end{equation}
and
\begin{equation}\label{eq:srep}
\begin{array}{cc@{\mbox{\hspace{2ex}and\hspace{2ex}}}c}
{\bm{S}}^{[N]} = \left( \begin{array}{l} {\bm{S}}_{\bar{\bm{r}}'}^{[N]} \\
{\bm{S}}_{\overline{\bm{\gamma}}'}^{[N]} \end{array} \right) \,, &
{\bm{S}}^{[N-1, \hspace{1ex} 1]} = \left( \begin{array}{l} {\bm{S}}_{\bar{\bm{r}}'}^{[N-1, \hspace{1ex} 1]} \\
{\bm{S}}_{\overline{\bm{\gamma}}'}^{[N-1, \hspace{1ex} 1]}
\end{array} \right) &
\end{array}
{\bm{S}}^{[N-2, \hspace{1ex} 2]} =
{\bm{S}}_{\overline{\bm{\gamma}}'}^{[N-2, \hspace{1ex} 2]} \,.
\end{equation}
In ${\bm{S}}$, symmetry coordinates of the same species are
grouped together.

Defining $W$ through the equation
\begin{equation}\label{eq:SWy}
{\bm{S}} = W \, {\bar{\bm{y}}'} \,,
\end{equation}
then Eqs.~(\ref{eq:FSbCV}), (\ref{eq:sbrep}), (\ref{eq:Wb}),
(\ref{eq:SbWby}), (\ref{eq:FSCV}), (\ref{eq:srep}) and
(\ref{eq:SWy}) imply that $W$ will be related to $\overline{W}$ by
the same transformation, i.e.\
\begin{equation} \label{eq:WUWb}
W = U \, \overline{W} \,,
\end{equation}
where
\begin{equation}
U = \left( \begin{array}{cc|cc|c}
    U_{\bar{\bm{r}}'}^{[N]} & 0 & 0 & 0 & 0 \\
    0 & U_{\overline{\bm{\gamma}}'}^{[N]} & 0 & 0 & 0 \\ \hline
    0 & 0 & U_{\bar{\bm{r}}'}^{[N-1, \hspace{1ex} 1]} & 0 & 0 \\
    0 & 0 & 0 & U_{\overline{\bm{\gamma}}'}^{[N-1, \hspace{1ex} 1]} & 0 \\ \hline
    0 & 0 & 0 & 0 & U_{\overline{\bm{\gamma}}'}^{[N-2, \hspace{1ex} 2]}
    \end{array} \right) \,,
\end{equation}
so that
\begin{equation} \label{eq:WUWbm}
W = \left( \begin{array}{cc} W_{\bar{\bm{r}}'}^{[N]} & \bm{0} \\
\bm{0} & W_{\overline{\bm{\gamma}}'}^{[N]} \\ \hline
W_{\bar{\bm{r}}'}^{[N-1, \hspace{1ex} 1]} & \bm{0} \\
\bm{0} & W_{\overline{\bm{\gamma}}'}^{[N-1, \hspace{1ex} 1]}
\\ \hline \bm{0} & W_{\overline{\bm{\gamma}}'}^{[N-2, \hspace{1ex} 2]}
\end{array}
\right) = \left( \begin{array}{cc} U_{\bar{\bm{r}}'}^{[N]} \, \overline{W}_{\bar{\bm{r}}'}^{[N]} & \bm{0} \\
\bm{0} & U_{\overline{\bm{\gamma}}'}^{[N]} \, \overline{W}_{\overline{\bm{\gamma}}'}^{[N]} \\
\hline
U_{\bar{\bm{r}}'}^{[N-1, \hspace{1ex} 1]} \, \overline{W}_{\bar{\bm{r}}'}^{[N-1, \hspace{1ex} 1]} & \bm{0} \\
\bm{0} & U_{\overline{\bm{\gamma}}'}^{[N-1, \hspace{1ex} 1]} \,
\overline{W}_{\overline{\bm{\gamma}}'}^{[N-1, \hspace{1ex} 1]}
\\ \hline \bm{0} & U_{\overline{\bm{\gamma}}'}^{[N-2, \hspace{1ex} 2]} \,
\overline{W}_{\overline{\bm{\gamma}}'}^{[N-2, \hspace{1ex} 2]}
\end{array} \right) \,.
\end{equation}
Thus we have
\begin{equation} \label{eq:WaXUaXWbaX}
W^\alpha_{\bm{X}'} = U^\alpha_{\bm{X}'} \,
\overline{W}^\alpha_{\bm{X}'} \,,
\end{equation}
where $\bm{X}'$ is $\bar{\bm{r}}'$ or $\overline{\bm{\gamma}}'$
(only $\overline{\bm{\gamma}}'$ for the $[N-2, \hspace{1ex} 2]$
sector). According to item~\alph{twostepsecb}).\ of
Sec.~\ref{subsec:ProgOutline} $W$ is an orthogonal transformation,
and so
\begin{equation}\label{eq:WaWaI}
W_{\bm{X}'}^\alpha [W_{\bm{X}'}^\alpha]^T = I_\alpha \,,
\end{equation}
where $I_\alpha$ is the unit matrix. \vspace{2em}

Equation~(\ref{eq:WaWaI}) is the {\em essential equation} for
$W_{\bm{X}'}^\alpha$ to satisfy.  \vspace{2em}

From Eqs.~(\ref{eq:ytransposeP}), (\ref{eq:FSCV}), (\ref{eq:SWy})
and (\ref{eq:WUWbm}), one derives
\begin{equation} \label{eq:SXp}
{\bm{S}}_{\bm{X}'}^\alpha = W_{\bm{X}'}^\alpha \, {\bm{X}}' \,,
\end{equation}
where ${\bm{X}}'$ is either of the $\bar{\bm{r}}'$ or
$\overline{\bm{\gamma}}'$ internal displacement coordinate vectors
and $\alpha \neq [N-2, \hspace{1ex} 2]$ when $\bm{X}' =
\bar{\bm{r}}'$\,.

According to Eqs.~(\ref{eq:WaXUaXWbaX}) and (\ref{eq:WaWaI}),
$U^\alpha_{\bm{X}'}$ must satisfy
\begin{equation}\label{eq:aaI}
U^\alpha_{\bm{X}'} \, \left\{\overline{W}^\alpha_{\bm{X}'}
[\overline{W}^\alpha_{\bm{X}'}]^T \right\} \,
[U^\alpha_{\bm{X}'}]^T = I_\alpha \,.
\end{equation}
Thus to determine $U^\alpha_{\bm{X}'}$ we need to know
$\overline{W}^\alpha_{\bm{X}'} [\overline{W}^\alpha_{\bm{X}'}]^T$
beforehand.

In accordance with item~\alph{twostepseca}).\ of
Sec.~\ref{subsec:ProgOutline},
$\overline{\bm{S}}_{\overline{\bm{\gamma}}'}^\alpha$, where
$\alpha$ is $[N]$ or $[N-1, \hspace{1ex} 1]$, transforms under
exactly the same representation of $S_N$ as
$\overline{\bm{S}}_{\bar{\bm{r}}'}^\alpha$\,. Likewise, according
to item~\alph{twostepsecb}).\ of Sec.~\ref{subsec:ProgOutline},
${\bm{S}}_{\overline{\bm{\gamma}}'}^\alpha$ transforms under
exactly the same representation of $S_N$ as
${\bm{S}}_{\bar{\bm{r}}'}^\alpha$\,. Thus we must have
\begin{equation} \label{eq:UgcUr}
U^\alpha_{\overline{\bm{\gamma}}'} = A_U^\alpha \,
U^\alpha_{\bar{\bm{r}}'} \,,
\end{equation}
where $\alpha$ is $[N]$ or $[N-1, \hspace{1ex} 1]$ and
$A_U^\alpha$ is a number (see below). Note that
Eq.~(\ref{eq:UgcUr}) substituted in Eq.~(\ref{eq:aaI}) yields:
\begin{equation} \label{eq:Wbg2eq1oa2Wbr2}
\overline{W}^\alpha_{\overline{\bm{\gamma}}'}
[\overline{W}^\alpha_{\overline{\bm{\gamma}}'}]^T =
\frac{1}{(A_U^\alpha)^2} \, \overline{W}^\alpha_{\bar{\bm{r}}'}
[\overline{W}^\alpha_{\bar{\bm{r}}'}]^T
\end{equation}
when $\alpha$ is $[N]$ or $[N-1, \hspace{1ex} 1]$ \,.

\subsubsection{The Motion Associated with the Symmetry Coordinates
about the Lewis Structure Configuration.}

Inverting Eq.~(\ref{eq:SWy}) we get
\begin{equation} \label{eq:XdecompSW}
{\bar{\bm{y}}'} = W^T \, {\bm{S}}\,,
\end{equation}
where ${\bar{\bm{y}}'}$ is defined in Eqs.~(\ref{eq:ytransposeP}),
(\ref{eq:bfrp}) and (\ref{eq:bfgammap}), ${\bm{S}}$ is given by
Eqs.~(\ref{eq:FSCV}) and (\ref{eq:srep}), and $W$ is given by
Eq.~(\ref{eq:WUWbm}). Equation~(\ref{eq:XdecompSW}) may be
directly derived from Eq.~(\ref{eq:SWy}) using the orthogonality
of the $W$ matrix.

From Eq.~(\ref{eq:XdecompSW}) we can derive
\begin{equation} \label{eq:yS}
{\bm{y}} = \left( \begin{array}{c} {\bm{r}} \\
\bm{\gamma} \end{array} \right) = \left(
\begin{array}{l} D^2 \overline{a}_{ho} \left( {\displaystyle
\bar{r}'_\infty {\bm{1}}_{\bar{\bm{r}}'} + \frac{1}{\sqrt{D}}
\sum_{\alpha} \sum_\xi \, \bar{\bm{r}}^{\prime \alpha}_\xi } \right) \\
{\displaystyle \gamma_\infty {\bm{1}}_{\overline{\bm{\gamma}}'} +
\frac{1}{\sqrt{D}} \sum_{\alpha} \sum_\xi \,
\overline{\bm{\gamma}}^{\prime \alpha}_\xi }
\end{array} \right)
\end{equation}
where
\begin{eqnarray}
\bar{\bm{r}}^{\prime \alpha}_\xi & = &
[{\bm{S}}_{\bar{\bm{r}}'}^\alpha]_\xi \,
[(W_{\bar{\bm{r}}'}^\alpha)_\xi]^T \,, \label{eq:rpaxi} \\
\overline{\bm{\gamma}}^{\prime \alpha}_\xi & = &
[{\bm{S}}_{\overline{\bm{\gamma}}'}^\alpha]_\xi \,
[(W_{\overline{\bm{\gamma}}'}^\alpha)_\xi]^T \,, \label{eq:gpaxi}
\end{eqnarray}
while
\begin{eqnarray} \label{eq:bf1i}
[{\bm{1}}_{\bar{\bm{r}}'}]_i= 1 & \forall & 1 \leq i \leq N
\end{eqnarray}
and
\begin{eqnarray} \label{eq:1eq1}
[{\bm{1}}_{\overline{\bm{\gamma}}'}]_{ij} = 1  & \forall & 1 \leq
i,j \leq N \,.
\end{eqnarray}
Equations~(\ref{eq:yS}), (\ref{eq:rpaxi}), (\ref{eq:gpaxi}),
(\ref{eq:bf1i}) and (\ref{eq:1eq1}) express the the motion
associated with the symmetry coordinate about the Lewis structure
configuration.

\subsection{The Reduction of the Eigensystem Equations in the Symmetry
Coordinate Basis\,.} \label{subsec:eigreduct}
\subsubsection{The Central Theorem.}
Defining $\bm{Q_W}$ to be a $\bm{Q}$ matrix in the symmetry
coordinate basis, i.e.\
\begin{equation} \label{eq:WQWT}
\bm{Q_W} = W \bm{Q} W^T \,,
\end{equation}
where $\bm{Q}$ is $\bm{F}$, $\bm{G}$ or $\bm{FG}$, and $\bm{Q_W}$
is $\bm{F_W}$, $\bm{G_W}$ or $\bm{(FG)_W}$\,. Since $\bm{F}$,
$\bm{G}$ and $\bm{FG}$ are invariant matrices (under
$S_N$)\cite{WDC}, then
\begin{equation} \label{eq:Qw}
\bm{Q_W} = \left( \begin{array}{ccc} \bm{\sigma_{[N]}^Q} \otimes
\bm{I_{[N]}} &
\bm{0} & \bm{0} \\
\bm{0} & \bm{\sigma_{[N-1, \hspace{1ex} 1]}^Q}
\otimes \bm{I_{[N-1, \hspace{1ex} 1]}} & \bm{0} \\
\bm{0} & \bm{0} & \bm{\sigma_{[N-2, \hspace{1ex} 2]}^Q} \otimes
\bm{I_{[N-2, \hspace{1ex} 2]}}
\end{array} \right) \,,
\end{equation}
where the symbol $\otimes$ indicates the direct product,
$\bm{\sigma_{[N]}^Q}$ and $\bm{\sigma_{[N-1, \hspace{1ex} 1]}^Q}$
are $2 \times 2$ dimensional matrices and $\bm{\sigma_{[N-2,
\hspace{1ex} 2]}^Q}$ is a single element (a number). The matrix
$\bm{I_{[N]}}$ is the $[N]$-sector identity matrix, simply the
number $1$, $\bm{I_{[N-1, \hspace{1ex} 1]}}$ is the $[N-1,
\hspace{1ex} 1]$-sector identity matrix, the $(N-1) \times
(N-1)$-dimensional unit matrix and $\bm{I_{[N-2, \hspace{1ex}
2]}}$ is the $[N-2, \hspace{1ex} 2]$-sector identity matrix, the
$N(N-3)/2 \times N(N-3)/2$-dimensional unit matrix. The matrix
elements
\begin{equation}\label{eq:sigmaQ}
[\bm{\sigma_\alpha^Q}]_{\bm{X}'_1,\,\bm{X}'_2} =
(W_{\bm{X}'_1}^\alpha)_\xi \, \bm{Q}_{\bm{X}'_1 \bm{X}'_2} \,
[(W_{\bm{X}'_2}^\alpha)_\xi]^T \,,
\end{equation}
where $\bm{X}'_1$ and $\bm{X}'_2$ are $\bar{\bm{r}}'$ or
$\overline{\bm{\gamma}}'$ (only $\overline{\bm{\gamma}}'$ for the
$[N-2, \hspace{1ex} 2]$ sector). The $\bm{Q}$-matrix quadrant,
$\bm{Q}_{\bm{X}'_1 \bm{X}'_2}$, is given by Eqs.~(\ref{eq:Q}),
(\ref{eq:Qrr}), (\ref{eq:Qrg}), (\ref{eq:Qgr}) and (\ref{eq:Qgg}).
Although we are {\em not} summing over the repeated
index $\xi$ in Eq.~(\ref{eq:sigmaQ}), the
$[\bm{\sigma_\alpha^Q}]_{\bm{X}'_1,\,\bm{X}'_2}$ are {\em
independent} of the $W_{\bm{X}'}^\alpha$ row label, $\xi$\,. If,
when we calculate $W_{\bm{X}'}^\alpha$\,, the matrix element
$[\bm{\sigma_\alpha^Q}]_{\bm{X}'_1,\,\bm{X}'_2}$ turns out to
depend on the $W_{\bm{X}'}^\alpha$ row label, $\xi$\,, then we
know that we have made a mistake calculating
$W_{\bm{X}'}^\alpha$\,. This is a strong check on the correctness
of our calculations in Sect.~\ref{sec:SNm1pNm2}.

\subsubsection{The Reduction of the Eigenvalue Equation in the
Symmetry Coordinate Basis\,.} As in Paper I, we
transform the basic eigenvalue
equation, Eq.~(\ref{eq:FGit}), to the symmetry coordinate basis:
\begin{equation} \label{eq:FwGwcb}
W \bm{F} W^T \, W \bm{G} W^T \, W {\bm{b}} = \bm{F_W} \, \bm{G_W}
\, {\bm{c}}^{(b)} = \lambda_b \, {\bm{c}}^{(b)} \,,
\end{equation}
where
\begin{eqnarray}\label{eq:FGcW}
\bm{F_W} = W \bm{F} W^T\,, & \bm{G_W} = W \bm{G} W^T\\, & {\bm{c}}^{(b)}
= W {\bm{b}} \,.
\end{eqnarray}
Using Eq.~(\ref{eq:qyt}) the $b^{\rm th}$ normal-mode coordinate can
be written under this transformation as:
\begin{equation} \label{eq:cTS}
[q']_b = {\bm{b}}^T W^T W {\bar{\bm{y}}'} = \left[
{{\bm{c}}^{(b)}} \right]^T {\bm{\cdot}} \,\, {\bm{S}} \,,
\end{equation}
where $[q']_b$ is now directly expressed in terms of the symmetry
coordinates.

The normal-coordinate coefficient vector, ${\bm{c}}^{(b)}$, has the
form:
\begin{equation} \label{eq:cb}
{\bm{c}}^{(b)} = \left(
\begin{array}{r@{\hspace{0.5ex}}c@{\hspace{0.5ex}}l}
\delta_{\alpha,\,[N]} \,\, {\mathsf{c}}^{[N]} & \otimes & 1 \\
\delta_{\alpha,\,[N-1, \hspace{1ex} 1]} \,\, {\mathsf{c}}^{[N-1,
\hspace{1ex} 1]} & \otimes &
{\bm{1}}^{[N-1, \hspace{1ex} 1]}_\xi\\
\delta_{\alpha,\,[N-2, \hspace{1ex} 2]} \,\, {\mathsf{c}}^{[N-2,
\hspace{1ex} 2]} & \otimes & {\bm{1}}^{[N-2, \hspace{1ex} 2]}_\xi
\end{array} \right) \,,
\end{equation}
where we have used  Eqs.~(\ref{eq:Qw}) and
(\ref{eq:FwGwcb}), and the ${\mathsf{c}}^{\alpha}$
satisfy the eigenvalue equations
\begin{equation} \label{eq:sceig}
\bm{\sigma_{\alpha}^{FG}} {\mathsf{c}}^{\alpha} = \lambda_\alpha
{\mathsf{c}}^{\alpha} \,.
\end{equation}
The $\bm{\sigma_{[N]}^{FG}}$ and $\bm{\sigma_{[N-1,
\hspace{1ex} 1]}^{FG}}$ are $2 \times 2$-dimensional matrices,
while $\bm{\sigma_{[N-2, \hspace{1ex} 2]}^{FG}}$ is a
one-dimensional matrix. Thus there are five solutions to
Eq.~(\ref{eq:sceig}) denoted by
$\{\lambda^\pm_{[N]},\,{\mathsf{c}}_\pm^{[N]}\}$\,,
$\{\lambda^\pm_{[N-1, \hspace{1ex} 1]},\,{\mathsf{c}}_\pm^{[N-1,
\hspace{1ex} 1]}\}$ and $\{\lambda_{[N-2, \hspace{1ex}
2]},\,{\mathsf{c}}^{[N-2, \hspace{1ex} 2]}\}$\,, where $\bm{X}'$
($\bar{\bm{r}}'$ or $\overline{\bm{\gamma}}'$\,, only
$\overline{\bm{\gamma}}'$ for the $[N-2, \hspace{1ex} 2]$ sector)
labels the rows of the elements of the column vector
${\mathsf{c}}^{\alpha}$. The normal-coordinate label, $b$, has
been replaced by the labels $\alpha$, $\xi$ and $\pm$ on the right
hand side of Eq.~(\ref{eq:cb}), while the elements of the column
vectors ${\bm{1}}^{\alpha}_\xi$ are
\begin{equation} 
[{\bm{1}}^{\alpha}_\xi]_\eta = \delta_{\xi \eta} \,,
\end{equation}
with $1 \leq \xi, \hspace{0.5ex} \eta \leq N-1$ when $\alpha=[N-1,
\hspace{1ex} 1]$, or $1 \leq \xi, \hspace{0.5ex} \eta \leq
N(N-3)/2$ when $\alpha=[N-2, \hspace{1ex} 2]$\,. The
${\mathsf{c}}^{\alpha}_\pm$ vectors for the $[N]$ and $[N-1,
\hspace{1ex} 1]$ sectors determine the amount of angular-radial
mixing between the symmetry coordinates in a normal coordinate of
a particular $\alpha$\,. For the $[N-2, \hspace{1ex} 2]$ sector,
the symmetry coordinates are the $[N-2, \hspace{1ex} 2]$ sector
normal coordinates up to a normalization constant,
$[{\mathsf{c}}^{[N-2, \hspace{1ex} 2]}]_{\overline{\bm{\gamma}}'}$
(see Sec.~\ref{subsubsec:norm}).

Writing ${\mathsf{c}}^{\alpha}$ as
\begin{equation} \label{eq:sfceqcthatc}
{\mathsf{c}}^{\alpha} = c^{\alpha} \times
\widehat{{\mathsf{c}}^{\alpha}}\,,
\end{equation}
where $\widehat{{\mathsf{c}}^{\alpha}}$ is a vector satisfying the
normalization condition
\begin{equation} \label{eq:hatcnorm}
[\widehat{{\mathsf{c}}^{\alpha}}]^T \,
\widehat{{\mathsf{c}}^{\alpha}} = 1\,,
\end{equation}
and $c^{\alpha}$ is a  normalization factor ensuring that
Eq.~(\ref{eq:normit}) is satisfied (see Sec.~\ref{subsubsec:norm}
below). Then the reduced eigenvalue equation,
Eq.~(\ref{eq:sceig}), determines $\widehat{{\mathsf{c}}^{\alpha}}$
alone.

With the $[N-2, \hspace{1ex} 2]$ sector, Eq.~(\ref{eq:hatcnorm})
yields
\begin{equation}  \label{eq:hatcnormeq1}
\widehat{{\mathsf{c}}^{[N-2, \hspace{1ex} 2]}} = 1\,,
\end{equation}
and so Eq.~(\ref{eq:sceig}) yields directly
\begin{equation} \label{eq:lNm2eqsig}
\lambda_{[N-2, \hspace{1ex} 2]} = \bm{\sigma_{[N-2, \hspace{1ex}
2]}^{FG}} \,.
\end{equation}
For the $[N]$ and $[N-1, \hspace{1ex} 1]$ sectors, if we write
\begin{equation} \label{eq:hatcpm}
\widehat{{\mathsf{c}}_\pm^{\alpha}} = \left(
\begin{array}{c} \cos{\theta^\alpha_\pm} \\
\sin{\theta^\alpha_\pm} \end{array} \right) \,,
\end{equation}
then from Eqs.~(\ref{eq:cTS}), (\ref{eq:cb}),
(\ref{eq:sfceqcthatc}) and (\ref{eq:hatcpm})
\begin{equation} \label{eq:qbeqSrSg}
[q']_b = c_\pm^{\alpha} \left( \cos{\theta^\alpha_\pm} \,
[{\bm{S}}_{\bar{\bm{r}}'}^{\alpha}]_\xi \, + \,
\sin{\theta^\alpha_\pm} \,
[{\bm{S}}_{\overline{\bm{\gamma}}'}^{\alpha}]_\xi \right) \,.
\end{equation}
Writing
\begin{equation} \label{eq:sigmamat}
\bm{\sigma_{\alpha}^{FG}} = \left( \begin{array}{cc}
\protect[\bm{\sigma_\alpha^{FG}}\protect]_{\bar{\bm{r}}',\,\bar{\bm{r}}'}
& \protect[\bm{\sigma_\alpha^{FG}}\protect]_{\bar{\bm{r}}',\,
\overline{\bm{\gamma}}'} \\
\protect[\bm{\sigma_\alpha^{FG}}\protect]_{\overline{\bm{\gamma}}',\,\bar{\bm{r}}'}
&
\protect[\bm{\sigma_\alpha^{FG}}\protect]_{\overline{\bm{\gamma}}',\,\overline{\bm{\gamma}}'}
\end{array} \right) \,,
\end{equation}
then Eq.~(\ref{eq:sceig}) may be written as
\begin{equation} \label{eq:sigma12}
\left( \begin{array}{cc}
(\protect[\bm{\sigma_\alpha^{FG}}\protect]_{\bar{\bm{r}}',\,\bar{\bm{r}}'}
- \lambda^\pm_\alpha) &
\protect[\bm{\sigma_\alpha^{FG}}\protect]_{\bar{\bm{r}}',\,
\overline{\bm{\gamma}}'} \\
\protect[\bm{\sigma_\alpha^{FG}}\protect]_{\overline{\bm{\gamma}}',\,\bar{\bm{r}}'}
&
(\protect[\bm{\sigma_\alpha^{FG}}\protect]_{\overline{\bm{\gamma}}',\,\overline{\bm{\gamma}}'}
- \lambda^\pm_\alpha)
\end{array} \right) \left( \begin{array}{c}
\cos{\theta^\alpha_\pm} \\ \sin{\theta^\alpha_\pm} \end{array}
\right) = 0 \,,
\end{equation}
from which we derive the following analytic formula for the frequencies:
\begin{equation} \label{eq:lambda12pm}
\lambda^\pm_\alpha =
\frac{(\protect[\bm{\sigma_\alpha^{FG}}\protect]_{\bar{\bm{r}}',\,\bar{\bm{r}}'}
+
\protect[\bm{\sigma_\alpha^{FG}}\protect]_{\overline{\bm{\gamma}}',\,\overline{\bm{\gamma}}'})
\pm
\sqrt{(\protect[\bm{\sigma_\alpha^{FG}}\protect]_{\bar{\bm{r}}',\,\bar{\bm{r}}'}
-
\protect[\bm{\sigma_\alpha^{FG}}\protect]_{\overline{\bm{\gamma}}',\,\overline{\bm{\gamma}}'})^2
+ 4 \protect[\bm{\sigma_\alpha^{FG}}\protect]_{\bar{\bm{r}}',\,
\overline{\bm{\gamma}}'}
\protect[\bm{\sigma_\alpha^{FG}}\protect]_{\overline{\bm{\gamma}}',\,\bar{\bm{r}}'}}}{2}
\end{equation}
Equations~(\ref{eq:sigma12}) and (\ref{eq:lambda12pm}) then yield
\begin{equation} \label{eq:tanthetaalphapm}
\tan{\theta^\alpha_\pm} = \frac{(\lambda^\pm_\alpha -
\protect[\bm{\sigma_\alpha^{FG}}\protect]_{\bar{\bm{r}}',\,\bar{\bm{r}}'})}
{\protect[\bm{\sigma_\alpha^{FG}}\protect]_{\bar{\bm{r}}',\,
\overline{\bm{\gamma}}'}} =
\frac{\protect[\bm{\sigma_\alpha^{FG}}\protect]_{\overline{\bm{\gamma}}',\,\bar{\bm{r}}'}}{(\lambda^\pm_\alpha
-
\protect[\bm{\sigma_\alpha^{FG}}\protect]_{\overline{\bm{\gamma}}',\,\overline{\bm{\gamma}}'})}\,,
\end{equation}

\subsubsection{The Normalization Condition.}
\label{subsubsec:norm} Using Eqs.~(\ref{eq:normit}),
(\ref{eq:FGcW}), (\ref{eq:Qw}), (\ref{eq:sigmaQ}) and
(\ref{eq:cb}), the ${\mathsf{c}}^{\alpha}$ also satisfy the
normalization condition
\begin{equation} \label{eq:cnormeq}
[{\mathsf{c}}^{\alpha}]^T \bm{\sigma_{\alpha}^G}
{\mathsf{c}}^{\alpha} = 1 \,.
\end{equation}
From Eqs.~(\ref{eq:sfceqcthatc}) and (\ref{eq:cnormeq}):
\begin{equation} \label{eq:cnorm}
c^{\alpha} = \frac{1}{\sqrt{[\widehat{{\mathsf{c}}^{\alpha}}]^T
\bm{\sigma_{\alpha}^G} \widehat{{\mathsf{c}}^{\alpha}}}} \,.
\end{equation}
For the $[N-2, \hspace{1ex} 2]$ sector,
Eqs.~(\ref{eq:hatcnormeq1}) and (\ref{eq:cnorm}) yield
\begin{equation} \label{eq:c2norm}
c^{[N-2, \hspace{1ex} 2]} = \frac{1}{\sqrt{\bm{\sigma_{[N-2,
\hspace{1ex} 2]}^G}}} \,.
\end{equation}
For the $[N]$
and $[N-1, \hspace{1ex} 1]$ sectors, Eqs.~(\ref{eq:hatcpm}) and
(\ref{eq:cnorm}) yield
for the normalization constant, $c_\pm^\alpha$
\begin{equation} \label{eq:calphapm}
c_\pm^\alpha = \frac{1}{\sqrt{\left( \begin{array}{c}
\cos{\theta^\alpha_\pm} \\ \sin{\theta^\alpha_\pm} \end{array}
\right)^T \bm{\sigma_{\alpha}^{G}} \left( \begin{array}{c}
\cos{\theta^\alpha_\pm} \\ \sin{\theta^\alpha_\pm} \end{array}
\right)}} \,\,.
\end{equation}

\subsubsection{The Normal Coordinates.}
The normal-coordinate vector, ${\bm{q}'}$\,, is given by
\renewcommand{\jot}{1em}
\begin{eqnarray}
{\bm{q}'} & = & \left( \begin{array}{l} {\bm{q}'}^{[N]} \\
{\bm{q}'}^{[N-1, \hspace{1ex} 1]} \\
{\bm{q}'}^{[N-2, \hspace{1ex} 2]} \end{array} \right) = \left( \begin{array}{l} {\bm{q}'}_+^{[N]} \\
{\bm{q}'}_-^{[N]}  \\ \hline {\bm{q}'}_+^{[N-1, \hspace{1ex} 1]} \\
{\bm{q}'}_-^{[N-1, \hspace{1ex} 1]} \\ \hline
{\bm{q}'}^{[N-2, \hspace{1ex} 2]} \end{array} \right)  \nonumber \\
& = & \left( \begin{array}{c} c^{[N]}_+ \cos{\theta^{[N]}_+}
{\bm{S}}_{\bar{\bm{r}}'}^{[N]} +
c^{[N]}_+ \sin{\theta^{[N]}_+} {\bm{S}}_{\overline{\bm{\gamma}}'}^{[N]} \\
c^{[N]}_- \cos{\theta^{[N]}_-} {\bm{S}}_{\bar{\bm{r}}'}^{[N]} +
c^{[N]}_- \sin{\theta^{[N]}_-}
{\bm{S}}_{\overline{\bm{\gamma}}'}^{[N]}
\\ \hline c^{[N-1, \hspace{1ex} 1]}_+ \cos{\theta^{[N-1,
\hspace{1ex} 1]}_+} {\bm{S}}_{\bar{\bm{r}}'}^{[N-1, \hspace{1ex}
1]} + c^{[N-1, \hspace{1ex} 1]}_+ \sin{\theta^{[N-1, \hspace{1ex}
1]}_+}
{\bm{S}}_{\overline{\bm{\gamma}}'}^{[N-1, \hspace{1ex} 1]} \\
c^{[N-1, \hspace{1ex} 1]}_- \cos{\theta^{[N-1, \hspace{1ex} 1]}_-}
{\bm{S}}_{\bar{\bm{r}}'}^{[N-1, \hspace{1ex} 1]} + c^{[N-1,
\hspace{1ex} 1]}_- \sin{\theta^{[N-1, \hspace{1ex} 1]}_-}
{\bm{S}}_{\overline{\bm{\gamma}}'}^{[N-1, \hspace{1ex} 1]} \\
\hline c^{[N-2, \hspace{1ex} 2]}
{\bm{S}}_{\overline{\bm{\gamma}}'}^{[N-2, \hspace{1ex} 2]}
\end{array} \right) \,. \label{eq:qvector}
\end{eqnarray}
\renewcommand{\jot}{0em}

\subsubsection{The Motion Associated with the Normal Coordinates
about the Lewis Structure Configuration.} In terms of the normal
coordinates, the unscaled internal coordinate vector is
\begin{equation} \label{eq:yq}
{\bm{y}} = \left( \begin{array}{c} {\bm{r}} \\
\bm{\gamma} \end{array} \right) = {\bm{y}}_\infty +
\frac{1}{\sqrt{D}} \left(
\begin{array}[t]{@{}l@{}} {\displaystyle \hspace{2ex} \sum} \\ {\scriptstyle \alpha= \left\{
\renewcommand{\arraystretch}{0.5}
\begin{array}{@{}c@{}} {\scriptstyle
\protect[N\protect]\,,} \\ {\scriptstyle \protect[N-1,
\hspace{1ex} 1\protect]}
\end{array} \renewcommand{\arraystretch}{1} \right\}
} \end{array} \sum_\xi \sum_{\tau=\pm} \, _{\tau}{\bm{y}}^{\prime
\, \alpha}_\xi + \sum_\xi \, {\bm{y}}^{\prime \, \protect[N-2,
\hspace{1ex} 2\protect]}_\xi \right)\,,
\end{equation}
where
\begin{equation} \label{eq:yqinfty}
{\bm{y}}_\infty = \left( \begin{array}{c} D^2
\overline{a}_{ho} \, \bar{r}'_\infty {\bm{1}}_{\bar{\bm{r}}'} \\
\gamma_\infty {\bm{1}}_{\overline{\bm{\gamma}}'} \end{array}
\right) \,,
\end{equation}
with ${\bm{1}}_{\bar{\bm{r}}'}$ and
${\bm{1}}_{\overline{\bm{\gamma}}'}$ given by Eqs.~(\ref{eq:bf1i})
and (\ref{eq:1eq1}) respectively. The vectors
$_{+}{\bm{y}}^{\prime \, \alpha}_\xi$, \,\,$_{-}{\bm{y}}^{\prime
\, \alpha}_\xi$ and ${\bm{y}}^{\prime \, \protect[N-2,
\hspace{1ex} 2\protect]}_\xi$ are
\begin{equation} \label{eq:pyqalphaxi}
_{+}{\bm{y}}^{\prime \, \alpha}_\xi = \frac{1}{s(\theta^{\alpha})}
\left(
\renewcommand{\arraystretch}{1.5}
\begin{array}{c}
{\displaystyle D^2 \overline{a}_{ho}
\frac{-\sin{\theta^{\alpha}_-}}{c^{\alpha}_+} \,
[{\bm{q}'}_+^{\alpha}]_\xi \, \protect[(W_{\bar{\bm{r}}'}^{\alpha})_\xi \protect]^T } \\
{\displaystyle \frac{\cos{\theta^{\alpha}_-}}{c^{\alpha}_+} \,
[{\bm{q}'}_+^{\alpha}]_\xi \,
\protect[(W_{\overline{\bm{\gamma}}'}^{\alpha})_\xi \protect]^T }
\end{array} \right) \,,
\renewcommand{\arraystretch}{1}
\end{equation}
\begin{equation} \label{eq:myqalphaxi}
_{-}{\bm{y}}^{\prime \, \alpha}_\xi =
\renewcommand{\arraystretch}{1.5}
\frac{1}{s(\theta^{\alpha})} \left(
\begin{array}{c}
{\displaystyle D^2 \overline{a}_{ho}
\frac{\sin{\theta^{\alpha}_+}}{c^{\alpha}_-} \,
[{\bm{q}'}_-^{\alpha}]_\xi \, \protect[(W_{\bar{\bm{r}}'}^{\alpha})_\xi \protect]^T } \\
{\displaystyle \frac{-\cos{\theta^{\alpha}_+}}{c^{\alpha}_-} \,
[{\bm{q}'}_-^{\alpha}]_\xi \,
\protect[(W_{\overline{\bm{\gamma}}'}^{\alpha})_\xi \protect]^T }
\end{array} \right)
\renewcommand{\arraystretch}{1}
\end{equation}
for $\alpha =[N]$ or $[N-1, \hspace{1ex} 1]$\,, and
\begin{equation} \label{eq:yqnm2xi}
{\bm{y}}^{\prime \, \protect[N-2, \hspace{1ex} 2\protect]}_\xi =
\left(
\renewcommand{\arraystretch}{1.5}
\begin{array}{c} \bm{0} \\
{\displaystyle \frac{1}{c^{[N-2, \hspace{1ex} 2]}} \,
[{\bm{q}'}^{[N-2, \hspace{1ex} 2]}]_\xi \,
\protect[(W_{\overline{\bm{\gamma}}'}^{[N-2, \hspace{1ex}
2]})_\xi]^T }
\end{array} \renewcommand{\arraystretch}{1} \right) \,.
\end{equation}
Equations~(\ref{eq:yq}), (\ref{eq:yqinfty}),
(\ref{eq:pyqalphaxi}), (\ref{eq:myqalphaxi}) and
(\ref{eq:yqnm2xi}) express, in terms of the internal coordinates
${\bm{r}}$ and $\bm{\gamma}$\,, the motion associated with the
normal coordinates, ${\bm{q}'}$\,, about the Lewis structure
configuration ${\bm{y}}_\infty$\,.

\section{Symmetry Coordinates Belonging to the $\bm{[N-1,} \hspace{1ex}
    \bm{1]}$ and \\
$\bm{[N-2,} \hspace{1ex} \bm{2]}$ Species.} \label{sec:SNm1pNm2}
The symmetry coordinates of $[N]$ species have been discussed in
Sec.~7.6 of Paper~I. Here we derive the symmetry coordinates of
$[N-1, \hspace{1ex} 1]$ and $[N-2, \hspace{1ex} 2]$ species.

\subsection{Symmetry Coordinates Belonging to the $\bm{[N-1,}
    \hspace{1ex} \bm{1]}$ Species.}

\subsubsection{The $\bar{\bm{r}}'$ Sector.}

\paragraph{The Primitive Irreducible Coordinate.}
Consider the $N(N-1)$ quantities $\bar{r}'_i - \bar{r}'_j$ where
$1 \leq i \neq j \leq N$. It's clear that under $S_N$ they
transform into themselves. However, they are not all linearly
independent. As we shall justify below, we identify the $N-1$
linearly independent elements of
$\overline{\bm{S}}_{\bar{\bm{r}}'}^{[N-1, \hspace{1ex} 1]}$ as
\begin{eqnarray} \label{eq:SbrNm1}
[\overline{\bm{S}}_{\bar{\bm{r}}'}^{[N-1, \hspace{1ex} 1]}]_i =
\bar{r}'_i - \bar{r}'_{i+1} & \mbox{where} & 1 \leq i \leq N-1 \,,
\end{eqnarray}
a subset of the $N(N-1)$ $\bar{r}'_i - \bar{r}'_j$\,. Obviously
\begin{equation}
\bar{r}'_j - \bar{r}'_i = - (\bar{r}'_i - \bar{r}'_j)
\end{equation}
so we can restrict ourselves to the $\bar{r}'_i - \bar{r}'_j$
for which $i \leq j$. Furthermore we can write
\begin{eqnarray}
\bar{r}'_i - \bar{r}'_{i+a} = \sum_{\kappa = 0}^{a-1}
(\bar{r}'_{i+\kappa} - \bar{r}'_{i+\kappa+1}) = \sum_{\kappa =
0}^{a-1} \, [\overline{\bm{S}}_{\bar{\bm{r}}'}^{[N-1, \hspace{1ex}
1]}]_{i+\kappa} & \forall & a \geq 1 \,,
\end{eqnarray}
where we have used Eq.~(\ref{eq:SbrNm1}) in the last step. Thus we
can write any of the $N(N-1)$\,\, $\bar{r}'_i - \bar{r}'_j$
in terms of the $N-1$\,\,
$[\overline{\bm{S}}_{\bar{\bm{r}}'}^{[N-1, \hspace{1ex} 1]}]_i$\,.
That the $N-1$\,\, $[\overline{\bm{S}}_{\bar{\bm{r}}'}^{[N-1,
\hspace{1ex} 1]}]_i$ form a minimal, irreducible, linearly
independent set transforming under the $[N-1, \hspace{1ex} 1]$
representation is assured since no linear combination of the
$[\overline{\bm{S}}_{\bar{\bm{r}}'}^{[N-1, \hspace{1ex} 1]}]_i$
can be formed to give $\overline{\bm{S}}_{\bar{\bm{r}}'}^{[N]}$ of
Eq.~(184) in Paper~I which transforms under the $[N]$
representation of $S_N$\,. Thus the $N-1$\,\,
$[\overline{\bm{S}}_{\bar{\bm{r}}'}^{[N-1, \hspace{1ex} 1]}]_i$
lie in the $(N-1)$-dimensional subspace orthogonal to the
one-dimensional subspace of the $[N]$ sector, which as we have
shown in Subsection~\ref{subsec:symCoor}, is the $[N-1,
\hspace{1ex} 1]$ irreducible carrier space of $S_N$\,. We have
thus proved what we set out to prove, that the $N-1$\,\,
$[\overline{\bm{S}}_{\bar{\bm{r}}'}^{[N-1, \hspace{1ex} 1]}]_i$
transform under the $[N-1, \hspace{1ex} 1]$ irreducible
representation of $S_N$\,.

As we have adverted in item~\alph{twostepseca}).\ of
Sec.~\ref{subsec:ProgOutline}, the
$[\overline{\bm{S}}_{\bar{\bm{r}}'}^{[N-1, \hspace{1ex} 1]}]_i$ of
Eq.~(\ref{eq:SbrNm1}) have the simplest functional form possible.
Any other choice will involve linear combinations of the
$[\overline{\bm{S}}_{\bar{\bm{r}}'}^{[N-1, \hspace{1ex} 1]}]_i$ of
Eq.~(\ref{eq:SbrNm1}) and will likely have a more complex
functional form. At the very best they will involve another
$(N-1)$-dimensional subset of terms of the form $\bar{r}'_i -
\bar{r}'_j$\,.

Furthermore, if we have two particles, we have only one primitive
irreducible coordinate $[\overline{\bm{S}}_{\bar{\bm{r}}'}^{[1,
\hspace{1ex} 1]}]_1 = \bar{r}'_1 - \bar{r}'_2$\,. If we have three
particles we have two primitive irreducible coordinates
$[\overline{\bm{S}}_{\bar{\bm{r}}'}^{[2, \hspace{1ex} 1]}]_1 =
\bar{r}'_1 - \bar{r}'_2$ and
$[\overline{\bm{S}}_{\bar{\bm{r}}'}^{[2, \hspace{1ex} 1]}]_2 =
\bar{r}'_2 - \bar{r}'_3$\,. Notice that adding one additional
particle does not entail a change in the primitive irreducible
coordinate, $[\overline{\bm{S}}_{\bar{\bm{r}}'}^{[2, \hspace{1ex}
1]}]_1$\,, involving only the first two particles. Quite generally
and by design, if we have $N-1$ particles and then add one more
particle to the system, according to Eq.~(\ref{eq:SbrNm1}) the
first primitive irreducible coordinates
$[\overline{\bm{S}}_{\bar{\bm{r}}'}^{[N-1, \hspace{1ex} 1]}]_i$\,,
$1 \leq i \leq N-2$\,, involving the first $N-1$ particles only,
are left unchanged by this addition. Thus we have
\begin{equation} \label{eq:SbNm1it}
\overline{\bm{S}}_{\bar{\bm{r}}'}^{[N-1, \hspace{1ex} 1]} = \left(
\begin{array}{c} \overline{\bm{S}}_{\bar{\bm{r}}'}^{[N-2, \hspace{1ex} 1]}
\\ \protect[\overline{\bm{S}}_{\bar{\bm{r}}'}^{[N-1, \hspace{1ex} 1]}
\protect]_{N-1}
\end{array} \right) \,,
\end{equation}
where $\overline{\bm{S}}_{\bar{\bm{r}}'}^{[N-2, \hspace{1ex} 1]}$
is the primitive irreducible coordinate vector of the first $N-1$
particles, and $[\overline{\bm{S}}_{\bar{\bm{r}}'}^{[N-1,
\hspace{1ex} 1]}]_{N-1}$ is the primitive irreducible coordinate
involving particles $N$ and $N-1$\,.

From Eq.~(\ref{eq:sbr}) we can identify $\overline{W}^{[N-1,
\hspace{1ex} 1]}_{\bar{\bm{r}}'}$ as
\begin{eqnarray}
[\overline{W}^{[N-1, \hspace{1ex} 1]}_{\bar{\bm{r}}'}]_{ij} =
\delta_{i,j} - \delta_{i+1,j} \,\,, & \mbox{where} & 1 \leq i \leq
N-1 \mbox{\hspace{1ex} and \hspace{1ex}} 1 \leq j \leq N \,.
\label{eq:WbNm1r}
\end{eqnarray}
In matrix form this is
\begin{equation} \label{eq:WbNm1rM}
\overline{W}^{[N-1, \hspace{1ex} 1]}_{\bar{\bm{r}}'} = \left(
\begin{array}{ccccccc} 1 & -1 & 0 & 0 & 0 & 0 & \cdots \\
0 & 1 & -1 & 0 & 0 & 0 & \cdots \\
0 & 0 & 1 & -1 & 0 & 0 & \cdots \\
0 & 0 & 0 & 1 & -1 & 0 & \cdots \\
0 & 0 & 0 & 0 & 1 & -1 & \cdots \\
\vdots & \vdots & \vdots & \vdots & \vdots & \ddots & \ddots
\end{array} \right) \,.
\end{equation}
Notice that if we have $N-1$ particles and then add one more
particle to the system, according to Eqs.~(\ref{eq:WbNm1r}) and
(\ref{eq:WbNm1rM}), the first $N-2$ rows of the matrix
$\overline{W}^{[N-1, \hspace{1ex} 1]}_{\bar{\bm{r}}'}$ are
unchanged by the addition of an additional row, aside, that is,
from the addition of an extra zero element at the end of the row,
c.f.\ the discussion of the paragraph prior to this one. To put it
more suscinctly we have
\begin{equation} 
\overline{W}^{[N-1, \hspace{1ex} 1]}_{\bar{\bm{r}}'} = \left(
\begin{array}{c|c} \overline{W}^{[N-2, \hspace{1ex} 1]}_{\bar{\bm{r}}'} &
\begin{array}{c} 0 \\ 0 \\ \vdots \\ 0 \end{array}  \\ \hline
\begin{array}{ccccc} 0 & 0 & \cdots & 0 & 1 \end{array} & -1 \\
\end{array} \right) \,.
\end{equation}

\paragraph{The Symmetry Coordinate}
To derive $W^{[N-1, \hspace{1ex} 1]}_{\bar{\bm{r}}'}$ and then the
the symmetry coordinate ${\bm{S}}_{\bar{\bm{r}}'}^{[N-1,
\hspace{1ex} 1]}$ via Eq.~(\ref{eq:SXp}), we use
Eq.~(\ref{eq:aaI}) to calculate $U^{[N-1, \hspace{1ex}
1]}_{\bar{\bm{r}}'}$ and then Eq.~(\ref{eq:WUWbm}) to arrive at
$W^{[N-1, \hspace{1ex} 1]}_{\bar{\bm{r}}'}$. The first step in
this process is to evaluate $\overline{W}^{[N-1, \hspace{1ex}
1]}_{\bar{\bm{r}}'} \, [\overline{W}^{[N-1, \hspace{1ex}
1]}_{\bar{\bm{r}}'}]^T$\,. From Eq.~(\ref{eq:WbNm1r}) we can
readily show that
\begin{equation} \label{eq:WNm1WNm1T}
\left(\overline{W}^{[N-1, \hspace{1ex} 1]}_{\bar{\bm{r}}'} \,
[\overline{W}^{[N-1, \hspace{1ex} 1]}_{\bar{\bm{r}}'}]^T
\right)_{i,k} = \sum_{j=1}^N [\overline{W}^{[N-1, \hspace{1ex}
1]}_{\bar{\bm{r}}'}]_{i,j} \, [\overline{W}^{[N-1, \hspace{1ex}
1]}_{\bar{\bm{r}}'}]_{k, j} = - \delta_{i+1,k} + 2\delta_{i,k}
-\delta_{i,k+1} \,,
\end{equation}
which in matrix form is
\begin{equation} \label{eq:WNm1WNm1TM}
\overline{W}^{[N-1, \hspace{1ex} 1]}_{\bar{\bm{r}}'} \,
[\overline{W}^{[N-1, \hspace{1ex} 1]}_{\bar{\bm{r}}'}]^T = \left(
\begin{array}{cccccc} 2 & -1 & 0 & 0 & 0 & \cdots \\
-1 & 2 & -1 & 0 & 0 & \cdots \\
0 & -1 & 2 & -1 & 0 & \cdots \\
0 & 0 & -1 & 2 & -1 & \cdots \\
0 & 0 & 0 & -1 & 2 & \ddots \\
\vdots & \vdots & \vdots & \vdots & \ddots & \ddots
\end{array} \right) = \left(
\begin{array}{c|c} \overline{W}^{[N-2, \hspace{1ex} 1]}_{\bar{\bm{r}}'} \,
[\overline{W}^{[N-2,
\hspace{1ex} 1]}_{\bar{\bm{r}}'}]^T & \begin{array}{c} 0 \\ 0 \\ \vdots \\ 0 \\
-1
\end{array}  \\ \hline
\begin{array}{ccccc} 0 & 0 & \cdots & 0 & -1 \end{array} & 2 \\
\end{array} \right) \,.
\end{equation}

Now Eqs.~(\ref{eq:aaI}) and (\ref{eq:WNm1WNm1T}) (or
Eq.~(\ref{eq:WNm1WNm1TM})) do not uniquely define $U^{[N-1,
\hspace{1ex} 1]}_{\bar{\bm{r}}'}$ as we can transform $U^{[N-1,
\hspace{1ex} 1]}_{\bar{\bm{r}}'}$ on the left by any orthogonal
matrix, and the result will still satisfy Eq.~(\ref{eq:aaI}).
However, as we have noted above in item~\alph{twostepsecb}).\ of
Sec.~\ref{subsec:ProgOutline}., according to Eq.~(\ref{eq:SUSb})
the symmetry coordinates are composed of linear combinations of
the primitive irreducible coordinates. Thus we choose the first
symmetry coordinate to be proportional to just the first
primitive irreducible coordinate, i.e.\
\begin{equation} \label{eq:SNm11proptoSbNm11}
[{\bm{S}}_{\bar{\bm{r}}'}^{[N-1, \hspace{1ex} 1]}]_1 \propto
[\overline{\bm{S}}_{\bar{\bm{r}}'}^{[N-1, \hspace{1ex} 1]}]_1 =
\bar{r}'_1 - \bar{r}'_2 \,.
\end{equation}
This involves only the first two particles in the simplest motion
possible under the requirement that the symmetry coordinate
transforms irreducibly under the $[N-1, \hspace{1ex} 1]$
representation of $S_N$. If $N=2$, $[{\bm{S}}_{\bar{\bm{r}}'}^{[1,
\hspace{1ex} 1]}]_1$ is the sole symmetry coordinate transforming
under the $[1, \hspace{1ex} 1]$ representation of $S_2$\,. For $N
> 2$, we require the symmetry coordinate vector to satisfy
\begin{equation} \label{eq:SNm1it}
{\bm{S}}_{\bar{\bm{r}}'}^{[N-1, \hspace{1ex} 1]} = \left(
\begin{array}{c} {\bm{S}}_{\bar{\bm{r}}'}^{[N-1, \hspace{1ex} 1]}
\\ \protect[{\bm{S}}_{\bar{\bm{r}}'}^{[N-1, \hspace{1ex} 1]}
\protect]_{N-1}
\end{array} \right) \,,
\end{equation}
c.f.\ Eq.~(\ref{eq:SbNm1it}).
Equations~(\ref{eq:SNm11proptoSbNm11}) and (\ref{eq:SNm1it}) mean
that $[{\bm{S}}_{\bar{\bm{r}}'}^{[N-1, \hspace{1ex} 1]}]_i$ is
formed from the first $i$ elements of the
$\overline{\bm{S}}_{\bar{\bm{r}}'}$ vector, which means that it is
formed from the first $i+1$ elements of the $\bar{\bm{r}}'$
vector, i.e.\ it involves the motion of only the first $i+1$
particles.

Thus from Eq.~(\ref{eq:WUWbm}) we see that $U^{[N-1, \hspace{1ex}
1]}_{\bar{\bm{r}}'}$ is a lower triangular matrix, and if we
further require all of its non-zero elements in the lower triangle
to be positive, these requirements, together with Eq.~(\ref{eq:aaI}),
uniquely specify $U^{[N-1, \hspace{1ex} 1]}_{\bar{\bm{r}}'}$\,.
Defining
\begin{eqnarray} \label{eq:thetaalphabeta}
\Theta_{\alpha-\beta} = \sum_{m=1}^{\alpha-1} \delta_{m, \, \beta}
\nonumber \\
& = & 1 \mbox{ when } \alpha - \beta > 0 \\
& = & 0 \mbox{ when } \alpha - \beta \leq 0 \,, \nonumber
\end{eqnarray}
and solving Eq.~(\ref{eq:aaI}) for $U^{[N-1, \hspace{1ex}
1]}_{\bar{\bm{r}}'}$\,, subject to the above conditions, yields
\begin{equation} \label{eq:UNm1ij}
[U^{[N-1, \hspace{1ex} 1]}_{\bar{\bm{r}}'}]_{ij} =
\frac{j}{\sqrt{i(i+1)}} \, \sum_{m=1}^i \delta_{mj} =
\frac{j}{\sqrt{i(i+1)}} \, \Theta_{i-j+1}
\end{equation}
where
\begin{equation}
\begin{array}{r@{\hspace{1ex}}l} {\displaystyle \Theta_{i-j+1} =
\sum_{m=1}^i \delta_{mj} } & =  1 \mbox{ when } j-i<1 \\
& = 0 \mbox{ when } j-i \geq 1 \,. \end{array}
\end{equation}
In matrix form Eq.~(\ref{eq:UNm1ij}) reads
\renewcommand{\jot}{1em}
\begin{eqnarray}
U^{[N-1, \hspace{1ex} 1]}_{\bar{\bm{r}}'} & = & \left(
\begin{array}{cccccc} \frac{1}{\sqrt{1}\sqrt{2}} & 0 & 0 & 0 & 0 & \cdots \\
\frac{1}{\sqrt{2}\sqrt{3}} & \frac{2}{\sqrt{2}\sqrt{3}} & 0 & 0 & 0 & \cdots \\
\frac{1}{\sqrt{3}\sqrt{4}} & \frac{2}{\sqrt{3}\sqrt{4}} &
\frac{3}{\sqrt{3}\sqrt{4}} & 0 & 0 & \cdots \\
\frac{1}{\sqrt{4}\sqrt{5}} & \frac{2}{\sqrt{4}\sqrt{5}} & \frac{3}{\sqrt{4}\sqrt{5}}
& \frac{4}{\sqrt{4}\sqrt{5}} & 0 & \cdots \\
\frac{1}{\sqrt{5}\sqrt{6}} & \frac{2}{\sqrt{5}\sqrt{6}} &
\frac{3}{\sqrt{5}\sqrt{6}} & \frac{4}{\sqrt{5}\sqrt{6}} &
\frac{5}{\sqrt{5}\sqrt{6}} & \cdots \\
\vdots & \vdots & \vdots & \vdots & \vdots & \ddots
\end{array} \right) \nonumber \\
& = & \left(
\begin{array}{c|c} U^{[N-2, \hspace{1ex} 1]}_{\bar{\bm{r}}'} & \begin{array}{c} 0 \\ 0 \\ \vdots
\\ 0 \\ 0
\end{array}  \\ \hline
\begin{array}{ccccc} \frac{1}{\sqrt{N-1}\sqrt{N}} & \frac{2}{\sqrt{N-1}\sqrt{N}}
& \cdots & \frac{N-3}{\sqrt{N-1}\sqrt{N}} & \frac{N-2}{\sqrt{N-1}\sqrt{N}} \end{array}
& \frac{N-1}{\sqrt{N-1}\sqrt{N}} \\
\end{array} \right) \,.  \label{eq:UNm1}
\end{eqnarray}
\renewcommand{\jot}{0em}

Thus from Eqs.~(\ref{eq:WUWbm}), (\ref{eq:WbNm1r}) and
(\ref{eq:UNm1ij}),
\begin{equation} \label{eq:WNm1r}
[W^{[N-1, \hspace{1ex} 1]}_{\bar{\bm{r}}'}]_{ik} =
\frac{1}{\sqrt{i(i+1)}} \left( \sum_{m=1}^i \delta_{mk} - i
\delta_{i+1,\, k} \right) = \frac{1}{\sqrt{i(i+1)}} \left(
\Theta_{i-k+1} - i \delta_{i+1,\, k} \right) \,\,,
\end{equation}
where $1 \leq i \leq N-1$ and $1 \leq k \leq N$\,. In matrix form
this is
\renewcommand{\jot}{1em}
\begin{eqnarray}
W^{[N-1, \hspace{1ex} 1]}_{\bar{\bm{r}}'} & = & \left(
\begin{array}{ccccccc} \frac{1}{\sqrt{1}\sqrt{2}} &
-\frac{1}{\sqrt{1}\sqrt{2}} & 0 & 0 & 0 & 0 & \cdots \\
\frac{1}{\sqrt{2}\sqrt{3}} & \frac{1}{\sqrt{2}\sqrt{3}} &
-\frac{2}{\sqrt{2}\sqrt{3}} & 0 & 0 & 0 & \cdots \\
\frac{1}{\sqrt{3}\sqrt{4}} & \frac{1}{\sqrt{3}\sqrt{4}} &
\frac{1}{\sqrt{3}\sqrt{4}} & -\frac{3}{\sqrt{3}\sqrt{4}} &
0 & 0 & \cdots \\
\frac{1}{\sqrt{4}\sqrt{5}} & \frac{1}{\sqrt{4}\sqrt{5}} &
\frac{1}{\sqrt{4}\sqrt{5}} & \frac{1}{\sqrt{4}\sqrt{5}} &
-\frac{4}{\sqrt{4}\sqrt{5}} & 0 & \cdots \\
\frac{1}{\sqrt{5}\sqrt{6}} & \frac{1}{\sqrt{5}\sqrt{6}} &
\frac{1}{\sqrt{5}\sqrt{6}} & \frac{1}{\sqrt{5}\sqrt{6}} &
\frac{1}{\sqrt{5}\sqrt{6}} & -\frac{5}{\sqrt{5}\sqrt{6}} &
\cdots \\
\vdots & \vdots & \vdots & \vdots & \vdots & \vdots & \ddots
\end{array} \right) \nonumber \\
& = & \left(
\begin{array}{c|c} W^{[N-2, \hspace{1ex} 1]}_{\bar{\bm{r}}'} & \begin{array}{c} 0 \\ 0 \\ \vdots \\ 0
\end{array}  \\ \hline
\begin{array}{cccccc} \frac{1}{\sqrt{N-1}\sqrt{N}} &
\frac{1}{\sqrt{N-1}\sqrt{N}} & \cdots & \frac{1}{\sqrt{N-1}\sqrt{N}}
& \frac{1}{\sqrt{N-1}\sqrt{N}} \end{array} &
-\frac{N-1}{\sqrt{N-1}\sqrt{N}} \\
\end{array} \right) \,. \label{eq:WNm1rM}
\end{eqnarray}
\renewcommand{\jot}{0em}

Thus the symmetry coordinate ${\bm{S}}_{\bar{\bm{r}}'}^{[N-1,
\hspace{1ex} 1]}$ from Eqs.~(\ref{eq:SXp}) and (\ref{eq:WNm1r}) is
\begin{eqnarray} \label{eq:SNm1}
[{\bm{S}}_{\bar{\bm{r}}'}^{[N-1, \hspace{1ex} 1]}]_i =
\frac{1}{\sqrt{i(i+1)}} \left( \sum_{k=1}^i \bar{r}'_k - i
\bar{r}'_{i+1} \right)\,, & \mbox{where} & 1 \leq i \leq N-1 \,.
\end{eqnarray}
In matrix form this reads
\begin{equation} \label{eq:SNm1M}
{\bm{S}}_{\bar{\bm{r}}'}^{[N-1, \hspace{1ex} 1]} = \left(
\begin{array}{l}
\frac{1}{\sqrt{1}\sqrt{2}} (\bar{r}'_1 - \bar{r}'_2) \\
\frac{1}{\sqrt{2}\sqrt{3}} (\bar{r}'_1 + \bar{r}'_2 -
2\bar{r}'_3) \\
\frac{1}{\sqrt{3}\sqrt{4}} (\bar{r}'_1 + \bar{r}'_2
+ \bar{r}'_3 - 3\bar{r}'_4) \\
\frac{1}{\sqrt{4}\sqrt{5}} (\bar{r}'_1 + \bar{r}'_2
 + \bar{r}'_3 + \bar{r}'_4 - 4\bar{r}'_5) \\
\multicolumn{1}{c}{\vdots} \\
{\displaystyle \frac{1}{\sqrt{N-1}\sqrt{N}} \left(
\sum_{k=1}^{N-1} \bar{r}'_k - (N-1)\,\bar{r}'_N \right) }
\end{array} \right)
= \left(
\begin{array}{c} {\bm{S}}_{\bar{\bm{r}}'}^{[N-2, \hspace{1ex} 1]}
\\ \protect[{\bm{S}}_{\bar{\bm{r}}'}^{[N-1, \hspace{1ex} 1]}
\protect]_{N-1}
\end{array} \right) \,,
\end{equation}
where ${\bm{S}}_{\bar{\bm{r}}'}^{[N-2, \hspace{1ex} 1]}$ is the
symmetry coordinate vector of the first $N-1$ particles and
$[{\bm{S}}_{\bar{\bm{r}}'}^{[N-1, \hspace{1ex} 1]}]_{N-1}$ is the
additional symmetry coordinate involving all $N$ particles. We see
from Eqs.~(\ref{eq:SNm1}) and (\ref{eq:SNm1M}) that the complexity
of the motions described by the symmetry coordinates is kept to a
minimum and builds up slowly as more particles become involved.
Referring again to Eqs.~(\ref{eq:SNm1it}) and (\ref{eq:SNm1M}), we
again advert that adding another particle to the system does not
cause widespread disruption to the symmetry coordinates. The
symmetry coordinates, and the motions they describe, remain the
same except for an additional symmetry coordinate which describes
a motion involving all of the particles.

\paragraph{Motions Associated with Symmetry
Coordinates $[{\bm{S}}_{\bar{\bm{r}}'}^{[N-1, \hspace{1ex}
1]}]_\xi$\,.} According to Eqs.~(\ref{eq:yS}), (\ref{eq:bf1i}),
(\ref{eq:rpaxi}) and (\ref{eq:WNm1r}), the motions associated with
symmetry coordinates $[{\bm{S}}_{\bar{\bm{r}}'}^{[N-1,
\hspace{1ex} 1]}]_\xi$ in the unscaled internal displacement
coordinates ${\bm{r}}$ about the unscaled Lewis structure
configuration
\begin{equation} \label{eq:rinftyup}
\bm{r}_\infty = D^2 \overline{a}_{ho} \bar{r}'_\infty
{\bm{1}}_{\bar{\bm{r}}'}
\end{equation}
are given by
\begin{equation} \label{eq:rNm1inr}
\begin{array}{r@{\hspace{0.5em}}c@{\hspace{0.5em}}l}
(r^{[N-1, \hspace{1ex} 1]}_\xi)_i & = & \overline{a}_{ho} \,
D^{3/2} \, (r^{\prime [N-1, \hspace{1ex} 1]}_\xi)_i =
\overline{a}_{ho} \, D^{3/2} \, [{\bm{S}}_{\bar{\bm{r}}'}^{[N-1,
\hspace{1ex} 1]}]_\xi \,
[(W_{\bar{\bm{r}}'}^{[N-1, \hspace{1ex} 1]})_\xi]_i  \\
[0.5em] & = & {\displaystyle \overline{a}_{ho} \,
\sqrt{\frac{D^3}{\xi(\xi+1)}} \, [{\bm{S}}_{\bar{\bm{r}}'}^{[N-1,
\hspace{1ex} 1]}]_\xi \, \left( \sum_{m=1}^\xi \delta_{mi} - \xi
\delta_{\xi+1,\, i} \right)  }
\\ [1.5em] & = & {\displaystyle \overline{a}_{ho} \,
\sqrt{\frac{D^3}{\xi(\xi+1)}} \, [{\bm{S}}_{\bar{\bm{r}}'}^{[N-1,
\hspace{1ex} 1]}]_\xi \, \left( \Theta_{\xi-i+1} - \xi
\delta_{\xi+1,\, i} \right)
 \,. } \end{array}
\end{equation}
Thus we see that the motion associated with symmetry coordinate
$[{\bm{S}}_{\bar{\bm{r}}'}^{[N-1, \hspace{1ex} 1]}]_1$ is an
antisymmetric stretch motion about the Lewis structure
configuration involving particles $1$ and $2$\,. As $\xi$ gets
larger, the motion involves more particles, $\xi+1$ particles,
where the first $\xi$ particles move one way while the
$(\xi+1)^{\rm th}$ moves in the other. Paradoxically though, as
$\xi$ increases the motion becomes more single-particle-like since
the $(\xi+1)^{\rm th}$ radius vector in Eq.~(\ref{eq:rNm1inr}) is
weighted by the quantity $\xi$\,.

\subsubsection{The $\overline{\bm{\gamma}}'$ Sector.}

\paragraph{The Primitive Irreducible Coordinate.}
Again according to item~\alph{twostepseca}).\ above,
$\overline{\bm{S}}_{\overline{\bm{\gamma}}'}^{[N-1, \hspace{1ex}
1]}$ should transforms under exactly the same non-orthogonal
irreducible $[N-1, \hspace{1ex} 1]$ representation of $S_N$ as
$\overline{\bm{S}}_{\bar{\bm{r}}'}^{[N-1, \hspace{1ex} 1]}$. If
$\widehat{{\bm{r}}_i}$ is the unit vector from the center of the
confining field to particle $i$\,, then $\widehat{{\bm{r}}_i} -
\widehat{{\bm{r}}_{i+1}}$, where $1 \leq i \leq N-1$\,, transform
under exactly the same $[N-1, \hspace{1ex} 1]$ irreducible
representation of $S_N$ as the primitive irreducible coordinate of
the $\bar{\bm{r}}'$ sector,
$[\overline{\bm{S}}_{\bar{\bm{r}}'}^{[N-1, \hspace{1ex} 1]}]_i$\,,
of Eq.~(\ref{eq:SbrNm1}). We also have that $\sum_{i=1}^N
\widehat{{\bm{r}}_i}$ is invariant under $S_N$, and so
$(\widehat{{\bm{r}}_i} - \widehat{{\bm{r}}_{i+1}}) \, {\bm{.}} \,
(\sum_{j=1}^N \widehat{{\bm{r}}_j}) = \sum_{j=1}^N
(\widehat{{\bm{r}}_i} \, {\bm{.}} \, \widehat{{\bm{r}}_j} -
\widehat{{\bm{r}}_{i+1}} \, {\bm{.}} \, \widehat{{\bm{r}}_j})$
will transform under exactly the same $[N-1, \hspace{1ex} 1]$
irreducible representation of $S_N$ as the primitive irreducible
coordinate of the $\bar{\bm{r}}'$ sector,
$[\overline{\bm{S}}_{\bar{\bm{r}}'}^{[N-1, \hspace{1ex} 1]}]_i$\,,
of Eq.~(\ref{eq:SbrNm1}). Upon using
\begin{equation} \label{eq:gij}
\gamma_{ij} = \widehat{{\bm{r}}_i} \, {\bm{.}} \,
\widehat{{\bm{r}}_j} \,,
\end{equation}
this becomes $\sum_{j=1}^N (\gamma_{ij} - \gamma_{(i+1)j})$. Since
$\gamma_{ii} = 1$, $\sum_{j=1}^N (\gamma_{ij} - \gamma_{(i+1)j}) =
\sum_{j \neq i} \gamma_{ij} - \sum_{j \neq i+1} \gamma_{(i+1)j}$
and upon using Eq.~(\ref{eq:taylor2}) we obtain $\delta^{1/2}
(\sum_{j \neq i} \overline{\gamma}'_{ij} - \sum_{j \neq i+1}
\overline{\gamma}'_{(i+1)j} )$\,. Thus we identify the primitive
irreducible coordinate of the $[N-1, \hspace{1ex} 1]$
representation in the $\overline{\bm{\gamma}}'$ sector as
\begin{equation}
[\overline{\bm{S}}_{\overline{\bm{\gamma}}'}^{[N-1, \hspace{1ex}
1]}]_i = \sum_{j \neq i} \overline{\gamma}'_{ij} - \sum_{j \neq
i+1} \overline{\gamma}'_{(i+1)j} = \left( \sqrt[4]{D}\,
\widehat{{\bm{r}}_i} - \sqrt[4]{D}\, \widehat{{\bm{r}}_{i+1}}
\right) \, {\bm{.}} \, \left(\sum_{j=1}^N \sqrt[4]{D}\,
\widehat{{\bm{r}}_j} \right) \,.
\end{equation}

So what is $\overline{W}^{[N-1, \hspace{1ex}
1]}_{\overline{\bm{\gamma}}'}$\,? From Eqs.~(184) of Paper~I,
and Eqs. (\ref{eq:SXp}) and (\ref{eq:SbrNm1})
\begin{equation} \label{eq:WbrNWbrNm1g}
[\overline{\bm{S}}_{\overline{\bm{\gamma}}'}^{[N-1, \hspace{1ex}
1]}]_i = \sum_{k,l=1}^N [\overline{W}^{[N-1, \hspace{1ex}
1]}_{\bar{\bm{r}}'}]_{ik} \,
[\overline{W}^{[N]}_{\bar{\bm{r}}'}]_l \, \left( \sqrt[4]{D}\,
\widehat{{\bm{r}}_k} \right) \, {\bm{.}} \, \left( \sqrt[4]{D}\,
\widehat{{\bm{r}}_l} \right) = \sum_{k,l=1}^N [\overline{W}^{[N-1,
\hspace{1ex} 1]}_{\bar{\bm{r}}'}]_{ik} \,
[\overline{W}^{[N]}_{\bar{\bm{r}}'}]_l \, \overline{\gamma}'_{kl}
\end{equation}
and since $\overline{\gamma}'_{kl}$ is symmetric
Eq.~(\ref{eq:WbrNWbrNm1g}) may be recast as
\begin{equation} \label{eq:WbrNWbrNm1gs}
[\overline{\bm{S}}_{\overline{\bm{\gamma}}'}^{[N-1, \hspace{1ex}
1]}]_i = \frac{1}{2} \sum_{k \neq l=1}^N \left(
[\overline{W}^{[N-1, \hspace{1ex} 1]}_{\bar{\bm{r}}'}]_{ik} \,
[\overline{W}^{[N]}_{\bar{\bm{r}}'}]_l + [\overline{W}^{[N-1,
\hspace{1ex} 1]}_{\bar{\bm{r}}'}]_{il} \,
[\overline{W}^{[N]}_{\bar{\bm{r}}'}]_k \right) \,
\overline{\gamma}'_{kl} \,.
\end{equation}
Using Eqs.~(185) of Paper~I, and Eqs. (\ref{eq:bfgammap}) and
(\ref{eq:bf1i}) in Eq.~(\ref{eq:WbrNWbrNm1gs}) we have
\begin{equation} \label{eq:WbrNWbrNm1gsl}
[\overline{\bm{S}}_{\overline{\bm{\gamma}}'}^{[N-1, \hspace{1ex}
1]}]_i = \sum_{l=2}^N \sum_{k=1}^{l-1} \left( [\overline{W}^{[N-1,
\hspace{1ex} 1]}_{\bar{\bm{r}}'}]_{ik} \,
[{\bm{1}}_{\bar{\bm{r}}'}]_l + [\overline{W}^{[N-1, \hspace{1ex}
1]}_{\bar{\bm{r}}'}]_{il} \, [{\bm{1}}_{\bar{\bm{r}}'}]_k \right)
\, \overline{\gamma}'_{kl} \,.
\end{equation}
Thus we identify
\begin{equation} \label{eq:WbNm1geqWbNm1r}
[\overline{W}^{[N-1, \hspace{1ex}
1]}_{\overline{\bm{\gamma}}'}]_{i,\,kl} = \left(
[\overline{W}^{[N-1, \hspace{1ex} 1]}_{\bar{\bm{r}}'}]_{ik} \,
[{\bm{1}}_{\bar{\bm{r}}'}]_l + [\overline{W}^{[N-1, \hspace{1ex}
1]}_{\bar{\bm{r}}'}]_{il} \, [{\bm{1}}_{\bar{\bm{r}}'}]_k \right)
\,,
\end{equation}
where $[\overline{W}^{[N-1, \hspace{1ex}
1]}_{\bar{\bm{r}}'}]_{ik}$ and $[{\bm{1}}_{\bar{\bm{r}}'}]_l$ are
given by Eqs.~(\ref{eq:WbNm1r}) (or Eq.~(\ref{eq:WbNm1rM})) and
(\ref{eq:bf1i}) respectively. Thus
\begin{equation}  \label{eq:WbNm1}
[\overline{W}^{[N-1, \hspace{1ex}
1]}_{\overline{\bm{\gamma}}'}]_{i,\,kl} = \left\{ \,\,
(\delta_{ik}[{\bm{1}}_{\bar{\bm{r}}'}]_l +
\delta_{il}[{\bm{1}}_{\bar{\bm{r}}'}]_k) - (
\delta_{i+1,\,k}[{\bm{1}}_{\bar{\bm{r}}'}]_l +
\delta_{i+1,\,l}[{\bm{1}}_{\bar{\bm{r}}'}]_k ) \,\, \right\} \,,
\end{equation}
where $1 \leq k < l \leq N$ and $1\leq i \leq N-1$\,.

\paragraph{The Symmetry Coordinate.}
To derive $W^{[N-1, \hspace{1ex} 1]}_{\overline{\bm{\gamma}}'}$
and then the symmetry coordinate
${\bm{S}}_{\overline{\bm{\gamma}}'}^{[N-1, \hspace{1ex} 1]}$ via
Eq.~(\ref{eq:SXp}), we use Eqs.~(\ref{eq:UgcUr}),
(\ref{eq:Wbg2eq1oa2Wbr2}) and (\ref{eq:UNm1ij}) to calculate
$U^{[N-1, \hspace{1ex} 1]}_{\overline{\bm{\gamma}}'}$ and then
Eq.~(\ref{eq:WaXUaXWbaX}) to arrive at $W^{[N-1, \hspace{1ex}
1]}_{\overline{\bm{\gamma}}'}$. The first step in this process is
to evaluate $\overline{W}^{[N-1, \hspace{1ex}
1]}_{\overline{\bm{\gamma}}'} \, [\overline{W}^{[N-1, \hspace{1ex}
1]}_{\overline{\bm{\gamma}}'}]^T$\,. From
Eq.~(\ref{eq:WbNm1geqWbNm1r}) we derive
\begin{equation}
[\overline{W}^{[N-1, \hspace{1ex} 1]}_{\overline{\bm{\gamma}}'} \,
[\overline{W}^{[N-1, \hspace{1ex}
1]}_{\overline{\bm{\gamma}}'}]^T]_{ij} = (N-2)\,
[\overline{W}^{[N-1, \hspace{1ex} 1]}_{\bar{\bm{r}}'} \,
[\overline{W}^{[N-1, \hspace{1ex} 1]}_{\bar{\bm{r}}'}]^T]_{ij}
\end{equation}
and from Eqs.~(\ref{eq:UgcUr}) and (\ref{eq:Wbg2eq1oa2Wbr2}):
\begin{equation} \label{eq:UNm1geqAUNm1r}
U^{[N-1, \hspace{1ex} 1]}_{\overline{\bm{\gamma}}'} =
\frac{1}{\sqrt{N-2}} \, U^{[N-1, \hspace{1ex} 1]}_{\bar{\bm{r}}'}
\,,
\end{equation}
where $U^{[N-1, \hspace{1ex} 1]}_{\bar{\bm{r}}'}$ is given by
Eqs.~(\ref{eq:UNm1ij}) (or Eq.~(\ref{eq:UNm1})).

Thus from Eqs.~(\ref{eq:WaXUaXWbaX}), (\ref{eq:WbNm1geqWbNm1r})
and (\ref{eq:UNm1geqAUNm1r}):
\begin{equation} \label{eq:WNm1geqWNm1r}
[W^{[N-1, \hspace{1ex} 1]}_{\overline{\bm{\gamma}}'}]_{i,\,kl} =
\frac{1}{\sqrt{N-2}} \,\left( [W^{[N-1, \hspace{1ex}
1]}_{\bar{\bm{r}}'}]_{ik} \, [{\bm{1}}_{\bar{\bm{r}}'}]_l +
[W^{[N-1, \hspace{1ex} 1]}_{\bar{\bm{r}}'}]_{il} \,
[{\bm{1}}_{\bar{\bm{r}}'}]_k \right) \,,
\end{equation}
where $[W^{[N-1, \hspace{1ex} 1]}_{\bar{\bm{r}}'}]_{ik}$ and
$[{\bm{1}}_{\bar{\bm{r}}'}]_l$ are given by Eqs.~(\ref{eq:WNm1r})
(or Eq.~(\ref{eq:WNm1rM})) and (\ref{eq:bf1i}) respectively. Upon
using Eq.~(\ref{eq:WNm1r}) in Eq.~(\ref{eq:WNm1geqWNm1r}) we
obtain
\renewcommand{\jot}{0.5em}
\begin{eqnarray}
\lefteqn{[W^{[N-1, \hspace{1ex} 1]}_{\overline{\bm{\gamma}}'}]_{i,\,kl} \hspace{1ex} =} \nonumber \\
& = & \frac{1}{\sqrt{i(i+1)(N-2)}} \, \left( \sum_{m=1}^i \big(
\delta_{mk} \, [{\bm{1}}_{\bar{\bm{r}}'}]_l + \delta_{ml} \,
[{\bm{1}}_{\bar{\bm{r}}'}]_k \big) - i \big( \delta_{i+1,\, k} \,
[{\bm{1}}_{\bar{\bm{r}}'}]_l + \delta_{i+1,\, l} \,
[{\bm{1}}_{\bar{\bm{r}}'}]_k \big)
\right) \nonumber \\
& = & \frac{1}{\sqrt{i(i+1)(N-2)}} \, \bigg( \big( \Theta_{i-k+1}
\, [{\bm{1}}_{\bar{\bm{r}}'}]_l + \Theta_{i-l+1} \,
[{\bm{1}}_{\bar{\bm{r}}'}]_k \big) - i \big( \delta_{i+1,\, k} \,
[{\bm{1}}_{\bar{\bm{r}}'}]_l + \delta_{i+1,\, l} \,
[{\bm{1}}_{\bar{\bm{r}}'}]_k \big) \bigg) \,, \label{eq:WNm1g}
\end{eqnarray}
\renewcommand{\jot}{0em}
where $1 \leq k < l \leq N$ and $1\leq i \leq N-1$\,.

Using Eqs.~(\ref{eq:bfgammap}) and (\ref{eq:WNm1g}) in
Eq.~(\ref{eq:SXp}), we obtain the symmetry coordinate
\renewcommand{\jot}{0.5em}
\begin{eqnarray}
[{\bm{S}}_{\overline{\bm{\gamma}}'}^{[N-1, \hspace{1ex} 1]}]_i & =
& \frac{1}{\sqrt{i(i+1)(N-2)}} \, \left( \sum_{k=1}^i \sum_{l=1}^N
\overline{\gamma}'_{kl} - i \sum_{l=1}^N \overline{\gamma}'_{i+1,\,l} \right) \nonumber \\
& = & \frac{1}{\sqrt{i(i+1)(N-2)}} \, \left( \sum_{k = 1}^i \,
\sum_{l \neq k =1}^N \hspace{-1ex} \overline{\gamma}'_{kl} - i
\hspace{-2ex}
\sum_{l \neq i+1 =1}^N \hspace{-2ex} \overline{\gamma}'_{i+1,\,l} \right) \label{eq:SgNm1i} \\
& = & \frac{1}{\sqrt{i(i+1)(N-2)}} \, \left( \left[ \sum_{l = 2}^i
\, \sum_{k=1}^{l-1} \hspace{-1ex} \overline{\gamma}'_{kl} +
\sum_{k = 1}^i \, \sum_{l=k+1}^{N} \hspace{-1ex}
\overline{\gamma}'_{kl} \right] - i \left[ \sum_{k=1}^i
\overline{\gamma}'_{k,\,i+1} + \sum_{l=i+2}^N \hspace{-0.5ex}
\overline{\gamma}'_{i+1,\,l} \right] \right) \,. \nonumber
\end{eqnarray}
\renewcommand{\jot}{0em}

\paragraph{Motions Associated with Symmetry
Coordinate ${\bm{S}}_{\overline{\bm{\gamma}}'}^{[N-1, \hspace{1ex}
1]}$\,.} According to Eqs.~(\ref{eq:yS}), (\ref{eq:gpaxi}),
(\ref{eq:1eq1}) and (\ref{eq:WNm1g}), the motions associated with
symmetry coordinates $[{\bm{S}}_{\overline{\bm{\gamma}}'}^{[N-1,
\hspace{1ex} 1]}]_\xi$ in the unscaled internal displacement
coordinates $\bm{\gamma}$ about the unscaled Lewis structure
configuration
\begin{equation} \label{eq:ginfty1}
\bm{\gamma}_\infty = \gamma_\infty
{\bm{1}}_{\overline{\bm{\gamma}}'}
\end{equation}
are given by
\begin{equation} \label{eq:gNm1ing}
\begin{array}{@{}r@{\hspace{0.5em}}c@{\hspace{0.5em}}l@{}}
\multicolumn{3}{l}{ \hspace{-0.5ex} (\gamma^{[N-1, \hspace{1ex}
1]}_\xi)_{ij} = {\displaystyle \frac{1}{\sqrt{D}} \,
(\gamma^{\prime [N-1, \hspace{1ex} 1]}_\xi)_{ij} =
\frac{1}{\sqrt{D}} \, [{\bm{S}}_{\overline{\bm{\gamma}}'}^{[N-1,
\hspace{1ex}
1]}]_\xi \, [(W_{\overline{\bm{\gamma}}'}^{[N-1, \hspace{1ex} 1]})_\xi]_{ij} } } \\
[1em] & = & {\displaystyle \frac{1}{\sqrt{\xi(\xi+1)(N-2)D}} \,
[{\bm{S}}_{\overline{\bm{\gamma}}'}^{[N-1, \hspace{1ex} 1]}]_\xi
\, \left( \sum_{m=1}^\xi \big( \delta_{mi} \,
[{\bm{1}}_{\overline{\bm{\gamma}}'}]_{ij} + \delta_{mj} \,
[{\bm{1}}_{\overline{\bm{\gamma}}'}]_{ij} \big) - \xi \big(
\delta_{\xi+1,\, i} \, [{\bm{1}}_{\overline{\bm{\gamma}}'}]_{ij} +
\delta_{\xi+1,\, j} \, [{\bm{1}}_{\overline{\bm{\gamma}}'}]_{ij}
\big) \right) }
\\ [1.5em] & = & {\displaystyle \frac{1}{\sqrt{\xi(\xi+1)(N-2)D}} \, [{\bm{S}}_{\overline{\bm{\gamma}}'}^{[N-1,
\hspace{1ex} 1]}]_\xi \, \bigg( \big( \Theta_{\xi-i+1} \,
[{\bm{1}}_{\overline{\bm{\gamma}}'}]_{ij} + \Theta_{\xi-j+1} \,
[{\bm{1}}_{\overline{\bm{\gamma}}'}]_{ij} \big) - \xi \big(
\delta_{\xi+1,\, i} \, [{\bm{1}}_{\overline{\bm{\gamma}}'}]_{ij} +
\delta_{\xi+1,\, j} \, [{\bm{1}}_{\overline{\bm{\gamma}}'}]_{ij}
\big) \bigg)
 \,. } \end{array}
\end{equation}
Thus we see that the motion associated with symmetry coordinate
$[{\bm{S}}_{\overline{\bm{\gamma}}'}^{[N-1, \hspace{1ex} 1]}]_1$
is an antisymmetric bending motion about the Lewis structure
configuration where as the angle cosines $\gamma_{13}$,
$\gamma_{14}$, $\gamma_{15}$, \ldots increase, $\gamma_{23}$,
$\gamma_{24}$, $\gamma_{25}$, \ldots decrease and vice versa while
$\gamma_{12}$, $\gamma_{34}$, $\gamma_{35}$, $\gamma_{45}$, \ldots
remain unchanged. Thus as for the $\bar{\bm{r}}'$ sector of the
$[N-1, \hspace{1ex} 1]$ species,
$[{\bm{S}}_{\overline{\bm{\gamma}}'}^{[N-1, \hspace{1ex} 1]}]_1$
involves the motions of particles $1$ and $2$ moving with opposite
phase to each other against the remaining particles. As $\xi$
gets larger, the motion involves more particles, $\xi+1$
particles, where the $(\xi+1)^{\rm th}$ particle moves with
opposite phase to the first $\xi$ particles. Paradoxically though,
and again like the $\bar{\bm{r}}'$ sector of the $[N-1,
\hspace{1ex} 1]$ species, as $\xi$ increases the motion becomes
more single-particle-like since the angle cosines involving the
$(\xi+1)^{\rm th}$ particle in Eq.~(\ref{eq:gNm1ing}) are weighted
by the quantity $\xi$\,.

\subsection{Symmetry Coordinates Belonging to the $\bm{[N-2,}
    \hspace{1ex} \bm{2]}$ Species.}

There is only one sector, the $\overline{\bm{\gamma}}'$ sector,
belonging to the $[N-2, \hspace{1ex} 2]$ species.

\paragraph{The Primitive Irreducible Coordinate.}
Consider the quantities
\begin{equation}
(\widehat{{\bm{r}}_i} - \widehat{{\bm{r}}_{j}}) \, {\bm{.}} \,
(\widehat{{\bm{r}}_k} - \widehat{{\bm{r}}_{l}})  =
\gamma_{ik}-\gamma_{il}+\gamma_{jl}-\gamma_{jk} \,,
\mbox{\hspace{2ex}for $i\neq j \neq k \neq l$ and $j>i$, $l>k$\,.}
\end{equation}
There are $N!/[2^3  (N-4)!]$ such elements\cite{Hamermesh_p._27}
which transform into themselves under $S_N$\,. According to
Eq.~(\ref{eq:taylor2})
\begin{equation}
\gamma_{ik}-\gamma_{il}+\gamma_{jl}-\gamma_{jk} =
\frac{1}{\sqrt{D}}
(\overline{\gamma}'_{ik}-\overline{\gamma}'_{il}+\overline{\gamma}'_{jl}-\overline{\gamma}'_{jk})
\end{equation}
so that
\begin{equation}
\overline{\gamma}'_{ik}-\overline{\gamma}'_{il}+\overline{\gamma}'_{jl}-\overline{\gamma}'_{jk}
= (\widehat{{\bm{r}}_i}' - \widehat{{\bm{r}}_j}') \, {\bm{.}} \,
(\widehat{{\bm{r}}_k}' - \widehat{{\bm{r}}_{l}}') \,,
\mbox{\hspace{2ex}where $i\neq j \neq k \neq l$ and $j>i$, $l>k$}
\end{equation}
and
\begin{equation}
\widehat{{\bm{r}}_i}' = \sqrt[4]{D} \widehat{{\bm{r}}_i} \,.
\end{equation}
The $N!/[2^3  (N-4)!]$ \, $(\widehat{{\bm{r}}_i}' -
\widehat{{\bm{r}}_j}') \, {\bm{.}} \, (\widehat{{\bm{r}}_k}' -
\widehat{{\bm{r}}_{l}}')$ also transform into themselves under
$S_N$\,. We wish to show that there are $N(N-3)/2$
linearly independent $(\widehat{{\bm{r}}_i}' -
\widehat{{\bm{r}}_j}') \, {\bm{.}} \, (\widehat{{\bm{r}}_k}' -
\widehat{{\bm{r}}_{l}}')$, a particular set of which we choose to
be the primitive irreducible coordinates of the $[N-2,
\hspace{1ex} 2]$ species.

There are two distinct sets of linear dependencies relating the
$N!/[2^3  (N-4)!]$ \, $(\widehat{{\bm{r}}_i}' -
\widehat{{\bm{r}}_j}') \, {\bm{.}} \, (\widehat{{\bm{r}}_k}' -
\widehat{{\bm{r}}_{l}}')$\,,
\begin{eqnarray}
(\widehat{{\bm{r}}_i}' - \widehat{{\bm{r}}_j}') \, {\bm{.}} \,
(\widehat{{\bm{r}}_m}' - \widehat{{\bm{r}}_{l}}') +
(\widehat{{\bm{r}}_i}' - \widehat{{\bm{r}}_j}') \, {\bm{.}} \,
(\widehat{{\bm{r}}_k}' - \widehat{{\bm{r}}_{m}}') & = &
(\widehat{{\bm{r}}_i}' - \widehat{{\bm{r}}_j}') \, {\bm{.}} \,
(\widehat{{\bm{r}}_k}' -
\widehat{{\bm{r}}_{l}}') \mbox{\hspace{2ex} and} \label{eq:lindep1} \\
(\widehat{{\bm{r}}_i}' - \widehat{{\bm{r}}_k}') \, {\bm{.}} \,
(\widehat{{\bm{r}}_j}' - \widehat{{\bm{r}}_{l}}') +
(\widehat{{\bm{r}}_i}' - \widehat{{\bm{r}}_l}') \, {\bm{.}} \,
(\widehat{{\bm{r}}_k}' - \widehat{{\bm{r}}_j}') & = &
(\widehat{{\bm{r}}_i}' - \widehat{{\bm{r}}_j}') \, {\bm{.}} \,
(\widehat{{\bm{r}}_k}' - \widehat{{\bm{r}}_{l}}') \,.
\label{eq:lindep2}
\end{eqnarray}
The number $N!/[2^3  (N-4)!]$ is huge when $N$ is large, but the
linear dependencies of Eqs.~(\ref{eq:lindep1}) and
(\ref{eq:lindep2}) however, drastically reduce the number to
$N(N-3)/2$ linearly independent terms.

Through Eq.~(\ref{eq:lindep1}) we can write
\begin{equation}
(\widehat{{\bm{r}}_i}' - \widehat{{\bm{r}}_j}') \, {\bm{.}} \,
(\widehat{{\bm{r}}_k}' - \widehat{{\bm{r}}_{l}}') =
\sum_{\kappa=0}^{j-i-1} (\widehat{{\bm{r}}_{i+\kappa}}' -
\widehat{{\bm{r}}_{i+1+\kappa}}') \, {\bm{.}} \,
(\widehat{{\bm{r}}_k}' - \widehat{{\bm{r}}_{l}}')\,.
\end{equation}
Thus we only need to consider the $(N-1)(N-2)(N-3)/2$ \,
$(\widehat{{\bm{r}}_i}' - \widehat{{\bm{r}}_{i+1}}') \, {\bm{.}}
\, (\widehat{{\bm{r}}_k}' - \widehat{{\bm{r}}_{l}}')$\,, where
$i\neq j \neq k \neq l$ and $l>k$\,.

Equation~(\ref{eq:lindep1}) may be used iteratively again to
reduce the number of $(\widehat{{\bm{r}}_i}' -
\widehat{{\bm{r}}_{i+1}}') \, {\bm{.}} \, (\widehat{{\bm{r}}_k}' -
\widehat{{\bm{r}}_{l}}')$ by writing
\begin{equation}
(\widehat{{\bm{r}}_i}' - \widehat{{\bm{r}}_{i+1}}') \, {\bm{.}} \,
(\widehat{{\bm{r}}_k}' - \widehat{{\bm{r}}_{l}}') =
\sum_{\zeta=0}^{l-k-1} (\widehat{{\bm{r}}_i}' -
\widehat{{\bm{r}}_{i+1}}') \, {\bm{.}} \,
(\widehat{{\bm{r}}_{k+\zeta}}' - \widehat{{\bm{r}}_{k+1+\zeta}}')
\mbox{\hspace{2ex}when $l<i$ or $k>i+1$\,,}
\end{equation}
and
\begin{eqnarray}
\lefteqn{(\widehat{{\bm{r}}_i}' - \widehat{{\bm{r}}_{i+1}}') \,
{\bm{.}} \,
(\widehat{{\bm{r}}_k}' - \widehat{{\bm{r}}_{l}}') = } \nonumber \\
& = & \hspace{-1ex} \begin{array}[t]{@{}r@{}l} & {\displaystyle
\sum_{\zeta=0}^{i-k-2} (\widehat{{\bm{r}}_i}' -
\widehat{{\bm{r}}_{i+1}}') \, {\bm{.}} \,
(\widehat{{\bm{r}}_{k+\zeta}}' - \widehat{{\bm{r}}_{k+1+\zeta}}')
+ (\widehat{{\bm{r}}_i}' - \widehat{{\bm{r}}_{i+1}}') \, {\bm{.}}
\, (\widehat{{\bm{r}}_{i-1}}' - \widehat{{\bm{r}}_{i+2}}') + } \\
+ & {\displaystyle \sum_{\zeta=0}^{l-i-3} (\widehat{{\bm{r}}_i}' -
\widehat{{\bm{r}}_{i+1}}') \, {\bm{.}} \,
(\widehat{{\bm{r}}_{i+2+\zeta}}' -
\widehat{{\bm{r}}_{i+3+\zeta}}')} \mbox{\hspace{2ex}when $k<i$ and
$l>i+1$\,.} \end{array}
\end{eqnarray}
Thus we only need to consider the $(\widehat{{\bm{r}}_i}' -
\widehat{{\bm{r}}_{i+1}}') \, {\bm{.}} \,
(\widehat{{\bm{r}}_{l-1}}' - \widehat{{\bm{r}}_l}')$ where $2\leq
l \leq i-1$ or $i+3 \leq l \leq N$ and $1 \leq i \leq N-1$, and
$(\widehat{{\bm{r}}_i}' - \widehat{{\bm{r}}_{i+1}}') \, {\bm{.}}
\, (\widehat{{\bm{r}}_{i-1}}' - \widehat{{\bm{r}}_{i+2}}')$, where
$1 \leq i \leq N-1$\,, i.e.\ $(N-1)(N-3)$ elements to consider.

Now let's apply Eq.~(\ref{eq:lindep2}) to this reduced subset.
Consider the case where $2\leq l \leq i-1$. In this case
Eq.~(\ref{eq:lindep2}) yields the unsurprising result
\begin{equation} \begin{array}{r@{\hspace{0.5ex}}c@{\hspace{0.5ex}}l}
(\widehat{{\bm{r}}_i}' - \widehat{{\bm{r}}_{i+1}}') \, {\bm{.}} \,
(\widehat{{\bm{r}}_{l-1}}' - \widehat{{\bm{r}}_l}') & = &
(\widehat{{\bm{r}}_{l-1}}' - \widehat{{\bm{r}}_i}') \, {\bm{.}} \,
(\widehat{{\bm{r}}_l}' - \widehat{{\bm{r}}_{i+1}}') -
(\widehat{{\bm{r}}_l}' - \widehat{{\bm{r}}_i}') \, {\bm{.}} \,
(\widehat{{\bm{r}}_{l-1}}' -
\widehat{{\bm{r}}_{i+1}}') = \\
& = & (\widehat{{\bm{r}}_{i}}' - \widehat{{\bm{r}}_{l-1}}') \,
{\bm{.}} \, (\widehat{{\bm{r}}_{i+1}}' - \widehat{{\bm{r}}_l}') -
(\widehat{{\bm{r}}_i}' - \widehat{{\bm{r}}_l}') \, {\bm{.}} \,
(\widehat{{\bm{r}}_{i+1}}' - \widehat{{\bm{r}}_{l-1}}') =
\\ & = & (\widehat{{\bm{r}}_{l-1}}' -
\widehat{{\bm{r}}_l}') \, {\bm{.}} \, (\widehat{{\bm{r}}_i}' -
\widehat{{\bm{r}}_{i+1}}') \,,
\end{array}
\end{equation}
i.e.\ we can write all of the $(\widehat{{\bm{r}}_i}' -
\widehat{{\bm{r}}_{i+1}}') \, {\bm{.}} \,
(\widehat{{\bm{r}}_{l-1}}' - \widehat{{\bm{r}}_l}')$ where $2\leq
l \leq i-1$ and $3 \leq i \leq N-1$ in terms of the
$(\widehat{{\bm{r}}_i}' - \widehat{{\bm{r}}_{i+1}}') \, {\bm{.}}
\, (\widehat{{\bm{r}}_{l-1}}' - \widehat{{\bm{r}}_l}')$ with $i+3
\leq l \leq N$ and $1 \leq i \leq N-3$\,.

Thus we reduce our $N!/[2^3  (N-4)!]$ elements of the form
$(\widehat{{\bm{r}}_i} - \widehat{{\bm{r}}_{j}}) \, {\bm{.}} \,
(\widehat{{\bm{r}}_k} - \widehat{{\bm{r}}_{l}})$, where $i\neq j
\neq k \neq l$, $j>i$ and $l>k$, down to the $(N-2)(N-3)/2$
elements $(\widehat{{\bm{r}}_i}' - \widehat{{\bm{r}}_{i+1}}') \,
{\bm{.}} \, (\widehat{{\bm{r}}_{l-1}}' - \widehat{{\bm{r}}_l}')$
with $i+3 \leq l \leq N$ and $1 \leq i \leq N-3$, and the $N-3$
elements of the form $(\widehat{{\bm{r}}_i}' -
\widehat{{\bm{r}}_{i+1}}') \, {\bm{.}} \,
(\widehat{{\bm{r}}_{i-1}}' - \widehat{{\bm{r}}_{i+2}}')$, where $2
\leq i \leq N-2$\,. Hence we have $N(N-3)/2$ linearly elements,
which is exactly the dimensionality of the $[N-2, \hspace{1ex} 2]$
representation. We identify these elements as the primitive
irreducible coordinates of the $[N-2, \hspace{1ex} 2]$
representation and write
\begin{equation}
\overline{\bm{S}}_{\overline{\bm{\gamma}}'}^{[N-2, \hspace{1ex}
2]} = \left(
\begin{array}{r@{\,}c@{\,}l} (\widehat{{\bm{r}}_1}' -
\widehat{{\bm{r}}_2}') & {\bm{.}} & (\widehat{{\bm{r}}_3}' -
\widehat{{\bm{r}}_4}') \\
(\widehat{{\bm{r}}_2}' - \widehat{{\bm{r}}_3}') & {\bm{.}}
& (\widehat{{\bm{r}}_1}' - \widehat{{\bm{r}}_4}') \\
\hline (\widehat{{\bm{r}}_1}' - \widehat{{\bm{r}}_2}') & {\bm{.}}
& (\widehat{{\bm{r}}_4}' -
\widehat{{\bm{r}}_5}') \\
(\widehat{{\bm{r}}_2}' - \widehat{{\bm{r}}_3}') & {\bm{.}} &
(\widehat{{\bm{r}}_4}' -
\widehat{{\bm{r}}_5}') \\
(\widehat{{\bm{r}}_3}' - \widehat{{\bm{r}}_4}') & {\bm{.}}
& (\widehat{{\bm{r}}_2}' - \widehat{{\bm{r}}_5}') \\
\hline (\widehat{{\bm{r}}_1}' - \widehat{{\bm{r}}_2}') & {\bm{.}}
&
(\widehat{{\bm{r}}_5}' - \widehat{{\bm{r}}_6}') \\
(\widehat{{\bm{r}}_2}' - \widehat{{\bm{r}}_3}') &
{\bm{.}} & (\widehat{{\bm{r}}_5}' - \widehat{{\bm{r}}_6}') \\
(\widehat{{\bm{r}}_3}' - \widehat{{\bm{r}}_4}') &
{\bm{.}} & (\widehat{{\bm{r}}_5}' - \widehat{{\bm{r}}_6}') \\
(\widehat{{\bm{r}}_4}' - \widehat{{\bm{r}}_5}') & {\bm{.}}
& (\widehat{{\bm{r}}_3}' - \widehat{{\bm{r}}_6}') \\
\hline (\widehat{{\bm{r}}_1}' - \widehat{{\bm{r}}_2}') & {\bm{.}}
&
(\widehat{{\bm{r}}_6}' - \widehat{{\bm{r}}_7}') \\
(\widehat{{\bm{r}}_2}' - \widehat{{\bm{r}}_3}') &
{\bm{.}} & (\widehat{{\bm{r}}_6}' - \widehat{{\bm{r}}_7}') \\
(\widehat{{\bm{r}}_3}' - \widehat{{\bm{r}}_4}') &
{\bm{.}} & (\widehat{{\bm{r}}_6}' - \widehat{{\bm{r}}_7}') \\
(\widehat{{\bm{r}}_4}' - \widehat{{\bm{r}}_5}') &
{\bm{.}} & (\widehat{{\bm{r}}_6}' - \widehat{{\bm{r}}_7}') \\
(\widehat{{\bm{r}}_5}' - \widehat{{\bm{r}}_6}') & {\bm{.}}
& (\widehat{{\bm{r}}_4}' - \widehat{{\bm{r}}_7}') \\
\hline (\widehat{{\bm{r}}_1}' - \widehat{{\bm{r}}_2}') & {\bm{.}}
& (\widehat{{\bm{r}}_7}' - \widehat{{\bm{r}}_8}')
\\ \multicolumn{1}{c}{\vdots} & & \multicolumn{1}{c}{\vdots}
\end{array} \right) \,,
\end{equation}
or
\begin{eqnarray}
[\overline{\bm{S}}_{\overline{\bm{\gamma}}'}^{[N-2, \hspace{1ex}
2]}]_{ij} & = & (\widehat{{\bm{r}}_i}' -
\widehat{{\bm{r}}_{i+1}}') \, {\bm{.}} \,
(\widehat{{\bm{r}}_{j-1}}' - \widehat{{\bm{r}}_j}')
\mbox{\hspace{2ex}when $i+3 \leq j \leq
N$ and $1 \leq i \leq N-3$} \label{eq:SgNm2eqa} \\
& = & (\widehat{{\bm{r}}_i}' - \widehat{{\bm{r}}_{i+1}}') \,
{\bm{.}} \, (\widehat{{\bm{r}}_{j-3}}' - \widehat{{\bm{r}}_j}')
\mbox{\hspace{2ex}when $j=i+2$ and $2 \leq i \leq N-2$} \,.
\label{eq:SgNm2eqb}
\end{eqnarray}
There are a myriad of ways to choose a set of $N(N-3)/2$ linearly
independent, primitive irreducible coordinates from the $N!/[2^3
(N-4)!]$\,\, $(\widehat{{\bm{r}}_i}' - \widehat{{\bm{r}}_j}') \,
{\bm{.}} \, (\widehat{{\bm{r}}_k}' - \widehat{{\bm{r}}_l}')$\,.
However, the set we have chosen has some significant advantages as
we shall see below.

What about $\overline{W}^{[N-2, \hspace{1ex}
2]}_{\overline{\bm{\gamma}}'}$\,? From Eqs.~(\ref{eq:gij}),
(\ref{eq:SgNm2eqa}) and (\ref{eq:SgNm2eqb}) we can write
\begin{equation} \label{eq:SgNm2eqexp}
[\overline{\bm{S}}_{\overline{\bm{\gamma}}'}^{[N-2, \hspace{1ex}
2]}]_{ij} = \sum_{m=1}^N \sum_{n=1}^N
(\delta_{im}-\delta_{i+1,\,m})(\delta_{kn}-\delta_{jn}) \,
\widehat{{\bm{r}}_m}' \, {\bm{.}} \, \widehat{{\bm{r}}_n}' =
\sum_{m=1}^N \sum_{n=1}^N
(\delta_{im}-\delta_{i+1,\,m})(\delta_{kn}-\delta_{jn}) \,
\overline{\gamma}'_{mn} \,,
\end{equation}
where $k=j-1$ when $i+3 \leq j \leq N$ and $1 \leq i \leq N-3$\,,
or $k=j-3$ when $j=i+2$ and $2 \leq i \leq N-2$\,. Since
$\overline{\gamma}'_{mn}$ is a symmetric matrix,
Eq.~(\ref{eq:SgNm2eqexp}) can be rewritten as
\begin{equation} \label{eq:SgNm2eqWg}
[\overline{\bm{S}}_{\overline{\bm{\gamma}}'}^{[N-2, \hspace{1ex}
2]}]_{ij} = \sum_{n=2}^N \sum_{m=1}^{n-1} \big[
(\delta_{im}-\delta_{i+1,\,m})(\delta_{kn}-\delta_{jn}) +
(\delta_{in}-\delta_{i+1,\,n})(\delta_{km}-\delta_{jm}) \big] \,
\overline{\gamma}'_{mn} \,,
\end{equation}
where $k=j-1$ when $i+3 \leq j \leq N$ and $1 \leq i \leq N-3$\,,
or $k=j-3$ when $j=i+2$ and $2 \leq i \leq N-2$\,. Thus from
Eq.~(\ref{eq:SXp}), we note that
\begin{equation} \label{eq:Wsum2}
W_{\overline{\bm{\gamma}}'}^\alpha \, \overline{\bm{\gamma}}' =
\sum_{j=1}^N \sum_{i < j} \,\,
[W_{\overline{\bm{\gamma}}'}^\alpha]_{ij} \,
\overline{\gamma}'_{ij} \,,
\end{equation}
and Eq.~(\ref{eq:SgNm2eqWg}) $\overline{W}^{[N-2, \hspace{1ex}
2]}_{\overline{\bm{\gamma}}'}$ may be identified as
\begin{equation} \label{eq:Wnm2eqdelta}
[\overline{W}^{[N-2, \hspace{1ex}
2]}_{\overline{\bm{\gamma}}'}]_{ij,\,mn} =
(\delta_{im}-\delta_{i+1,\,m})(\delta_{kn}-\delta_{jn}) +
(\delta_{in}-\delta_{i+1,\,n})(\delta_{km}-\delta_{jm}) \,,
\end{equation}
where $2 \leq n \leq N$ and $1 \leq m \leq n-1$\,, $k=j-1$ when
$i+3 \leq j \leq N$ and $1 \leq i \leq N-3$\,, or $k=j-3$ when
$j=i+2$ and $2 \leq i \leq N-2$\,. Equation~(\ref{eq:Wnm2eqdelta})
and the above conditions on $k$ can be written as
\begin{equation} \label{eq:Wnm2eqdeltakinc}
[\overline{W}^{[N-2, \hspace{1ex}
2]}_{\overline{\bm{\gamma}}'}]_{ij,\,mn} =
\renewcommand{\arraystretch}{1.5}
\begin{array}[t]{r@{\hspace{0.5ex}}l} & (\delta_{im}-\delta_{i+1,\,m}) \biggl( (1-\delta_{i,\,j-2})
\delta_{j-1,\,n} + \delta_{i,\,j-2} \, \delta_{j-3,\,n} -
\delta_{jn} \biggr) \\ + & (\delta_{in}-\delta_{i+1,\,n}) \biggl(
(1-\delta_{i,\,j-2}) \delta_{j-1,\,m} + \delta_{i,\,j-2} \,
\delta_{j-3,\,m} - \delta_{jm} \biggr) \,, \end{array}
\renewcommand{\arraystretch}{1}
\end{equation}
where $1 \leq i \leq j-2$\,, $i+2 \leq j \leq N$\,, $2 \leq n \leq
N$ and $1 \leq m \leq n-1$\,.

\paragraph{The Symmetry Coordinate.}
We derive the symmetry coordinate
${\bm{S}}_{\overline{\bm{\gamma}}'}^{[N-2, \hspace{1ex} 2]}$ from
Eq.~(\ref{eq:SUSb}), via Eq.~(\ref{eq:aaI}) to calculate $U^{[N-2,
\hspace{1ex} 2]}_{\overline{\bm{\gamma}}'}$, and
Eqs.~(\ref{eq:SgNm2eqa}) and (\ref{eq:SgNm2eqb}) for
$\overline{\bm{S}}_{\overline{\bm{\gamma}}'}^{[N-2, \hspace{1ex}
2]}$. The first step in this process is to evaluate
$\overline{W}^{[N-2, \hspace{1ex} 2]}_{\overline{\bm{\gamma}}'} \,
[\overline{W}^{[N-2, \hspace{1ex}
2]}_{\overline{\bm{\gamma}}'}]^T$\,. Using
Eq.~(\ref{eq:Wnm2eqdelta}) we derive
\begin{eqnarray}
\lefteqn{\bigl[ \overline{W}^{[N-2, \hspace{1ex}
2]}_{\overline{\bm{\gamma}}'} \, [\overline{W}^{[N-2, \hspace{1ex}
2]}_{\overline{\bm{\gamma}}'}]^T \bigr]_{ij,\,i'j'} = \sum_{n=2}^N
\sum_{m=1}^{n-1} [\overline{W}^{[N-2, \hspace{1ex}
2]}_{\overline{\bm{\gamma}}'}]_{ij,\,mn} \,
[\overline{W}^{[N-2, \hspace{1ex} 2]}_{\overline{\bm{\gamma}}'}]_{i'j',\,mn} } \nonumber \\
& = & \sum_{n=2}^N \sum_{m=1}^{n-1} \begin{array}[t]{r@{}l} &
\bigl[ (\delta_{im}-\delta_{i+1,\,m})(\delta_{kn}-\delta_{jn}) +
(\delta_{in}-\delta_{i+1,\,n})(\delta_{km}-\delta_{jm}) \bigr]
\times
\\ \times & \bigl[ (\delta_{i'm}-\delta_{i'+1,\,m})(\delta_{k'n}-\delta_{j'n})
+ (\delta_{i'n}-\delta_{i'+1,\,n})(\delta_{k'm}-\delta_{j'm})
\bigr]
 \end{array} \nonumber \\
& = & \sum_{n=1}^N \sum_{m=1}^{N} \begin{array}[t]{r@{}l} &
(\delta_{im}-\delta_{i+1,\,m})(\delta_{kn}-\delta_{jn}) \times
\\ \times & \bigl[ (\delta_{i'm}-\delta_{i'+1,\,m})(\delta_{k'n}-\delta_{j'n})
+ (\delta_{i'n}-\delta_{i'+1,\,n})(\delta_{k'm}-\delta_{j'm})
\bigr] \,, \end{array}
 \label{WbWbT1}
\end{eqnarray}
where the $m=n$ contribution is zero since $i \neq i+1 \neq k \neq
j$ and $i' \neq i'+1 \neq k' \neq j'$\,. Thus
\renewcommand{\jot}{0.5em}
\begin{eqnarray}
\bigl[ \overline{W}^{[N-2, \hspace{1ex}
2]}_{\overline{\bm{\gamma}}'} \, [\overline{W}^{[N-2, \hspace{1ex}
2]}_{\overline{\bm{\gamma}}'}]^T \bigr]_{ij,\,i'j'} & = &
\hspace{-1.5em}
\begin{array}[t]{r@{}l} & (\delta_{ii'}-\delta_{i,\, i'+1} -
\delta_{i+1,\, i'} + \delta_{i+1,\,
i'+1})(\delta_{kk'}-\delta_{kj'} - \delta_{jk'} +
\delta_{jj'}) + \\
+ & (\delta_{ik'}-\delta_{ij'} - \delta_{i+1,\, k'} +
\delta_{i+1,\, j'})(\delta_{ki'}-\delta_{k,\, i'+1} - \delta_{ji'}
+ \delta_{j,\,i'+1}) \end{array} \nonumber \\
& = & R_{i(i+1)kj,\,i'(i'+1)k'j'} + R_{i(i+1)kj,\,k'j'i'(i'+1)}
\,, \label{eq:WbWbT2}
\end{eqnarray}
\renewcommand{\jot}{0em}
where
\begin{equation}
R_{abcd,\,a'b'c'd'} = (\delta_{aa'}-\delta_{ab'} - \delta_{ba'} +
\delta_{bb'})(\delta_{cc'}-\delta_{cd'} - \delta_{dc'} +
\delta_{dd'}) \,,
\end{equation}
$k=j-1$ when $i+3 \leq j \leq N$ and $1 \leq i \leq N-3$\,, or
$k=j-3$ when $j=i+2$ and $2 \leq i \leq N-2$\,. Likewise,
$k'=j'-1$ when $i'+3 \leq j' \leq N$ and $1 \leq i' \leq N-3$\,,
or $k'=j'-3$ when $j'=i'+2$ and $2 \leq i' \leq N-2$\,.

We will find $U^{[N-2, \hspace{1ex} 2]}_{\overline{\bm{\gamma}}'}$
of Eqs.~(\ref{eq:WaXUaXWbaX}) and (\ref{eq:aaI}) in a two step
process. First of all we will introduce a transformation
${\mathcal{J}}^{[N-2, \hspace{1ex} 2]}_{\overline{\bm{\gamma}}'}$
which block diagonalizes Eq.~(\ref{eq:WbWbT2}) in the $j$ and $j'$
indices. Subsequently we will then introduce a further
transformation ${\mathcal{I}}^{[N-2, \hspace{1ex}
2]}_{\overline{\bm{\gamma}}'}$ which diagonalizes the $j$ blocks
in the indices $i$ and $i'$ to yield Eq.~(\ref{eq:aaI}). Thus we
have that
\begin{equation} \label{eq:UNm2eqIJ}
U^{[N-2, \hspace{1ex} 2]}_{\overline{\bm{\gamma}}'} =
{\mathcal{I}}^{[N-2, \hspace{1ex} 2]}_{\overline{\bm{\gamma}}'} \,
{\mathcal{J}}^{[N-2, \hspace{1ex} 2]}_{\overline{\bm{\gamma}}'}
\,.
\end{equation}
Let's work with the transformation ${\mathcal{J}}^{[N-2,
\hspace{1ex} 2]}_{\overline{\bm{\gamma}}'}$ first. If we take
${\mathcal{J}}^{[N-2, \hspace{1ex} 2]}_{\overline{\bm{\gamma}}'}$
to be
\begin{equation} \label{eq:JNm2g}
[{\mathcal{J}}^{[N-2, \hspace{1ex}
2]}_{\overline{\bm{\gamma}}'}]_{ij,\,i'j'} =
[{\bm{1}}_{\bar{\bm{r}}'}]_j \, \delta_{i+1,\, j'} \,
(1-\delta_{i-1,\, i'}) i' + \delta_{ii'} \, \Theta_{j-j'+1} \,
\Theta_{j'-i-1} (j'-3) \,,
\end{equation}
then after some calculation we find
\begin{eqnarray}
\lefteqn{ \Bigl[ {\mathcal{J}}^{[N-2, \hspace{1ex}
2]}_{\overline{\bm{\gamma}}'} \, \overline{W}^{[N-2, \hspace{1ex}
2]}_{\overline{\bm{\gamma}}'} \, [ {\mathcal{J}}^{[N-2,
\hspace{1ex} 2]}_{\overline{\bm{\gamma}}'} \, \overline{W}^{[N-2,
\hspace{1ex} 2]}_{\overline{\bm{\gamma}}'}]^T \Bigr]_{ij,\,i'j'} =} \nonumber \\
& = & \sum_{l=4}^N \sum_{k=1}^{l-2} \,\, \sum_{l'=4}^N
\sum_{k'=1}^{l'-2} \, [{\mathcal{J}}^{[N-2, \hspace{1ex}
2]}_{\overline{\bm{\gamma}}'}]_{ij,\,kl} \,\Bigl[
\overline{W}^{[N-2, \hspace{1ex} 2]}_{\overline{\bm{\gamma}}'} \,
[ \overline{W}^{[N-2, \hspace{1ex}
2]}_{\overline{\bm{\gamma}}'}]^T \Bigr]_{kl,\,k'l'} \,
[{\mathcal{J}}^{[N-2, \hspace{1ex}
2]}_{\overline{\bm{\gamma}}'}]_{i'j',\,k'l'} \nonumber \\ & = &
\vphantom{\sum_{l=4}^N \sum_{k=1}^{l-2}}
(2\delta_{ii'}-\delta_{i,\,i'+1}-\delta_{i+1,\,i'})(j-3)(j-2)
\delta_{jj'} \label{eq:JWbWbTJT}
\end{eqnarray}
Thus the transformation ${\mathcal{J}}^{[N-2, \hspace{1ex}
2]}_{\overline{\bm{\gamma}}'}$ has block diagonalized
Eq.~(\ref{eq:WbWbT2}) in the $j$ and $j'$ indices as advertised.

To understand what the transformation ${\mathcal{J}}^{[N-2,
\hspace{1ex} 2]}_{\overline{\bm{\gamma}}'}$ actually achieves,
let's apply it to the primitive irreducible coordinate
$\overline{\bm{S}}_{\overline{\bm{\gamma}}'}^{[N-2, \hspace{1ex}
2]}$\,. Defining
\begin{eqnarray}
\lefteqn{\bigl[ {\bm{T}}_{\overline{\bm{\gamma}}'}^{[N-2,
\hspace{1ex} 2]}\bigr]_{ij} = \bigl[{\mathcal{J}}^{[N-2,
\hspace{1ex} 2]}_{\overline{\bm{\gamma}}'} \,
\overline{\bm{S}}_{\overline{\bm{\gamma}}'}^{[N-2, \hspace{1ex}
2]} \bigr]_{ij}  }
\nonumber \\
 & = & \sum_{j'=4}^N
\sum_{i'=1}^{j'-2} [{\mathcal{J}}^{[N-2, \hspace{1ex}
2]}_{\overline{\bm{\gamma}}'}]_{ij,\,i'j'}
\,[\overline{\bm{S}}_{\overline{\bm{\gamma}}'}^{[N-2, \hspace{1ex}
2]}]_{i'
j'}  \nonumber \\
& = & \sum_{j'=4}^N \sum_{i'=1}^{j'-2} \Bigl[
[{\bm{1}}_{\bar{\bm{r}}'}]_j \, \delta_{i+1,\, j'} \,
(1-\delta_{i-1,\, i'}) i' + \delta_{ii'} \, \Theta_{j-j'+1} \,
\Theta_{j'-i-1} (j'-3) \Bigr]
\,[\overline{\bm{S}}_{\overline{\bm{\gamma}}'}^{[N-2, \hspace{1ex}
2]}]_{i' j'}  \nonumber \\
& = & \hspace{-1ex} \begin{array}[t]{@{}r@{\hspace{0.5ex}}l} &
{\displaystyle \sum_{j'=4}^N \sum_{i'=1}^{j'-2}
[{\bm{1}}_{\bar{\bm{r}}'}]_j \, \delta_{i+1,\, j'} \,
(1-\delta_{i-1,\, i'}) i' \,
[\overline{\bm{S}}_{\overline{\bm{\gamma}}'}^{[N-2, \hspace{1ex}
2]}]_{i' j'} \hspace{1ex} +} \\ + & {\displaystyle \sum_{j'=4}^{N}
\sum_{i'=1}^{j'-2} \delta_{ii'} \, \Theta_{j-j'+1} \,
\Theta_{j'-i-1}
(j'-3)\,[\overline{\bm{S}}_{\overline{\bm{\gamma}}'}^{[N-2,
\hspace{1ex} 2]}]_{i' j'} }  \end{array} \nonumber \\
& = & \sum_{i'=1}^{i-2}
i'\,[\overline{\bm{S}}_{\overline{\bm{\gamma}}'}^{[N-2,
\hspace{1ex} 2]}]_{i',\, i+1} + \sum_{j'=i+2}^j
(j'-3)\,[\overline{\bm{S}}_{\overline{\bm{\gamma}}'}^{[N-2,
\hspace{1ex}
2]}]_{i j'}  \nonumber \\
& = & \sum_{i'=1}^{i-2}
i'\,[\overline{\bm{S}}_{\overline{\bm{\gamma}}'}^{[N-2,
\hspace{1ex} 2]}]_{i',\, i+1} +
(i-1)\,[\overline{\bm{S}}_{\overline{\bm{\gamma}}'}^{[N-2,
\hspace{1ex} 2]}]_{i,\, i+2} + \sum_{j'=i+3}^{j}
(j'-3)\,[\overline{\bm{S}}_{\overline{\bm{\gamma}}'}^{[N-2,
\hspace{1ex}
2]}]_{i j'}  \nonumber \\
& = & (\widehat{{\bm{r}}_i}' - \widehat{{\bm{r}}_{i+1}}') \,
{\bm{.}} \, (\widehat{{\bm{r}}_1}' + \widehat{{\bm{r}}_2}' +
\widehat{{\bm{r}}_3}' + \cdots + \widehat{{\bm{r}}_{i-1}}' +
\widehat{{\bm{r}}_{i+2}}' + \widehat{{\bm{r}}_{i+3}}' + \cdots +
\widehat{{\bm{r}}_{j-1}}' - (j-3) \widehat{{\bm{r}}_j}') 
\nonumber \\ & = & (\widehat{{\bm{r}}_i}' -
\widehat{{\bm{r}}_{i+1}}') \, {\bm{.}} \, \left( \sum_{j'=1}^{j-1}
\widehat{{\bm{r}}_{j'}}' - (j-3) \widehat{{\bm{r}}_j}' -
(\widehat{{\bm{r}}_i}' + \widehat{{\bm{r}}_{i+1}}') \right) \,,
\label{eq:JSb}
\end{eqnarray}
where ${\bm{1}}_{\bar{\bm{r}}'}$ is defined in
Eq.~(\ref{eq:bf1i}), $\Theta_{i-j}$ is defined in
Eq.~(\ref{eq:thetaalphabeta}), and Eqs~(\ref{eq:SgNm2eqa}) and
(\ref{eq:SgNm2eqb}) have been used in the last step. The last term
in parenthesis in the last line of Eq.~(\ref{eq:JSb}) should be
compared with Eq.~(\ref{eq:SNm1}). Not surprisingly, it has the
same form as Eq.~(\ref{eq:SNm1}), except that it includes $N-2$
particles, the $i^{\rm th}$ and $(i+1)^{\rm th}$ particles are not
included. We further note that since
\begin{equation}
(\widehat{{\bm{r}}_i}' - \widehat{{\bm{r}}_{i+1}}') \, {\bm{.}} \,
(\widehat{{\bm{r}}_i}' + \widehat{{\bm{r}}_{i+1}}') = 0 \,,
\end{equation}
\begin{equation} \label{eq:Tg}
\bigl[ {\bm{T}}_{\overline{\bm{\gamma}}'}^{[N-2, \hspace{1ex}
2]}\bigr]_{ij} = (\widehat{{\bm{r}}_i}' -
\widehat{{\bm{r}}_{i+1}}') \, {\bm{.}} \, \left( \sum_{j'=1}^{j-1}
\widehat{{\bm{r}}_{j'}}' - (j-3) \widehat{{\bm{r}}_j}' \right) \,.
\end{equation}

Looking now at each $j$ block in Eq.~(\ref{eq:JWbWbTJT}), it is
tridiagonal in the $i$ index and has exactly the same form as
Eq.~(\ref{eq:WNm1WNm1TM}), aside from the $(j-3)(j-2)$ multiplier.
From Eq.~(\ref{eq:UNm1ij}) we see that ${\mathcal{I}}^{[N-2,
\hspace{1ex} 2]}_{\overline{\bm{\gamma}}'}$ is
\begin{eqnarray}
[{\mathcal{I}}^{[N-2, \hspace{1ex}
2]}_{\overline{\bm{\gamma}}'}]_{ij,\,i'j'} & = &
\frac{1}{\sqrt{(j-3)(j-2)}} \, [U^{[j-2, \hspace{1ex}
1]}_{\bar{\bm{r}}'}]_{ii'} \, \delta_{jj'} = 
\frac{i'}{\sqrt{i(i+1)(j-3)(j-2)}} \, \sum_{m=1}^i \delta_{mi'} \,
\delta_{jj'}  \nonumber \\
& = & \frac{i'}{\sqrt{i(i+1)(j-3)(j-2)}} \, \Theta_{i-i'+1} \,
\delta_{jj'} \,, \label{eq:INm2g}
\end{eqnarray}
where $\Theta_{i-j}$ is defined in Eq.~(\ref{eq:thetaalphabeta}).
Thus from Eqs.~(\ref{eq:UNm2eqIJ}), (\ref{eq:JNm2g}) and
(\ref{eq:INm2g}),
\begin{eqnarray}
\lefteqn{ [U^{[N-2, \hspace{1ex}
2]}_{\overline{\bm{\gamma}}'}]_{ij,\,i'j'}  } \nonumber \\ & = &
\sum_{l=4}^N \sum_{k=1}^{l-2} \, [{\mathcal{I}}^{[N-2,
\hspace{1ex} 2]}_{\overline{\bm{\gamma}}'}]_{ij,\,kl} \,
[{\mathcal{J}}^{[N-2,
\hspace{1ex} 2]}_{\overline{\bm{\gamma}}'}]_{kl,\,i'j'}  \nonumber \\
& = & \sum_{k=1}^{j-2} \, \frac{1}{\sqrt{(j-3)(j-2)}} \, [U^{[j-2,
\hspace{1ex} 1]}_{\bar{\bm{r}}'}]_{ik} \, [{\mathcal{J}}^{[N-2,
\hspace{1ex}
2]}_{\overline{\bm{\gamma}}'}]_{kj,\,i'j'}  \nonumber \\
& = & \sum_{k=1}^{j-2} \, \frac{k}{\sqrt{i(i+1)(j-3)(j-2)}} \,
\Theta_{i-k+1} \, \Bigl( [{\bm{1}}_{\bar{\bm{r}}'}]_j \,
\delta_{k+1,\, j'} \, (1-\delta_{k-1,\,
i'}) i' + \delta_{ki'} \, \Theta_{j-j'+1} \, \Theta_{j'-k-1} (j'-3) \Bigr)  \nonumber \\
& = & \frac{1}{\sqrt{i(i+1)(j-3)(j-2)}} \, \Bigl(
[{\bm{1}}_{\bar{\bm{r}}'}]_j \,  i'(j'-1) \, (1-\delta_{i',\,
j'-2}) \, \Theta_{i-j'+2} + i' (j'-3) \, \Theta_{i-i'+1} \,
\Theta_{j-j'+1} \,
\Theta_{j'-i'-1} \Bigr) \nonumber \\
& = & \frac{1}{\sqrt{i(i+1)(j-3)(j-2)}} \, \Bigl(
[{\bm{1}}_{\bar{\bm{r}}'}]_j \,  i'(j'-1) \, (1-\delta_{i',\,
j'-2}) \, \Theta_{i-j'+2} + i' (j'-3) \,
\Theta_{i-i'+1} \, \Theta_{j-j'+1} \Bigr) \,. \nonumber \\
\label{eq:UNm2geq}
\end{eqnarray}

The symmetry coordinate ${\bm{S}}_{\overline{\bm{\gamma}}'}^{[N-2,
\hspace{1ex} 2]}$ is thus
\begin{equation}
[{\bm{S}}_{\overline{\bm{\gamma}}'}^{[N-2, \hspace{1ex} 2]}]_{ij}
= \sum_{l=4}^N \sum_{k=1}^{l-2} \, [{\mathcal{I}}^{[N-2,
\hspace{1ex} 2]}_{\overline{\bm{\gamma}}'}]_{ij,\,kl} \, \bigr[
{\bm{T}}_{\overline{\bm{\gamma}}'}^{[N-2, \hspace{1ex}
2]}\bigr]_{kl} \,,
\end{equation}
which from Eqs.~(\ref{eq:Tg}) and (\ref{eq:INm2g}) is
\begin{equation} \label{eq:SNm2gip1j}
[{\bm{S}}_{\overline{\bm{\gamma}}'}^{[N-2, \hspace{1ex} 2]}]_{ij}
= \frac{1}{\sqrt{i(i+1)(j-3)(j-2)}} \, \left( \sum_{k=1}^i
\widehat{{\bm{r}}_k}' - i \widehat{{\bm{r}}_{i+1}}' \right) \,
{\bm{.}} \, \left( \sum_{j'=1}^{j-1} \widehat{{\bm{r}}_{j'}}' -
(j-3) \widehat{{\bm{r}}_j}' \right) \,,
\end{equation}
where $1 \leq i \leq j-2$ and $i+2 \leq j \leq N$\,. In matrix
form ${\bm{S}}_{\overline{\bm{\gamma}}'}^{[N-2, \hspace{1ex} 2]}$
may be written
\begin{equation}
{\bm{S}}_{\overline{\bm{\gamma}}'}^{[N-2, \hspace{1ex} 2]} =
\left(
\begin{array}{c}
\protect[{\bm{S}}_{\overline{\bm{\gamma}}'}^{[N-2, \hspace{1ex} 2]}\protect]_{14} \\
\protect[{\bm{S}}_{\overline{\bm{\gamma}}'}^{[N-2, \hspace{1ex} 2]}\protect]_{24} \\
\hline
\protect[{\bm{S}}_{\overline{\bm{\gamma}}'}^{[N-2, \hspace{1ex} 2]}\protect]_{15} \\
\protect[{\bm{S}}_{\overline{\bm{\gamma}}'}^{[N-2, \hspace{1ex} 2]}\protect]_{25} \\
\protect[{\bm{S}}_{\overline{\bm{\gamma}}'}^{[N-2, \hspace{1ex} 2]}\protect]_{35} \\
\hline
\protect[{\bm{S}}_{\overline{\bm{\gamma}}'}^{[N-2, \hspace{1ex} 2]}\protect]_{16} \\
\protect[{\bm{S}}_{\overline{\bm{\gamma}}'}^{[N-2, \hspace{1ex} 2]}\protect]_{26} \\
\protect[{\bm{S}}_{\overline{\bm{\gamma}}'}^{[N-2, \hspace{1ex} 2]}\protect]_{36} \\
\protect[{\bm{S}}_{\overline{\bm{\gamma}}'}^{[N-2, \hspace{1ex} 2]}\protect]_{46} \\
\hline
\protect[{\bm{S}}_{\overline{\bm{\gamma}}'}^{[N-2, \hspace{1ex} 2]}\protect]_{17} \\
\protect[{\bm{S}}_{\overline{\bm{\gamma}}'}^{[N-2, \hspace{1ex} 2]}\protect]_{27} \\
\protect[{\bm{S}}_{\overline{\bm{\gamma}}'}^{[N-2, \hspace{1ex} 2]}\protect]_{37} \\
\protect[{\bm{S}}_{\overline{\bm{\gamma}}'}^{[N-2, \hspace{1ex} 2]}\protect]_{47} \\
\protect[{\bm{S}}_{\overline{\bm{\gamma}}'}^{[N-2, \hspace{1ex} 2]}\protect]_{57} \\
\hline \protect[{\bm{S}}_{\overline{\bm{\gamma}}'}^{[N-2,
\hspace{1ex} 2]}\protect]_{18} \\ \multicolumn{1}{c}{\vdots}
\end{array}
\right) \,.
\end{equation}

\paragraph{Motions Associated with Symmetry
Coordinate ${\bm{S}}_{\overline{\bm{\gamma}}'}^{[N-2, \hspace{1ex}
2]}$\,.} To calculate the motions of the unscaled internal
displacement coordinates $\bm{\gamma}$  about the unscaled Lewis
structure configuration $\bm{\gamma}_\infty$ of
Eq.~(\ref{eq:ginfty1}) engendered by the symmetry coordinates
$[{\bm{S}}_{\overline{\bm{\gamma}}'}^{[N-2, \hspace{1ex}
2]}]_{ij}$, we use Eqs.~(\ref{eq:yS}), (\ref{eq:gpaxi}),
(\ref{eq:1eq1}) and (\ref{eq:WNm1g}). The first step in this
process is the evaluation of $[W^{[N-2, \hspace{1ex}
2]}_{\overline{\bm{\gamma}}'}]_{ij,\,mn}$\,. From
Eqs.~(\ref{eq:WaXUaXWbaX}), (\ref{eq:Wnm2eqdeltakinc}) and
(\ref{eq:UNm2geq}) we find that
\begin{eqnarray}
\lefteqn{[W^{[N-2, \hspace{1ex}
2]}_{\overline{\bm{\gamma}}'}]_{ij,\,mn} = \sum_{j'=4}^N
\sum_{i'=1}^{j'-2} \, [U^{[N-2, \hspace{1ex}
2]}_{\overline{\bm{\gamma}}'}]_{ij,\,i'j'} [\overline{W}^{[N-2,
\hspace{1ex} 2]}_{\overline{\bm{\gamma}}'}]_{i'j',\,mn}  }
\nonumber \\ & = &
\renewcommand{\arraystretch}{1.5}
\begin{array}[t]{r@{\hspace{0.5ex}}l} \multicolumn{2}{l}{ {\displaystyle
\frac{1}{\sqrt{i(i+1)(j-3)(j-2)}} \, \sum_{j'=4}^N \sum_{i'=1}^{j'-2} \,
\begin{array}[t]{r@{}l}
\Bigl( & [{\bm{1}}_{\bar{\bm{r}}'}]_j \,  i'(j'-1) \,
(1-\delta_{i',\, j'-2}) \, \Theta_{i-j'+2} + \\ & + i' (j'-3) \,
\Theta_{i-i'+1} \,
\Theta_{j-j'+1} \Bigr) \times \end{array} } } \\
\times \Biggl[ & (\delta_{i'm}-\delta_{i'+1,\,m}) \biggl(
(1-\delta_{i',\,j'-2}) \delta_{j'-1,\,n} + \delta_{i',\,j'-2} \,
\delta_{j'-3,\,n} - \delta_{j'n} \biggr) + \\ & +
(\delta_{i'n}-\delta_{i'+1,\,n}) \biggl( (1-\delta_{i',\,j'-2})
\delta_{j'-1,\,m} + \delta_{i',\,j'-2} \, \delta_{j'-3,\,m} -
\delta_{j'm} \biggr) \Biggr] \end{array}
\renewcommand{\arraystretch}{1}
\nonumber \\ & = & \frac{1}{\sqrt{i(i+1)(j-3)(j-2)}} \,
\begin{array}[t]{r@{\hspace{0.5ex}}l} \Bigl( & (\Theta_{i-m+1} - i
\delta_{i+1,\,m})(\Theta_{j-n} -(j-3)\delta_{jn}) + \\ & +
(\Theta_{i-n+1} - i \delta_{i+1,\,n})(\Theta_{j-m}
-(j-3)\delta_{jm}) \Bigr)  \,, \end{array} \label{eq:WNm2gijmn}
\end{eqnarray}
c.f.\ Eq.~(\ref{eq:SNm2gip1j}). Thus
\begin{equation} 
\begin{array}{r@{\hspace{0.5em}}c@{\hspace{0.5em}}l}
\multicolumn{3}{l}{ \hspace{-0.5ex} (\gamma^{[N-2, \hspace{1ex}
2]}_{ij})_{mn} = {\displaystyle \frac{1}{\sqrt{D}} \,
(\gamma^{\prime [N-2, \hspace{1ex} 2]}_{ij})_{mn} =
\frac{1}{\sqrt{D}} \, [{\bm{S}}_{\overline{\bm{\gamma}}'}^{[N-2,
\hspace{1ex}
2]}]_{ij} \, [W_{\overline{\bm{\gamma}}'}^{[N-2, \hspace{1ex} 2]}]_{ij,\,mn} }  } \\
[1em] & = & {\displaystyle \frac{1}{\sqrt{i(i+1)(j-3)(j-2)D}} \,
\, \, [{\bm{S}}_{\overline{\bm{\gamma}}'}^{[N-2, \hspace{1ex}
2]}]_{ij}
\begin{array}[t]{r@{\hspace{0.5ex}}l} \Bigl( & (\Theta_{i-m+1} - i
\delta_{i+1,\,m})(\Theta_{j-n} -(j-3)\delta_{jn}) + \\ & +
(\Theta_{i-n+1} - i \delta_{i+1,\,n})(\Theta_{j-m}
-(j-3)\delta_{jm}) \Bigr)  \,, \end{array} } \end{array}
\end{equation}
where $1 \leq m < n \leq N$, and $1 \leq i \leq j-2$ and $i+2 \leq
j \leq N$\,.

\section{The Frequencies and Normal-Mode Coordinates of the
System.} \label{sec:SysFreqNorm}
\subsection{The $\bm{G}$ and $\bm{FG}$ matrices in the
Symmetry-Coordinate Basis.} We can use the $W_{\bm{X}'}^\alpha$
matrices of Eqs.~(\ref{eq:WNm1r}), (\ref{eq:WNm1g}) and
(\ref{eq:WNm2gijmn}) to calculate the reduced $\bm{G}$ and
$\bm{FG}$ matrix elements,
$[\bm{\sigma_\alpha^{G}}]_{\bm{X}'_1,\,\bm{X}'_2}$ and
$[\bm{\sigma_\alpha^{FG}}]_{\bm{X}'_1,\,\bm{X}'_2}$\,, of
Eq.~(\ref{eq:sigmaQ}) for the $\alpha = [N-1, \hspace{1ex} 1]$ and
$[N-2, \hspace{1ex} 2]$ species. The reduced matrix elements,
$[\bm{\sigma_{[N]}^{G}}]_{\bm{X}'_1,\,\bm{X}'_2}$ and
$[\bm{\sigma_{[N]}^{FG}}]_{\bm{X}'_1,\,\bm{X}'_2}$ for the $[N]$
sector are calculated in Paper~I. An outline of this calculation
is to be found in Appendix~\ref{app:sigmacalc}, with the following
results.

\subsubsection{The $[N-1, \hspace{1ex} 1]$ Species.}
\paragraph{The Matrix Elements $[\bm{\sigma_{[N-1,
\hspace{1ex} 1]}^G}]_{\bm{X}'_1,\,\bm{X}'_2}$\,.} Using
Eqs.~(\ref{eq:Q}), (\ref{eq:Qrr}), (\ref{eq:Qrg}), (\ref{eq:Qgr}),
(\ref{eq:Qgg}), (\ref{eq:Gsub}), (\ref{eq:Gsym}),
(\ref{eq:Goneorzero}), (\ref{eq:sigmaQ}), (\ref{eq:WNm1r}) and
(\ref{eq:WNm1g}) we derive
\begin{equation} \label{eq:sigmaNm1G}
\bm{\sigma_{[N-1, \hspace{1ex} 1]}^G} = \left(
\begin{array}{l@{\hspace{1.5em}}l} \protect[\bm{\sigma_{[N-1, \hspace{1ex}
1]}^G}\protect]_{\bar{\bm{r}}',\,\bar{\bm{r}}'} = \tilde{a}' &
{\displaystyle \protect[\bm{\sigma_{[N-1, \hspace{1ex}
1]}^G}\protect]_{\bar{\bm{r}}',\, \overline{\bm{\gamma}}'} =
0} \\
{\displaystyle \protect[\bm{\sigma_{[N-1, \hspace{1ex}
1]}^G}\protect]_{\overline{\bm{\gamma}}',\,\bar{\bm{r}}'} = 0 } &
{\displaystyle \protect[\bm{\sigma_{[N-1, \hspace{1ex}
1]}^G}\protect]_{\overline{\bm{\gamma}}',\,\overline{\bm{\gamma}}'}
= (\tilde{g}' + (N-2)\tilde{h}')} \end{array} \right) \,.
\end{equation}

\paragraph{The Matrix Elements $[\bm{\sigma_{[N-1,
\hspace{1ex} 1]}^{FG}}]_{\bm{X}'_1,\,\bm{X}'_2}$\,.} Using
Eqs.~(\ref{eq:Q}), (\ref{eq:Qrr}), (\ref{eq:Qrg}), (\ref{eq:Qgr}),
(\ref{eq:Qgg}), (\ref{GFsub}), (\ref{GFsym}), (\ref{eq:sigmaQ}),
(\ref{eq:sigmamat}), (\ref{eq:WNm1r}) and (\ref{eq:WNm1g}) we
derive
\begin{equation}  \label{eq:sigmaNm1FG}
\bm{\sigma_{[N-1, \hspace{1ex} 1]}^{FG}} = \left(
\begin{array}{l@{\hspace{1.5em}}l} \protect[\bm{\sigma_{[N-1, \hspace{1ex}
1]}^{FG}}\protect]_{\bar{\bm{r}}',\,\bar{\bm{r}}'} = \tilde{a} &
{\displaystyle \protect[\bm{\sigma_{[N-1, \hspace{1ex}
1]}^{FG}}\protect]_{\bar{\bm{r}}',\, \overline{\bm{\gamma}}'} =
\sqrt{N-2} \,\, \tilde{e} } \\
{\displaystyle \protect[\bm{\sigma_{[N-1, \hspace{1ex}
1]}^{FG}}\protect]_{\overline{\bm{\gamma}}',\,\bar{\bm{r}}'} =
\sqrt{N-2} \,\, \tilde{c} } & {\displaystyle
\protect[\bm{\sigma_{[N-1, \hspace{1ex}
1]}^{FG}}\protect]_{\overline{\bm{\gamma}}',\,\overline{\bm{\gamma}}'}
= (\tilde{g}+(N-2)\tilde{h})} \end{array} \right) \,.
\end{equation}

\subsubsection{The $[N-2, \hspace{1ex} 2]$ Species.}
\paragraph{The Matrix Element $[\bm{\sigma_{[N-2,
\hspace{1ex} 2]}^G}]$\,.} Using Eqs.~(\ref{eq:Qgg}),
(\ref{eq:Gsub}), (\ref{eq:Gsym}), (\ref{eq:Goneorzero}),
(\ref{eq:sigmaQ}) and (\ref{eq:WNm2gijmn}) we derive
\begin{equation}  \label{eq:sigmaNm2G}
\bm{\sigma_{[N-2, \hspace{1ex} 2]}^G} = [\bm{\sigma_{[N-2,
\hspace{1ex}
2]}^G}]_{\overline{\bm{\gamma}}',\,\overline{\bm{\gamma}}'} =
\tilde{g}' \,.
\end{equation}

\paragraph{The Matrix Element $[\bm{\sigma_{[N-2,
\hspace{1ex} 2]}^{FG}}]$\,.} Using Eqs.~(\ref{eq:Qgg}),
(\ref{GFsub}), (\ref{GFsym}), (\ref{eq:sigmaQ}) and
(\ref{eq:WNm2gijmn}) we derive
\begin{equation} \label{eq:sigmaNm2FG}
\bm{\sigma_{[N-2, \hspace{1ex} 2]}^{FG}} = [\bm{\sigma_{[N-2,
\hspace{1ex}
2]}^{FG}}]_{\overline{\bm{\gamma}}',\,\overline{\bm{\gamma}}'} =
\tilde{g} \,.
\end{equation}

\subsection{The Frequencies and Normal Modes.} \label{subsec:FNM}
Using Eq.~(\ref{eq:sigmaNm1FG}) in Eq.~(\ref{eq:lambda12pm}) we
obtain $\lambda^\pm_{[N-1, \hspace{1ex} 1]}$ and from
Eqs.~(\ref{eq:sigmaNm2FG}) and (\ref{eq:lNm2eqsig}) we obtain
$\lambda_{[N-2, \hspace{1ex} 2]}$\,. The frequencies are then
determined from Eq.~(\ref{eq:omega_p}). The $\lambda^\pm_{[N]}$
for the $[N]$ species are discussed in Paper~I.

Likewise the normal modes ${\bm{q}'}_+^{[N]}$ and
${\bm{q}'}_-^{[N]}$ for the $[N]$ species are also calculated in
Paper~I (Eqs.~(203) and (204)). In regards to the $[N-1,
\hspace{1ex} 1]$ species, the
$\bar{\bm{r}}'$-$\overline{\bm{\gamma}}'$ mixing angles,
$\theta^{[N-1, \hspace{1ex} 1]}_\pm$\,, are determined from
Eq.~(\ref{eq:tanthetaalphapm}). The normalization constant
$c^{[N-1, \hspace{1ex} 1]}$ of the reduced normal-coordinate
coefficient vector, ${\mathsf{c}}^{[N-1, \hspace{1ex} 1]}$, of
Eqs.~(\ref{eq:cb}) and (\ref{eq:sfceqcthatc}) is determined from
Eqs.~(\ref{eq:c2norm}) and (\ref{eq:calphapm}). One then
determines the normal mode vector, ${\bm{q}'}$\,, through
Eqs.~(\ref{eq:qvector}), (\ref{eq:SNm1}), (\ref{eq:SgNm1i}) and
(\ref{eq:SNm2gip1j}). Thus we arrive at
\renewcommand{\jot}{0.5em}
\begin{eqnarray}
\protect[{\bm{q}'}_+^{[N-1, \hspace{1ex} 1]}\protect]_i & = &
c^{[N-1, \hspace{1ex} 1]}_+ \cos{\theta^{[N-1, \hspace{1ex} 1]}_+}
\frac{1}{\sqrt{i(i+1)}} \left( \sum_{i=1}^i \bar{r}'_i - i
\bar{r}'_{i+1} \right)  \nonumber \\
&& + c^{[N-1, \hspace{1ex} 1]}_+ \sin{\theta^{[N-1, \hspace{1ex}
1]}_+} \frac{1}{\sqrt{i(i+1)(N-2)}}\left( \vphantom{\left[ \sum_{l
= 2}^i \, \sum_{k=1}^{l-1} \hspace{-1ex} \overline{\gamma}'_{kl} +
\sum_{k = 1}^i \, \sum_{l=k+1}^{N} \hspace{-1ex}
\overline{\gamma}'_{kl} \right] -} \right. \hspace{-0.5ex}
\renewcommand{\arraystretch}{2} \begin{array}[t]{@{}l} {\displaystyle \left[ \sum_{l =
2}^i \, \sum_{k=1}^{l-1} \hspace{-1ex} \overline{\gamma}'_{kl} +
\sum_{k = 1}^i \,
\sum_{l=k+1}^{N} \hspace{-1ex} \overline{\gamma}'_{kl} \right]  } \\
{\displaystyle \left. - i \left[ \sum_{k=1}^i
\overline{\gamma}'_{k,\,i+1} + \sum_{l=i+2}^N \hspace{-0.5ex}
\overline{\gamma}'_{i+1,\,l} \right] \hspace{0.3ex} \right) } \end{array} \renewcommand{\arraystretch}{1} \\
\protect[{\bm{q}'}_-^{[N-1, \hspace{1ex} 1]}\protect]_i & = &
c^{[N-1, \hspace{1ex} 1]}_- \cos{\theta^{[N-1, \hspace{1ex} 1]}_-}
\frac{1}{\sqrt{i(i+1)}} \left( \sum_{i=1}^i \bar{r}'_i - i
\bar{r}'_{i+1} \right)  \nonumber \\
&& + c^{[N-1, \hspace{1ex} 1]}_- \sin{\theta^{[N-1, \hspace{1ex}
1]}_-} \frac{1}{\sqrt{i(i+1)(N-2)}} \left( \vphantom{\left[
\sum_{l = 2}^i \, \sum_{k=1}^{l-1} \hspace{-1ex}
\overline{\gamma}'_{kl} + \sum_{k = 1}^i \, \sum_{l=k+1}^{N}
\hspace{-1ex} \overline{\gamma}'_{kl} \right] -} \right.
\hspace{-0.5ex}
\renewcommand{\arraystretch}{2} \begin{array}[t]{@{}l} {\displaystyle \left[ \sum_{l =
2}^i \, \sum_{k=1}^{l-1} \hspace{-1ex} \overline{\gamma}'_{kl} +
\sum_{k = 1}^i \,
\sum_{l=k+1}^{N} \hspace{-1ex} \overline{\gamma}'_{kl} \right]  } \\
{\displaystyle \left. - i \left[ \sum_{k=1}^i
\overline{\gamma}'_{k,\,i+1} + \sum_{l=i+2}^N \hspace{-0.5ex}
\overline{\gamma}'_{i+1,\,l} \right] \hspace{0.3ex} \right) } \end{array} \renewcommand{\arraystretch}{1} \\
\protect[{\bm{q}'}^{[N-2, \hspace{1ex} 2]}\protect]_{ij} & = &
c^{[N-2, \hspace{1ex} 2]} \frac{1}{\sqrt{i(i+1)(j-3)(j-2)}} \,
\left( \sum_{k=1}^i \widehat{{\bm{r}}_k}' - i
\widehat{{\bm{r}}_{i+1}}' \right) \, {\bm{.}} \, \left(
\sum_{j'=1}^{j-1} \widehat{{\bm{r}}_{j'}}' - (j-3)
\widehat{{\bm{r}}_j}' \right)\,.\hspace{4ex}
\end{eqnarray}
\renewcommand{\jot}{0em}

\subsection{The Motions Associated with the Normal Modes.}
\label{subsec:MotNormM} From Eqs.~(\ref{eq:yq}),
(\ref{eq:yqinfty}), (\ref{eq:pyqalphaxi}), (\ref{eq:myqalphaxi}),
(\ref{eq:yqnm2xi}), (\ref{eq:WNm1r}), (\ref{eq:WNm1g}) and
(\ref{eq:WNm2gijmn})
\begin{equation} \label{eq:yqf}
{\bm{y}} = \left( \begin{array}{c} {\bm{r}} \\
\bm{\gamma} \end{array} \right) = \,\, {\bm{y}}_\infty \,\, +
\begin{array}[t]{@{}l@{}} {\displaystyle \hspace{2ex} \sum} \\ {\scriptstyle \alpha= \left\{
\renewcommand{\arraystretch}{0.5}
\begin{array}{@{}c@{}} {\scriptstyle
\protect[N\protect]\,,} \\ {\scriptstyle \protect[N-1,
\hspace{1ex} 1\protect]}
\end{array} \renewcommand{\arraystretch}{1} \right\}
} \end{array} \sum_\xi \sum_{\tau=\pm} \, \left( \begin{array}{c} _{\tau}{\bm{r}}^{\alpha}_\xi \\
_{\tau}\bm{\gamma}^{\alpha}_\xi
\end{array} \right) \,\,\, + \,\,\, \sum_\xi \,
\left( \begin{array}{c} {\bm{0}} \\
\bm{\gamma}^{\protect[N-2, \hspace{1ex} 2\protect]}_\xi
\end{array} \right) \,,
\end{equation}
where ${\bm{y}}_\infty$ is given by Eq.~(\ref{eq:yqinfty}). The
$\xi$ sum for the $[N]$ species only includes one term and
$_+{\bm{r}}^{[N]}$,\, $_+\bm{\gamma}^{[N]}$,\, $_-{\bm{r}}^{[N]}$
and $_-\bm{\gamma}^{[N]}$ are identified in Paper~I.  For the
$[N-1, \hspace{1ex} 1]$ species $1 \leq \xi \leq N-1$, and so we
have
\begin{eqnarray} 
( _+r^{[N-1, \hspace{1ex} 1]}_\xi)_i & = & - \, \overline{a}_{ho}
\, \sqrt{\frac{D^3}{\xi(\xi+1)}} \,\, \frac{\sin{\theta^{[N-1,
\hspace{1ex} 1]}_-}}{s(\theta^{[N-1, \hspace{1ex} 1]}) \, c^{[N-1,
\hspace{1ex} 1]}_+} \,\, [{\bm{q}'}_+^{[N-1, \hspace{1ex} 1]}]_\xi
\, \left( \Theta_{\xi-i+1} - \xi \delta_{\xi+1,\, i}
\right) \,, \\
( _+\gamma^{[N-1, \hspace{1ex} 1]}_\xi)_{ij} & = & {\displaystyle
\frac{1}{\sqrt{\xi(\xi+1)(N-2)D}} \, \frac{\cos{\theta^{[N-1,
\hspace{1ex} 1]}_-}}{s(\theta^{[N-1, \hspace{1ex} 1]}) \, c^{[N-1,
\hspace{1ex} 1]}_+}
\,\, [{\bm{q}'}_+^{[N-1, \hspace{1ex} 1]}]_\xi \, \times } \nonumber \\
& & {\displaystyle \times \bigg( \big( \Theta_{\xi-i+1} \,
[{\bm{1}}_{\bar{\bm{r}}'}]_{ij} + \Theta_{\xi-j+1} \,
[{\bm{1}}_{\bar{\bm{r}}'}]_{ij} \big) - \xi \big( \delta_{\xi+1,\,
i} \, [{\bm{1}}_{\bar{\bm{r}}'}]_{ij} + \delta_{\xi+1,\, j} \,
[{\bm{1}}_{\bar{\bm{r}}'}]_{ij} \big) \bigg)
 \,, }
\end{eqnarray}
\begin{eqnarray} 
( _-r^{[N-1, \hspace{1ex} 1]}_\xi)_i & = & \overline{a}_{ho} \,
\sqrt{\frac{D^3}{\xi(\xi+1)}} \, \frac{\sin{\theta^{[N-1,
\hspace{1ex} 1]}_+}}{s(\theta^{[N-1, \hspace{1ex} 1]}) \, c^{[N-1,
\hspace{1ex} 1]}_-} \,\, [{\bm{q}'}_-^{[N-1, \hspace{1ex} 1]}]_\xi
\, \left( \Theta_{\xi-i+1} - \xi
\delta_{\xi+1,\, i} \right) \,, \\
( _-\gamma^{[N-1, \hspace{1ex} 1]}_\xi)_{ij} & = & {\displaystyle
- \, \frac{1}{\sqrt{\xi(\xi+1)(N-2)D}} \, \frac{\cos{\theta^{[N-1,
\hspace{1ex} 1]}_+}}{s(\theta^{[N-1, \hspace{1ex} 1]}) \, c^{[N-1,
\hspace{1ex} 1]}_-} \,\,
[{\bm{q}'}_-^{[N-1, \hspace{1ex} 1]}]_\xi \, \times } \nonumber \\
& & {\displaystyle \times \bigg( \big( \Theta_{\xi-i+1} \,
[{\bm{1}}_{\bar{\bm{r}}'}]_{ij} + \Theta_{\xi-j+1} \,
[{\bm{1}}_{\bar{\bm{r}}'}]_{ij} \big) - \xi \big( \delta_{\xi+1,\,
i} \, [{\bm{1}}_{\bar{\bm{r}}'}]_{ij} + \delta_{\xi+1,\, j} \,
[{\bm{1}}_{\bar{\bm{r}}'}]_{ij} \big) \bigg) \,. }
\end{eqnarray}
For the $[N-2, \hspace{1ex} 2]$ species $\xi$ runs over the set
$\{ k,\hspace{1ex}l: \hspace{1ex} \forall \hspace{1ex} 1 \leq k
\leq l-2 \mbox{\hspace{1ex} and $k+2 \leq l \leq N$} \} $ and so
\begin{eqnarray} 
(\gamma^{[N-2, \hspace{1ex} 2]}_{kl})_{ij} & = & {\displaystyle
\frac{1}{\sqrt{k(k+1)(l-3)(l-2)D}} \, \, \, \frac{1}{c^{[N-2,
\hspace{1ex} 2]}} \,\,
[{\bm{q}'}^{[N-2, \hspace{1ex} 2]}]_{kl} \, \times } \nonumber \\
& & {\displaystyle \times
\begin{array}[t]{r@{\hspace{0.5ex}}l} \Bigl( & (\Theta_{k-i+1} - k
\delta_{k+1,\,i})(\Theta_{l-j} -(l-3)\delta_{lj}) + \\ & +
(\Theta_{k-j+1} - k \delta_{k+1,\,j})(\Theta_{l-i}
-(l-3)\delta_{li}) \Bigr)  \,. \end{array} }
\end{eqnarray}

\section{Summary.} \label{sec:Sum}
In this paper we have completed the derivation, begun in Paper I,
of the lowest-order DPT $S$-wave wave function of a correlated
quantum confined system under spherical confinement  with weak,
intermediate or strong interparticle interactions. Remarkably, it
is a largely analytic approach, yielding analytic expressions for
energies, frequencies and wave functions (also density profiles).
The achievement of analytic results is not due to a simplified
description of the two-body interaction, but rather to the high
degree of symmetry of the zeroth-order configuration in which
every particle is equidistant and equiangular from every other
particle in a high dimensional space. This symmetry allows us to
use group theoretic methods rather than numerical techniques to
account individually for each two-body interaction.  The resulting
simplification is remarkable resulting in analytic formulas with
$N$ as a simple parameter as well as a stunning reduction in the
number of normal modes.  Even the zeroth-order result contains
correlation effects as demonstrated by the interparticle angle
cosine of the Lewis structure, $\overline{\gamma}_{\infty}$, which
differs significantly from its mean-field, i.e. uncorrelated,
value of zero.


We have obtained the normal modes and wave functions to lowest
order using the FG method\cite{dcw} which
directly relates the structure of the Schr\"odinger equation to
normal-mode coordinates describing the fundamental motions of the
system (Eqs.~(\ref{Gham}) - (\ref{eq:Phi_0})).

However, since DPT is a beyond mean-field
treatment of the interacting $N$-body problem, where $N$ may be
large and where the $N(N+1)/2$ interactions can be strong, this is
still a formidable problem. In particular there are $P=N(N+1)/2$
normal modes, and up to $P$ distinct frequencies to calculate.
However, the symmetry of the Lewis structure allows  an enormous
simplification in the calculation of frequencies and normal modes.
Since the Lewis structure is invariant under interchange of
particles, it satisfies an $S_N$ symmetry. The full
$D$-dimensional Hamiltonian as well as the Hamiltonian for the
zeroth-order wave function are invariant under particle
interchange, and thus
invariant under $S_N$\,. This symmetry results, in fact,
in only five distinct frequencies, a remarkable reduction in
the number $P$ of possible distinct frequencies. It also provides
an immense simplification in the calculation of the normal mode
coordinates.

We note that the $S_N$ invariance of Eq.~(\ref{Gham}) means that
the $\bm{F}$, $\bm{G}$ and $\bm{F}\bm{G}$ matrices of
Eqs.~(\ref{Gham} - \ref{eq:normit}) are invariant under
$S_N$\,. This implies that the eigenvectors ${\bm{b}}$ and so the
normal mode coordinates transform under irreducible
representations of $S_N$. The coordinates $\bar{r}'_i$ and
 $\overline{\gamma}'_{ij}$\, transform {\em reducibly}
under $S_N$; however using the
theory of group characters one can show that the $\bar{r}'_i$ are
reducible to one one-dimensional $[N]$ irreducible representation
of $S_N$ and one (N-1)-dimensional $[N-1, \hspace{1ex} 1]$
irreducible representation of $S_N$\,, while the
$\overline{\gamma}'_{ij}$ are reducible to one one-dimensional
$[N]$ irreducible representation of $S_N$\,, one (N-1)-dimensional
$[N-1, \hspace{1ex} 1]$ irreducible representation of $S_N$ and
one N(N-3)/2-dimensional $[N-2, \hspace{1ex} 2]$ irreducible
representation of $S_N$\,. Since the normal coordinates are linear
combinations of the internal coordinates $\bar{r}'_i$ and
$\overline{\gamma}'_{ij}$ (Eqs.~(\ref{eq:ytransposeP}) -
(\ref{eq:bfgammap}) and (\ref{eq:qyt})), the
normal coordinate set is comprised of two one-dimensional $[N]$
irreducible representations of $S_N$\, two (N-1)-dimensional
$[N-1, \hspace{1ex} 1]$ irreducible representations of $S_N$ and
one entirely angular N(N-3)/2-dimensional $[N-2, \hspace{1ex} 2]$
irreducible representation of $S_N$\,.

This group theoretic information
provides an immense simplification in the calculation of the
normal mode coordinates through the use of
symmetry coordinates\cite{dcw}. We have determined the
normal coordinates and distinct frequencies in a three-step
process:
\newcounter{twostepC}
\newcounter{twostepCseca}
\newcounter{twostepCsecb}
\newcounter{twostepCsecc}
\begin{list}{\alph{twostepC}).}{\usecounter{twostepC}\setlength{\rightmargin}{\leftmargin}}
\item \setcounter{twostepCseca}{\value{twostepC}} We define sets
of primitive irreducible coordinates having the simplest
functional form possible subject to the requirement of
transforming under particular non-orthogonal irreducible
representations of $S_N$. For the $\bar{\bm{r}}'$ sector we define
two sets of linear combinations of elements of the $\bar{\bm{r}}'$
vector which transform under non-orthogonal $[N]$ and
$[N-\nolinebreak 1,1]$ irreducible representations of $S_N$\,. We
then derive two sets of linear combinations of elements of the
$\overline{\bm{\gamma}}'$ vector which transform under exactly
these same two irreducible representations of $S_N$. Finally we
define a set of linear combinations of elements of
$\overline{\bm{\gamma}}'$ which transform under a particular
non-orthogonal $[N-\nolinebreak 2,2]$ irreducible representation
of $S_N$.
\item \setcounter{twostepCsecb}{\value{twostepC}} Using linear
combinations within each set  of primitive irreducible
coordinates, we determine symmetry coordinates defined to
transform under {\em orthogonal} irreducible representations of $S_N$\,. Care is
taken to ensure that this transformation to the symmetry
coordinates preserves the identity of equivalent representations
in the $\bar{\bm{r}}'$ and $\overline{\bm{\gamma}}'$ sectors. We
choose one of the symmetry coordinates to be a single primitive
irreducible coordinate, the simplest functional form possible
under the requirement that it transforms irreducibly under $S_N$.
The next symmetry coordinate is chosen to be composed of two
primitive irreducible coordinates and so on. Thus the complexity
of the symmetry coordinates is minimized, building up slowly as
symmetry coordinates are added.
\item  \setcounter{twostepCsecc}{\value{twostepC}} The $\bm{FG}$
matrix, originally expressed in the $\bar{\bm{r}}'$ and
$\overline{\bm{\gamma}}'$ basis, is now expressed in symmetry
coordinates resulting in a stunning simplification. The $P \times
P$ eigenvalue equation of Eq.~(\ref{eq:FGit}) is reduced to one $2
\times 2$ eigenvalue equation for the $[N]$ sector,
$N-\nolinebreak 1$ identical $2 \times 2$ eigenvalue equations for
the $[N-\nolinebreak 1,1]$ sector and $N(N-\nolinebreak 3)/2$
identical $1 \times 1$ eigenvalue equations for the
$[N-\nolinebreak 2,2]$ sector. For the $[N]$ and $[N-\nolinebreak
1,1]$ sectors, the $2 \times 2$ structure allows for mixing of the
$\bar{\bm{r}}'$ and $\overline{\bm{\gamma}}'$ symmetry coordinates
in the normal coordinates (see Eq.~(\ref{eq:qbeqSrSg})). The $1
\times 1$ structure of the equations in the $[N-\nolinebreak 2,2]$
sector reflects the absence of $\bar{\bm{r}}'$ symmetry
coordinates in this sector, i.e.\ the $[N-\nolinebreak 2,2]$
normal modes are entirely angular.
\end{list}

\section{Conclusions.} \label{sec:Conc}

The increasing interest in creating systems controlled by
quantum confinement is resulting in new interest and new demands
on the $N$-body techniques of quantum physics and chemistry,
originally developed for atoms and molecules. Mean-field
treatments, such as the Hartree-Fock method in atomic physics and
the Gross-Pitaevskii method for Bose-Einstein condensates, do not
include correlation effects, and therefore fail for systems under tight
confinement or strong interaction. These new systems, with a few
hundred to millions of particles, present serious challenges for
existing $N$-body methods, most of which were developed with small
systems in mind.

Dimensional perturbation theory directly addresses these issues.
Since the perturbation parameter is the dimensionality of space it
is not limited to weak or strong interparticle interactions. Its
formulation  allows the symmetry of the problem to be
exploited to the highest degree, enabling solutions for large $N$
systems to be obtained with a minimum of numerical computation.
In fact, dimensional perturbation theory
for such systems produces analytic results that are a function of
$N$\,. This means essentially that results for any $N$ are
obtained from a single calculation\cite{FGpaper,energy}.

Almost all past work using dimensional perturbation theory has
focused on {\em energies} with little attention given to the
difficult task of obtaining wave
functions, even at lowest order. In Paper I we began the
derivation of the normal mode coordinates of the lowest-order wave
function. The current paper completes this work. With the
normal-mode coordinates of the zeroth-order wave function at hand,
many other properties of the system beyond low-order energies
become accessible. For macroscopic quantum confined systems the
density profile is an experimentally accessible observable, and to
lowest order may be calculated analytically from the zeroth-order
wave function. The fact that it is an
analytic function of $N$, means that the density profile is
obtained for any $N$ in a single calculation. The normal mode
coordinates directly relate to the nature of the motions of
excited states of the system which are also
experimentally accessible for macroscopic quantum confined
systems. Expectation values may be calculated, as may transition
matrix elements. The fact that dimensional perturbation theory is
a beyond-mean-field method means that such interesting properties
may be derived for weakly, intermediate and strongly interacting
systems, the latter two appearing in experiments with low-$D$ or
large-$N$ systems as well as systems with strong interparticle
interactions such as is the case with Feshbach resonances in
Bose-Einstein condensates\cite{wieman}. Knowing
the zeroth-order wave function is a precurser
to calculating higher-order results.

Almost all confining potentials for quantum confined systems in
the lab currently have cylindrical symmetry. These include
condensates confined in a cylindrical potential as well as axially
symmetric quantum dots, two-dimensional electronic systems in a
corbino disk geometry and rotating superfluid helium
systems. One can extend the dimensional
perturbation approach to handle such systems with axial, as
opposed to spherical symmetry.

\section{Acknowledgments}
We would like to thank the Army Research Office for ongoing
support. This work was also supported in part by the Office of
Naval Research.

\appendix
\renewcommand{\theequation}{A\arabic{equation}}
\setcounter{equation}{0}
\section{Calculation of $\bm{[
\sigma_{[N-1, \hspace{1ex} 1]}^{G}]_{\bm{X}'_1,\,\bm{X}'_2}}$,\,
$\bm{[\sigma_{[N-2, \hspace{1ex}
2]}^{G}]_{\overline{\bm{\gamma}}',\,\overline{\bm{\gamma}}'}}$,\,
$\bm{[\sigma_{[N-1, \hspace{1ex}
1]}^{FG}]_{\bm{X}'_1,\,\bm{X}'_2}}\,$ and $\bm{[\sigma_{[N-2,
\hspace{1ex}
2]}^{FG}]_{\overline{\bm{\gamma}}',\,\overline{\bm{\gamma}}'}}$\,,
the reduced $\bm{G}$ and $\bm{FG}$ matrix elements in the
symmetry-coordinate basis.}\label{app:sigmacalc} In this appendix
we use the  $W_{\bm{X}'}^\alpha$ matrices ($\alpha= [N-1,
\hspace{1ex} 1]\,,\,\, [N-2, \hspace{1ex} 2]\,,\,\,  \bm{X}'=
\bar{\bm{r}}'$ or $\overline{\bm{\gamma}}'$) to calculate the
reduced $\bm{G}$ and $\bm{FG}$ matrix elements,
$[\bm{\sigma_\alpha^{G}}]_{\bm{X}'_1,\,\bm{X}'_2}$ and
$[\bm{\sigma_\alpha^{FG}}]_{\bm{X}'_1,\,\bm{X}'_2}$\,, using
Eq.~(\ref{eq:sigmaQ}):
\begin{equation}\label{eq:sigmaQ1}
[\bm{\sigma_\alpha^Q}]_{\bm{X}'_1,\,\bm{X}'_2} =
(W_{\bm{X}'_1}^\alpha)_\xi \, \bm{Q}_{\bm{X}'_1 \bm{X}'_2} \,
[(W_{\bm{X}'_2}^\alpha)_\xi]^T \,,
\end{equation}
where $\bm{Q} = \bm{G}$ or $\bm{FG}$ and $\xi$ is a row label.
Notice that $[\bm{\sigma_\alpha^Q}]_{\bm{X}'_1,\,\bm{X}'_2}$ on
the left-hand side of Eq.~(\ref{eq:sigmaQ1}) should be independent
of the row label $\xi$ on the right hand side of the equation.
This is a strong check on the correctness of our calculations.

\subsection{The Matrix Elements
$\bm{[\sigma_{[N-1, \hspace{1ex}
1]}^{G}]_{\bm{X}'_1,\,\bm{X}'_2}\,.}$ and $\bm{[\sigma_{[N-1,
\hspace{1ex} 1]}^{FG}]_{\bm{X}'_1,\,\bm{X}'_2}\,}$. }
Equations~(\ref{eq:WNm1r}) and Eq.~(\ref{eq:WNm1g}) read
\begin{equation} \label{eq:WNm1rD}
[W^{[N-1, \hspace{1ex} 1]}_{\bar{\bm{r}}'}]_{ik} =
\frac{1}{\sqrt{i(i+1)}} \left( \sum_{m=1}^i \delta_{mk} - i
\delta_{i+1,\, k} \right) = \frac{1}{\sqrt{i(i+1)}} \left(
\Theta_{i-k+1} - i \delta_{i+1,\, k} \right)
\end{equation}
and
\renewcommand{\jot}{0.5em}
\begin{eqnarray}
\lefteqn{[W^{[N-1, \hspace{1ex} 1]}_{\overline{\bm{\gamma}}'}]_{i,\,kl} \hspace{1ex}} \nonumber \\
& = & \frac{1}{\sqrt{i(i+1)(N-2)}} \, \left( \sum_{m=1}^i \big(
\delta_{mk} \, [{\bm{1}}_{\bar{\bm{r}}'}]_l + \delta_{ml} \,
[{\bm{1}}_{\bar{\bm{r}}'}]_k \big) - i \big( \delta_{i+1,\, k} \,
[{\bm{1}}_{\bar{\bm{r}}'}]_l + \delta_{i+1,\, l} \,
[{\bm{1}}_{\bar{\bm{r}}'}]_k \big)
\right) \,\,, 
\label{eq:WNm1gD}
\end{eqnarray}
\renewcommand{\jot}{0em}
where $1 \leq k < l \leq N$ and $1\leq i \leq N-1$\,. In this case
the row index $\xi=i$\,. Thus from Eqs.~(\ref{eq:Gsub}),
(\ref{eq:sigmaQ1}), (\ref{eq:WNm1rD}) and (\ref{eq:WNm1gD}) we
obtain:
\renewcommand{\jot}{0.5em}
\begin{eqnarray}
[\bm{\sigma_{[N-1, \hspace{1ex}
1]}^G}]_{\bar{\bm{r}}',\,\bar{\bm{r}}'} & = & \sum_{j,k=1}^N
[W_{\bar{\bm{r}}'}^{[N-1, \hspace{1ex} 1]}]_{ij}
\, [{\bm{G}}_{\bar{\bm{r}}' \bar{\bm{r}}'}]_{jk} \, [(W_{\bar{\bm{r}}'}^{[N-1, \hspace{1ex} 1]})^{T}]_{ki} \, \nonumber \\
& = & \frac{1}{i(i+1)} \sum_{j,k=1}^N \Bigl( \sum_{p=1}^i
\delta_{pj} - \delta_{i+1,j} \Bigr) (\tilde{a}' \, \delta_{jk})
\Bigl( \sum_{p'=1}^i \delta_{p'k} - i \delta_{i+1,k} \Bigr) \nonumber \\
& = & \frac{\tilde{a}'}{i(i+1)} \sum_{j}^N \Bigl( \sum_{p=1}^i
\delta_{pj} - i \delta_{i+1,j} \Bigr) \Bigl( \sum_{p'=1}^i
\delta_{p'j} -i \delta_{i+1,j} \Bigr) \nonumber \\
& = & \frac{\tilde{a}'}{i(i+1)} \Bigl[ \sum_{p=1}^i \sum_{p'=1}^i
\delta_{pp'} - i \sum_{p=1}^i \delta_{p,i+1} -i \sum_{p'}^i
\delta_{i+1,p'}
+ i^2 \delta_{i+1,i+1} \Bigr] \nonumber \\
& = & \frac{\tilde{a}'}{i(i+1)} [i + i^2] \nonumber \\
& = & \tilde{a}' \label{sigmaG11N-1}
\end{eqnarray}
\begin{eqnarray}
[\bm{\sigma_{[N-1, \hspace{1ex}
1]}^G}]_{\bar{\bm{r}}',\,\overline{\bm{\gamma}}'} & = &
\sum_{j=1}^N \sum_{l=2}^N \sum_{k=1}^{l-1}
[W_{\bar{\bm{r}}'}^{[N-1, \hspace{1ex} 1]}]_{ij} \,
[\bm{G}_{\bar{\bm{r}}' \overline{\bm{\gamma}}'}]_{j,kl} \,
[(W_{\overline{\bm{\gamma}}'}^{[N-1, \hspace{1ex} 1]})^{T}]_{kl,i}
\, \nonumber \\
& = & 0 \label{sigmaG12N-1}
\end{eqnarray}
\begin{eqnarray}
[\bm{\sigma_{[N-1, \hspace{1ex}
1]}^G}]_{\overline{\bm{\gamma}}',\, \bar{\bm{r}}'} & = &
\sum_{k=2}^N \sum_{j=1}^{k-1} \sum_{l=1}^{N}
[W_{\overline{\bm{\gamma}}'}^{[N-1, \hspace{1ex} 1]}]_{i,jk} \,
[\bm{G}_{\overline{\bm{\gamma}}' \bar{\bm{r}}'}]_{jk,l} \,
[(W_{\bar{\bm{r}}'}^{[N-1, \hspace{1ex} 1]})^{T}]_{li}
\, \nonumber \\
& = & 0 \label{sigmaG21N-1}
\end{eqnarray}
and
\begin{eqnarray}
[\bm{\sigma_{[N-1, \hspace{1ex}
1]}^G}]_{\overline{\bm{\gamma}}',\, \overline{\bm{\gamma}}'} & = &
[W_{\overline{\bm{\gamma}}'}^{[N-1, \hspace{1ex} 1]}] \, [\bm{G}_{\overline{\bm{\gamma}}' \overline{\bm{\gamma}}'}] \, [W_{\overline{\bm{\gamma}}'}^{[N-1, \hspace{1ex} 1]}]^{T} \nonumber \\
& = & [W_{\overline{\bm{\gamma}}'}^{[N-1, \hspace{1ex} 1]}] \,
[\tilde{g'} I_{M} +
    \tilde{h'} R^T R ] \, [W_{\overline{\bm{\gamma}}'}^{[N-1, \hspace{1ex} 1]}]^{T} \nonumber \\
& = & \tilde{g'} [W_{\overline{\bm{\gamma}}'}^{[N-1, \hspace{1ex}
1]}] \, [W_{\overline{\bm{\gamma}}'}^{[N-1, \hspace{1ex} 1]}]^{T}
+ \tilde{h'} [W_{\overline{\bm{\gamma}}'}^{[N-1, \hspace{1ex} 1]}]
R^T R [W_{\overline{\bm{\gamma}}'}^{[N-1, \hspace{1ex} 1]}]^{T}
\nonumber \\
& = & \tilde{g'} + \tilde{h'} \sum_{k=2}^N \sum_{j=1}^{k-1}
\sum_{m=2}^{N} \sum_{l=1}^{m-1}
[W_{\overline{\bm{\gamma}}'}^{[N-1, \hspace{1ex} 1]}]_{i,jk}
\, [R^T R]_{jk,lm} \, [(W_{\overline{\bm{\gamma}}'}^{[N-1, \hspace{1ex} 1]})^{T}]_{lm,i} \, \nonumber \\
& = & \tilde{g'} + \frac{\tilde{h'}}{i(i+1)(N-2)} \sum_{k=2}^N
\sum_{j=1}^{k-1} \sum_{m=2}^{N} \sum_{l=1}^{m-1} \Bigl(
\sum_{p=1}^i (\delta_{pj} + \delta_{pk})
- i (\delta_{i+1,j} + \delta_{i+1,k}) \Bigr) \nonumber \\
&  & \Bigl( (\delta_{jl} + \delta_{kl}) + (\delta_{jm} +
\delta_{km}) \Bigr)
 \Bigl( \sum_{p'=1}^{i} ( \delta_{lp'} + \delta_{mp'}) - i (\delta_{l,i+1}
+ \delta_{m,i+1}) \Bigr) \,. \label{sigmaG22N-1}
\end{eqnarray}
Using $ \sum_{m=2}^N \sum_{l=1}^{m-1} = \frac{1}{2} \sum_{m=1}^N
\sum_{l=1}^N - \frac{1}{2} \sum_{l=1}^N \delta_{ml}$ in
Eq.~(\ref{sigmaG22N-1}), and summing over $m$ yeilds:
\begin{eqnarray}
[\bm{\sigma_{[N-1, \hspace{1ex}
1]}^G}]_{\overline{\bm{\gamma}}',\, \overline{\bm{\gamma}}'} & = &
\tilde{g'} + \frac{\tilde{h'}}{2 i(i+1)(N-2)} \sum_{k=2}^N
\sum_{j=1}^{k-1} \sum_{l=1}^{N} \Bigl( \sum_{p=1}^i (\delta_{pj} +
\delta_{pk}) - i (\delta_{i+1,j}
  + \delta_{i+1,k}) \Bigr) \nonumber \\
&  & \Bigl( 2 (\delta_{jl} + \delta_{kl}) \Bigr) \bigg(  N
\Bigl(\sum_{p'=1}^{i} \delta_{lp'} - i \delta_{l,i+1} \Bigr)
+  \sum_{m=1}^{N} \Bigl( \sum_{p'=1}^{i} \delta_{mp'} - i \delta_{m,i+1} \Bigr)  \nonumber \\
&  &  - 2 ( \sum_{p'=1}^{i} \delta_{lp'} - i \delta_{l, i+1})
\bigg). \label{sigmaG22cN-1}
\end{eqnarray}
Then using $ \sum_{m=1}^N ( \sum_{p'=1}^{i} \delta_{mp'} - i
\delta_{m,i+1}) = 0$  and $ \sum_{k=2}^N \sum_{j=1}^{k-1} =
\frac{1}{2} \sum_{k=1}^N \sum_{j=1}^N - \frac{1}{2} \sum_{k=1}^N
\delta_{jk}$ in Eq.~(\ref{sigmaG22cN-1}), we find:
\begin{eqnarray}
[\bm{\sigma_{[N-1, \hspace{1ex}
1]}^G}]_{\overline{\bm{\gamma}}',\, \overline{\bm{\gamma}}'} & = &
\tilde{g'} + \frac{\tilde{h'}}{4 i(i+1)(N-2)} \biggl[
\sum_{k,j,l=1}^N \biggl( \Bigl( \sum_{p=1}^i \delta_{pj} - i
\delta_{i+1,j} \Bigr)
 + \Bigl( \sum_{p=1}^{i} \delta_{pk} - i \delta_{i+1,k} \Bigr) \biggr) \nonumber \\
&  &  ( 4 \delta_{jl}) \biggl( (N-2) \Bigl( \sum_{p'=1}^{i}
\delta_{lp'}
- i \delta_{l,i+1}\Bigr) \biggr)  \nonumber \\
&  &- 2 \sum_{j,l=1}^N \Bigl( \sum_{p=1}^{i} \delta_{pj} - i
\delta_{i+1,j}\Bigr) (4 \delta_{jl}) \biggl( (N-2) \Bigl(
\sum_{p'=1}^{i} \delta_{lp'} - i \delta_{l,i+1}\Bigr) \biggr)
\biggr] \nonumber \\
& = & \tilde{g'} + \frac{\tilde{h'}}{ i(i+1)} \biggl[
\sum_{j,l=1}^{N} \biggl( N \Bigl( \sum_{p=1}^i \delta_{pj} - i
\delta_{i+1,j} \Bigr)
 + \sum_{k=1}^N \Bigl( \sum_{p=1}^{i} \delta_{pk}
 - i \delta_{i+1,k}\Bigr) \nonumber \\
&  & - 2 \Bigl(  \sum_{p=1}^{i} \delta_{pj}
 - i \delta_{i+1,j}\Bigr) \biggr) (\delta_{jl})
 \Bigl( \sum_{p'=1}^{i} \delta_{lp'}
 - i \delta_{l,i+1} \Bigr) \biggr] \nonumber
 \\
& = & \tilde{g'} + \frac{\tilde{h'}(N-2)}{ i(i+1)}
\sum_{j,l=1}^{N} \Bigl( \sum_{p=1}^i \delta_{pj} - i
\delta_{i+1,j}\Bigr) (\delta_{jl}) \Bigl( \sum_{p'=1}^{i}
\delta_{lp'} - i \delta_{l,i+1}\Bigr)
\nonumber \\
& = & \tilde{g'} + \frac{\tilde{h'}(N-2)}{ i(i+1)} (i(i+1)) \nonumber \\
& = & \tilde{g'} + (N-2) \tilde{h'} \label{sigmaG22c1N-1}
\end{eqnarray}
\renewcommand{\jot}{0em}
Thus we obtain for the reduced $\bm{G}$ matrix in the $[N-1,
\hspace{1ex} 1]$ symmetry-coordinate basis:
\begin{equation} \label{eq:sigmaNm1GD}
\bm{\sigma_{[N-1, \hspace{1ex} 1]}^G} = \left(
\begin{array}{l@{\hspace{1.5em}}l} \protect[\bm{\sigma_{[N-1, \hspace{1ex}
1]}^{G}}\protect]_{\bar{\bm{r}}',\,\bar{\bm{r}}'} = \tilde{a}' &
{\displaystyle \protect[\bm{\sigma_{[N-1, \hspace{1ex}
1]}^{FG}}\protect]_{\bar{\bm{r}}',\, \overline{\bm{\gamma}}'} =
0} \\
{\displaystyle \protect[\bm{\sigma_{[N-1, \hspace{1ex}
1]}^{G}}\protect]_{\overline{\bm{\gamma}}',\,\bar{\bm{r}}'} = 0 }
& {\displaystyle \protect[\bm{\sigma_{[N-1, \hspace{1ex}
1]}^{G}}\protect]_{\overline{\bm{\gamma}}',\,\overline{\bm{\gamma}}'}
= (\tilde{g}' + (N-2)\tilde{h}')} \end{array} \right) \,.
\end{equation}

Now letting $\bm{Q} = \bm{FG}$ and using Eqs.~(\ref{GFsub}),
(\ref{eq:sigmaQ1}), (\ref{eq:WNm1rD}) and (\ref{eq:WNm1gD}) we can
derive:
\renewcommand{\jot}{0.5em}
\begin{eqnarray}
[\bm{\sigma_{[N-1, \hspace{1ex}
1]}^{FG}}]_{\bar{\bm{r}}',\,\bar{\bm{r}}'} & = & \sum_{j,k=1}^N
[W_{\bar{\bm{r}}'}^{[N-1, \hspace{1ex} 1]}]_{ij}
\, [\bm{FG}_{\bar{\bm{r}}' \bar{\bm{r}}'}]_{jk} \, [(W_{\bar{\bm{r}}'}^{[N-1, \hspace{1ex} 1]})^{T}]_{ki} \, \nonumber \\
& = & \frac{1}{i(i+1)} \sum_{j,k=1}^N \Bigl( \sum_{p=1}^i
\delta_{pj} - \delta_{i+1,j} \Bigr) (\tilde{a} \delta_{jk} +
\tilde{b} [ 1_{\bar{\bm{r}}'\bar{\bm{r}}'}]_{jk})
\Bigl( \sum_{p'=1}^i \delta_{p'k} - i \delta_{i+1,k} \Bigr) \nonumber \\
& = & \frac{1}{i(i+1)} \biggl[ \renewcommand{\arraystretch}{1.5}
\begin{array}[t]{@{}l@{}} {\displaystyle \hphantom{+} \sum_{j}^N
\Bigl( \sum_{p=1}^i \delta_{pj} - i \delta_{i+1,j} \Bigr) \,
\tilde{a} \, \Bigl( \sum_{p'=1}^i
\delta_{p'j} -i \delta_{i+1,j} \Bigr) } \\
{\displaystyle + \sum_{j}^N \Bigl( \sum_{p=1}^i \delta_{pj} - i
\delta_{i+1,j} \Bigr) \, \tilde{b} \, \sum_{k=1}^N \Bigl(
\sum_{p'=1}^i
\delta_{p'k} -i \delta_{i+1,k} \Bigr) \biggr] } \end{array}
\renewcommand{\arraystretch}{1.5} \nonumber \\
& = & \tilde{a} \label{sigmaFG11N-1}
\end{eqnarray}
\begin{eqnarray}
[\bm{\sigma_{[N-1, \hspace{1ex}
1]}^{FG}}]_{\bar{\bm{r}}',\,\overline{\bm{\gamma}}'} & = &
\sum_{j=1}^N
  \sum_{l=2}^N \sum_{k=1}^{l-1} [W_{\bar{\bm{r}}'}^{[N-1, \hspace{1ex} 1]}]_{ij}
\, [\bm{FG}_{\bar{\bm{r}}' \overline{\bm{\gamma}}'}]_{j,kl} \,
[(W_{\overline{\bm{\gamma}}'}^{[N-1, \hspace{1ex} 1]})^{T}]_{kl,i}
\, \nonumber \\
& = & \frac{1}{i(i+1) \sqrt{N-2}} \sum_{j=1}^N \sum_{l=2}^N
\sum_{k=1}^{l-1} \Bigl( \sum_{p=1}^i \delta_{pj} - i
\delta_{i+1,j}\Bigr)
( \tilde{e} (\delta_{jk} + \delta_{jl}) + \tilde{f} )
\nonumber \\
&  & \hspace{14ex} \times \, \Bigl( \sum_{p'=1}^i ( \delta_{kp'} +
\delta_{lp'}) - i ( \delta_{k,i+1}
+ \delta_{l,i+1}) \Bigr) \nonumber \\
& = & \frac{1}{2 i(i+1) \sqrt{N-2}} \biggl[ \sum_{j,l,k=1}^N
\Bigl( \sum_{p=1}^i \delta_{pj} - i \delta_{i+1,j}\Bigr)
( \tilde{e} (\delta_{jk} + \delta_{jl}) + \tilde{f}) \nonumber \\
&  & \hspace{14ex} \times \, \Bigl( \sum_{p'=1}^i ( \delta_{kp'} +
\delta_{lp'})
- i ( \delta_{k,i+1} + \delta_{l,i+1}) \Bigr)  \nonumber \\
&  & \hspace{6ex} - \, 2 \biggl( \sum_{j,k=1}^N \Bigl(
\sum_{p=1}^i \delta_{pj} - i \delta_{i+1,j}\Bigr) (2 \tilde{e}
\delta_{jk} + \tilde{f}) \Bigl(\sum_{p'=1}^i
\delta_{kp'} - i \delta_{k,i+1}\Bigr) \biggr) \biggr] \,. \nonumber \\
\label{sigmaFG12N-1}
\end{eqnarray}
Using $ \sum_{j=1}^N ( \sum_{p=1}^i \delta_{pj} - i
\delta_{i+1,j}) = 0$ in Eq.~(\ref{sigmaFG12N-1}), the term
involving $\tilde{f}$ is zero. Summing over $j$ then yields:
\begin{eqnarray}
[\bm{\sigma_{[N-1, \hspace{1ex}
1]}^{FG}}]_{\bar{\bm{r}}',\,\overline{\bm{\gamma}}'} & = &
\frac{1}{2} \frac{1}{i(i+1) \sqrt{N-2}} \biggl[ \sum_{k,l=1}^N
\biggl( \Bigl( \sum_{p=1}^i \delta_{pk} - i \delta_{i+1,k} \Bigr)
 + \Bigl( \sum_{p=1}^N \delta_{pl} - i \delta_{i+1,l} \Bigr) \biggr)  \nonumber \\
&  & \hspace{9em} \times \, \tilde{e} \,
 \biggl( \Bigl( \sum_{p'=1}^i  \delta_{kp'} - i \delta_{k,i+1} \Bigr)
 + \Bigl( \sum_{p'=1}^N \delta_{lp'}  - i \delta_{l,i+1} \Bigr) \biggr) \nonumber \\
&  & \hspace{7em} - \, 2 \biggl( \sum_{k=1}^N  \Bigl( \sum_{p=1}^i
     \delta_{pk} - i \delta_{i+1,k}\Bigr) \, (2 \tilde{e}) \,
\Bigl( \sum_{p'=1}^i \delta_{kp'}
     - i \delta_{k,i+1}\Bigr) \biggr)  \biggr] \,. \nonumber \\
\label{sigmaFG12cN-1}
\end{eqnarray}
The term involving $2\tilde{e}$ in Eq.~(\ref{sigmaFG12cN-1}) gives
$4\tilde{e} i(i+1)$ (see Eq.~(\ref{sigmaG11N-1}) ) leaving
\begin{eqnarray}
[\bm{\sigma_{[N-1, \hspace{1ex}
1]}^{FG}}]_{\bar{\bm{r}}',\,\overline{\bm{\gamma}}'} & = &
\frac{1}{2} \frac{1}{i(i+1) \sqrt{N-2}} \biggl[ \tilde{e}
\sum_{k,l=1}^N \bigg(
      \Bigl( \sum_{p=1}^i \delta_{pk} - i \delta_{i+1,k} \Bigr)
      \Bigl( \sum_{p'=1}^i \delta_{kp'} - i \delta_{k,i+1} \Bigr) \nonumber \\
&  & \hspace{6em} + \Bigl( \sum_{p=1}^i \delta_{pk} - i
\delta_{i+1,k} \Bigr)
      \Bigl( \sum_{p'=1}^i \delta_{lp'} - i \delta_{l,i+1} \Bigr) \nonumber \\
&  & \hspace{6em} + \Bigl( \sum_{p=1}^i \delta_{pl} - i
\delta_{i+1,l} \Bigr)
      \Bigl( \sum_{p'=1}^i \delta_{kp'} - i \delta_{k,i+1} \Bigr) \nonumber \\
&  & \hspace{6em} + \Bigl( \sum_{p=1}^i \delta_{pl} - i
\delta_{i+1,l} \Bigr)
      \Bigl( \sum_{p'=1}^i \delta_{lp'} - i \delta_{l,i+1} \Bigr)
      - 4 \, \tilde{e} \, i(i+1)  \biggr] \nonumber \\
& = & \frac{1}{2} \frac{\tilde{e}}{i(i+1) \sqrt{N-2}} \bigg[ N
\sum_{k=1}^N
      \Bigl( \sum_{p=1}^i \delta_{pk} - i \delta_{i+1,k} \Bigr)
      \Bigl( \sum_{p'=1}^i \delta_{kp'} - i \delta_{k,i+1} \Bigr) \nonumber  \\
&  & \hspace{6em} + N \sum_{l=1}^N \Bigl( \sum_{p=1}^i \delta_{pl}
- i \delta_{i+1,l}
      \Bigr)
      \Bigl( \sum_{p'=1}^i \delta_{lp'} - i \delta_{l,i+1} \Bigr)
 - 4 \,  \, i(i+1)  \bigg] \nonumber \\
& = & \frac{1}{2} \frac{\tilde{e}}{i(i+1) \sqrt{N-2}} \,\, [ 2 \,
N i(i+1)
    - 4 \,  i(i+1)] \nonumber \\
& = & \frac{\tilde{e}}{\sqrt{N-2}} \,\, [N-2] \nonumber \\
& = & \sqrt{N-2} \,\, \tilde{e} \,. \label{sigmaFG12c1N-1}
\end{eqnarray}
\renewcommand{\jot}{0em}
An analogous calculation yields
\begin{equation}
[\bm{\sigma_{[N-1, \hspace{1ex}
1]}^{FG}}]_{\overline{\bm{\gamma}}',\, \bar{\bm{r}}'} = \sqrt{N-2}
\,\, \tilde{c} \,. \label{sigmaFG21N-1}
\end{equation}
Finally from Eq.~(\ref{GFsub})
\renewcommand{\jot}{0.5em}
\begin{eqnarray}
[\bm{\sigma_{[N-1, \hspace{1ex}
1]}^{FG}}]_{\overline{\bm{\gamma}}',\, \overline{\bm{\gamma}}'} &
= & [W_{\overline{\bm{\gamma}}'}^{[N-1, \hspace{1ex} 1]}
[\bm{FG}_{\overline{\bm{\gamma}}' \overline{\bm{\gamma}}'}] \, [W_{\overline{\bm{\gamma}}'}^{[N-1, \hspace{1ex} 1]}]^{T} \nonumber \\
& = & [W_{\overline{\bm{\gamma}}'}^{[N-1, \hspace{1ex} 1]}] \,
[\tilde{g} I_{M} +
    \tilde{h} R^T R + \tilde{\iota} J_{M}]]
    [W_{\overline{\bm{\gamma}}'}^{[N-1, \hspace{1ex} 1]}]^{T}\,.
\label{sigmaFG22N-1}
\end{eqnarray}
The first two terms in the center brackets are evaluated
identically to similar terms for $[\bm{\sigma_{[N-1, \hspace{1ex}
1]}^G}]_{\overline{\bm{\gamma}}',\, \overline{\bm{\gamma}}'}$ and
yield $ \tilde{g} + (N-2) \tilde{h}$\,. The term involving
$\tilde{\iota} J_{M}$ is zero as shown below:
\begin{eqnarray}
[W_{\overline{\bm{\gamma}}'}^{[N-1, \hspace{1ex} 1]} \,
\tilde{\iota} \, J_{M} \, [W_{\overline{\bm{\gamma}}'}^{[N-1,
\hspace{1ex} 1]}]^{T} & = & \frac{1}{i(i+1)(N-2)} \bigg[ \,
\sum_{k=2}^N \sum_{j=1}^{k-1} \sum_{m=2}^{N} \sum_{l=1}^{m-1}
\Bigl( \sum_{p=1}^i (\delta_{pj} + \delta_{pk}) - i
(\delta_{i+1,j}
  + \delta_{i+1,k}) \Bigr) \nonumber \\
&  & \hspace{10em} \times \, \tilde{\iota} \, \Bigl(
\sum_{p'=1}^{i} ( \delta_{lp'} + \delta_{mp'})
  - i (\delta_{l,i+1} + \delta_{m,i+1})  \Bigr) \bigg] \nonumber \\
& = & \frac{1}{2} \frac{1}{i(i+1)(N-2)} \bigg[ \Bigl[ \sum_{k,j =
1}^N
\biggl( \Bigl( \sum_{p=1}^i \delta_{pj} - i \delta_{i+1,j} \Bigr)
+ \Bigl( \sum_{p=1}^N \delta_{pk} - i \delta_{i+1,k} \Bigr) \biggr) \nonumber \\
&  & \hspace{4em} - 2 \sum_{j=1}^N
\Bigl( \sum_{p=1}^i \delta_{pj} -i \delta_{i+1,j} \Bigr)  \Bigr]
\, \tilde{\iota} \, \biggl( \sum_{m=2}^{N} \sum_{l=1}^{m-1} \Big(
\sum_{p'=1}^i
\delta_{lp'} - i \delta_{l,i+1} \Bigr) \nonumber \\
&  & \hspace{18.5em} + \Bigl( \sum_{p'=1}^i
\delta_{mp'} - i \delta_{m,i+1} \Bigr) \biggr) \bigg] \nonumber \\
& = & 0 \,, \label{sigmaFG22cN-1}
\end{eqnarray}
\renewcommand{\jot}{0em}
where we have used $\sum_{j=1}^N (\sum_{p=1}^i \delta_{pj} - i
\delta_{i+1,j}) = 0$\,.

Thus we obtain
\begin{equation}  \label{eq:sigmaNm1FG_App}
\bm{\sigma_{[N-1, \hspace{1ex} 1]}^{FG}} = \left(
\begin{array}{l@{\hspace{1.5em}}l} \protect[\bm{\sigma_{[N-1, \hspace{1ex}
1]}^{FG}}\protect]_{\bar{\bm{r}}',\,\bar{\bm{r}}'} = \tilde{a} &
{\displaystyle \protect[\bm{\sigma_{[N-1, \hspace{1ex}
1]}^{FG}}\protect]_{\bar{\bm{r}}',\, \overline{\bm{\gamma}}'} =
\sqrt{N-2} \,\, \tilde{e} } \\
{\displaystyle \protect[\bm{\sigma_{[N-1, \hspace{1ex}
1]}^{FG}}\protect]_{\overline{\bm{\gamma}}',\,\bar{\bm{r}}'} =
\sqrt{N-2} \,\, \tilde{c} } & {\displaystyle
\protect[\bm{\sigma_{[N-1, \hspace{1ex}
1]}^{FG}}\protect]_{\overline{\bm{\gamma}}',\,\overline{\bm{\gamma}}'}
= (\tilde{g}+(N-2)\tilde{h})} \end{array} \right) \,,
\end{equation}
for the reduced $\bm{FG}$ matrix in the $[N-1, \hspace{1ex} 1]$
symmetry-coordinate basis.

\subsection{The Matrix Elements
$\bm{[\sigma_{[N-2, \hspace{1ex}
2]}^{G}]_{\overline{\bm{\gamma}}',\,\overline{\bm{\gamma}}'}}$ and
$\bm{[\sigma_{[N-2, \hspace{1ex}
2]}^{FG}]_{\overline{\bm{\gamma}}',\,\overline{\bm{\gamma}}'}}$\,.}
From Eq.~(\ref{eq:WNm2gijmn})
\begin{equation}
\begin{array}[b]{@{}r@{}c@{}l@{}c@{}} {\displaystyle [W^{[N-2, \hspace{1ex} 2]}_{\overline{\bm{\gamma}}'}]_{ij,\,mn} =
\frac{1}{\sqrt{i(i+1)(j-3)(j-2)}} \, \Bigl( }
 & & (\Theta_{i-m+1} - i
\delta_{i+1,\,m})(\Theta_{j-n} -(j-3)\delta_{jn}) & \\
& + & (\Theta_{i-n+1} - i \delta_{i+1,\,n})(\Theta_{j-m}
-(j-3)\delta_{jm}) & \Bigr)  \,, \end{array} \label{eq:WNm2gijmnD}
\end{equation}
where $2 \leq n \leq N, \,\,\, 1 \leq m \leq n-1, \,\,\, 1 \leq i
\leq j-2, \,\,\, 4 \leq j \leq N $ and
(Eq.~(\ref{eq:thetaalphabeta}))
\begin{eqnarray} \label{eq:thetaalphabeta_App}
\Theta_{\alpha-\beta} = \sum_{m=1}^{\alpha-1} \delta_{m, \, \beta}
& = & 1 \mbox{ when } \alpha - \beta > 0 \nonumber \\
& = & 0 \mbox{ when } \alpha - \beta \leq 0 \,.
\end{eqnarray}
In this case the row index $\xi$ is the pair $ij$\,.
Equations~(\ref{eq:WNm2gijmnD}) and (\ref{eq:thetaalphabeta_App})
give
\begin{equation}
\renewcommand{\arraystretch}{1.5}
\begin{array}[b]{@{}r@{}c@{}l@{}c@{}} {\displaystyle [W^{[N-2, \hspace{1ex} 2]}_{\overline{\bm{\gamma}}'}]_{ij,\,mn} =
\frac{1}{\sqrt{i(i+1)(j-3)(j-2)}} \, \bigg( }
 & & {\displaystyle \Bigl( \sum_{p=1}^i
\delta_{pm} - i\delta_{i+1,\,m} \Bigr)
\Bigl( \sum_{p'=1}^{j-1} \delta_{p'n} - (j-3)\delta_{jn} \Bigr) } & \\
& + & {\displaystyle \Bigl( \sum_{p=1}^i \delta_{pn} - i
\delta_{i+1,\,n} \Bigr)
 \Bigl( \sum_{p'=1}^{j-1} \delta_{p'm} - (j-3)\delta_{jm} \Bigr) } & \bigg) \,,
 \end{array} \label{eq:WNm2gijmnDc}
 \renewcommand{\arraystretch}{1}
\end{equation}
and so from Eqs.~(\ref{eq:Gsub}), (\ref{eq:sigmaQ1}) and
(\ref{eq:WNm2gijmnDc}) we obtain
\renewcommand{\jot}{0.5em}
\begin{eqnarray}
[\bm{\sigma_{[N-2, \hspace{1ex}
2]}^G}]_{\overline{\bm{\gamma}}',\, \overline{\bm{\gamma}}'} & = &
[W_{\overline{\bm{\gamma}}'}^{[N-2, \hspace{1ex} 2]}] \, [\bm{G}_{\overline{\bm{\gamma}}' \overline{\bm{\gamma}}'}] \, [W_{\overline{\bm{\gamma}}'}^{[N-2, \hspace{1ex} 2]}]^{T} \nonumber \\
& = & [W_{\overline{\bm{\gamma}}'}^{[N-2, \hspace{1ex} 2]}] \,
[\tilde{g'} I_{M} +
    \tilde{h'} R^T R ] \, [W_{\overline{\bm{\gamma}}'}^{[N-2, \hspace{1ex} 2]}]^{T} \nonumber  \\
& = & \tilde{g'} [W_{\overline{\bm{\gamma}}'}^{[N-2, \hspace{1ex}
2]}] \, [W_{\overline{\bm{\gamma}}'}^{[N-2, \hspace{1ex} 2]}]^{T}
+ \tilde{h'} [W_{\overline{\bm{\gamma}}'}^{[N-2, \hspace{1ex} 2]}]
R^T R [W_{\overline{\bm{\gamma}}'}^{[N-2, \hspace{1ex} 2]}]^{T}
\nonumber \\
& = & \tilde{g'} + \tilde{h'} [W_{\overline{\bm{\gamma}}'}^{[N-2,
\hspace{1ex} 2]}] R^T R [W_{\overline{\bm{\gamma}}'}^{[N-2,
\hspace{1ex} 2]}]^{T}
\nonumber \\
& = & \tilde{g'} + \tilde{h'} \, \sum_{l=2}^N \sum_{k=1}^{l-1}
\sum_{n=2}^{N} \sum_{m=1}^{n-1}
[W_{\overline{\bm{\gamma}}'}^{[N-2, \hspace{1ex} 2]}]_{ij,kl}
\, [R^T R]_{kl,mn} \, [(W_{\overline{\bm{\gamma}}'}^{[N-2, \hspace{1ex} 2]})^{T}]_{mn,ij} \, \nonumber \\
& = & \tilde{g'} + \tilde{h'} \sum_{r=1}^N \Bigl( \sum_{l=2}^{N}
\sum_{k=1}^{l-1}  [W_{\overline{\bm{\gamma}}'}^{[N-2, \hspace{1ex}
2]}]_{ij;kl} (R^T)_{kl,r} \Bigr) \Bigl( \sum_{n=2}^N
\sum_{m=1}^{n-1} (R)_{r,mn}
     [(W_{\overline{\bm{\gamma}}'}^{[N-2, \hspace{1ex} 2]})^T]_{mn;ij} \Bigr)
     \,. \nonumber \\
\label{sigmaG22N-2D}
\end{eqnarray}
We can show that the term involving $\tilde{h'}$ is zero
by performing the double sum $\sum_{l=2}^N \sum_{k=1}^{l-1}$:
\begin{eqnarray}
\lefteqn{\sum_{l=2}^{N} \sum_{k=1}^{l-1}
[W_{\overline{\bm{\gamma}}'}^{[N-2, \hspace{1ex} 2]}]_{ij;kl} (R^T)_{kl,r} } \hspace*{4em} \nonumber \\
& = & \frac{1}{\sqrt{i(i+1)(j-3)(j-2)}} \, \sum_{l=2}^N
\sum_{k=1}^{l-1} \bigg( \Bigl( \sum_{p=1}^i \delta_{pk} -
i\delta_{i+1,\,k} \Bigr)
\Bigl( \sum_{p'=1}^{j-1} \delta_{p'l} - (j-3)\delta_{jl} \Bigr)
\nonumber  \\
&  & + \Bigl( \sum_{p=1}^i \delta_{pl} - i \delta_{i+1,\,l} \Bigr)
\Bigl( \sum_{p'=1}^{j-1} \delta_{p'k} - (j-3)\delta_{jk} \Bigr)
\bigg) \,
\Bigl(\delta_{rk} + \delta_{rl}\Bigr) \nonumber \\
& = & \frac{1}{2 \sqrt{i(i+1)(j-3)(j-2)}} \, \bigg[
  \sum_{k,l=1}^N
\bigg( \Bigl( \sum_{p=1}^i \delta_{pk} - i\delta_{i+1,\,k} \Bigr)
\Bigl( \sum_{p'=1}^{j-1} \delta_{p'l} - (j-3)\delta_{jl} \Bigr)
\nonumber \\
&  & + \Bigl( \sum_{p=1}^i \delta_{pl} - i \delta_{i+1,\,l} \Bigr)
\Bigl( \sum_{p'=1}^{j-1} \delta_{p'k} - (j-3)\delta_{jk} \Bigr)
\bigg) \,
\Bigl(\delta_{rk} + \delta_{rl}\Bigr) \nonumber \\
&  & -2 \sum_{k=1}^N \Bigl( \sum_{p=1}^i \delta_{pk} -i
\delta_{i+1,k} \Bigr) \Bigl( \sum_{p'=1}^{j-1} \delta_{p'k} -
(j-3) \delta_{jk} \Bigr) \, \Bigl(\delta_{rk}\Bigr) \, \bigg] \,.
\label{sigmaG22N-2Dc}
\end{eqnarray}
Summing over $k$ and using the fact that $i \leq j-2$ in
the last term we obtain:
\begin{eqnarray}
\sum_{l=2}^{N} \sum_{k=1}^{l-1}
[W_{\overline{\bm{\gamma}}'}^{[N-2, \hspace{1ex} 2]}]_{ij;kl}
(R^T)_{kl,r} & = & \frac{1}{\sqrt{i(i+1)(j-3)(j-2)}} \, \bigg[
\sum_{l=1}^N \Bigl( \sum_{p=1}^i \delta_{pr} - i\delta_{i+1,\,r}
\Bigr)
\Bigl( \sum_{p'=1}^{j-1} \delta_{p'l} - (j-3)\delta_{jl} \Bigr)
\nonumber \\
&  & + \Bigl( \sum_{p=1}^i \delta_{pl} - i \delta_{i+1,\,l} \Bigr)
\Bigl( \sum_{p'=1}^{j-1} \delta_{p'r} - (j-3)\delta_{jr} \Bigr) \,
-2 \Bigl( \sum_{p=1}^i \delta_{pr} -i \delta_{i+1,r} \Bigr) \bigg]
\nonumber \label{sigmaG22N-2Dd}
\end{eqnarray}
Then summing over $l$ and using $\sum_{l=1}^N ( \sum_{p=1}^i
\delta_{pl} - i \delta_{i+1,l} ) = 0$ \,  along with \,
$\sum_{l=1}^N ( \sum_{p'=1}^{j-1} \delta_{p'l} - (j-3)
\delta_{jl}) = 2$ \, one obtains
\begin{eqnarray}
\sum_{l=2}^{N} \sum_{k=1}^{l-1}
[W_{\overline{\bm{\gamma}}'}^{[N-2, \hspace{1ex} 2]}]_{ij;kl}
(R^T)_{kl,r} & = & \frac{1}{\sqrt{i(i+1)(j-3)(j-2)}} \, \bigg[
 2 \Bigl( \sum_{p=1}^i \delta_{pr} - i\delta_{i+1,\,r} \Bigr)
- 2 \Bigl( \sum_{p=1}^i \delta_{pr} - i\delta_{i+1,\,r} \Bigr) \bigg] \nonumber \\
& = & 0 \,. \label{sigmaG22N-2De}
\end{eqnarray}
Thus from Eqs.~(\ref{sigmaG22N-2D}) and (\ref{sigmaG22N-2De})
\begin{eqnarray}
[\bm{\sigma_{[N-2, \hspace{1ex}
2]}^{G}}]_{\overline{\bm{\gamma}}',\, \overline{\bm{\gamma}}'} & =
& \, \tilde{g'} \,. \label{sigmaG22N-2p}
\end{eqnarray}

Similarly for the matrix element $[\bm{\sigma_{[N-2, \hspace{1ex}
2]}^{FG}}]_{\overline{\bm{\gamma}}',\, \overline{\bm{\gamma}}'}$
we obtain
\begin{eqnarray}
[\bm{\sigma_{[N-2, \hspace{1ex}
2]}^{FG}}]_{\overline{\bm{\gamma}}',\, \overline{\bm{\gamma}}'} &
= &
[W_{\overline{\bm{\gamma}}'}^{[N-2, \hspace{1ex} 2]}] \, [\bm{FG}_{\overline{\bm{\gamma}}' \overline{\bm{\gamma}}'}] \, [W_{\overline{\bm{\gamma}}'}^{[N-2, \hspace{1ex} 2]}]^{T} \nonumber \\
& = & [W_{\overline{\bm{\gamma}}'}^{[N-2, \hspace{1ex} 2]}] \,
[\tilde{g} I_{M} +
    \tilde{h} R^T R  + \tilde{\iota} J_{M}] \, [W_{\overline{\bm{\gamma}}'}^{[N-2, \hspace{1ex} 2]}]^{T}
\end{eqnarray}
The first two terms in the center bracket are evaluated
identically to the analogous terms for $[\bm{\sigma_{[N-2,
\hspace{1ex} 2]}^{G}}]_{\overline{\bm{\gamma}}',\,
\overline{\bm{\gamma}}'}$ and from Eq.~(\ref{sigmaG22N-2p}) yield
$\tilde{g}$. The term involving $\tilde{\iota} J_{M}$ is zero as
shown below:
\begin{eqnarray}
\lefteqn{[W_{\overline{\bm{\gamma}}'}^{[N-2, \hspace{1ex} 2]}] \,
\tilde{\iota} \, J_{M} \,
[W_{\overline{\bm{\gamma}}'}^{[N-2, \hspace{1ex} 2]}]^{T}} \nonumber \\
 & = & \sum_{l=2}^N \sum_{k=1}^{l-1}
\sum_{n=2}^{N} \sum_{m=1}^{n-1}
[W_{\overline{\bm{\gamma}}'}^{[N-2, \hspace{1ex} 2]}]_{ij,kl}
\,\, \tilde{\iota} \,\, [(W_{\overline{\bm{\gamma}}'}^{[N-2, \hspace{1ex} 2]})^{T}]_{mn,ij} \, \nonumber \\
& = & \frac{1}{i(i+1)(j-3)(j-2)} \,
\sum_{l=2}^N \sum_{k=1}^{l-1} \sum_{n=2}^N \sum_{m=1}^{n-1} \nonumber \\
&  & \hphantom{\times \, \tilde{\iota} \,} \bigg[ \Bigl(
\sum_{p=1}^i \delta_{pk} - i\delta_{i+1,\,k} \Bigr) \Bigl(
\sum_{p'=1}^{j-1} \delta_{p'l} - (j-3)\delta_{jl} \Bigr)
 + \Bigl( \sum_{p=1}^i \delta_{pl} - i \delta_{i+1,\,l} \Bigr)
\Bigl( \sum_{p'=1}^{j-1} \delta_{p'k} - (j-3)\delta_{jk} \Bigr)
\bigg] \,
\nonumber \\
&  & \times \, \tilde{\iota} \, \bigg[ \Bigl( \sum_{p=1}^i
\delta_{pm} - i\delta_{i+1,\,m} \Bigr) \Bigl( \sum_{p'=1}^{j-1}
\delta_{p'n} - (j-3)\delta_{jn} \Bigr)
 + \Bigl( \sum_{p=1}^i \delta_{pn} - i \delta_{i+1,\,n} \Bigr)
\Bigl( \sum_{p'=1}^{j-1} \delta_{p'm} - (j-3)\delta_{jm} \Bigr)
\bigg] \,
\nonumber \\
& = & \renewcommand{\arraystretch}{1.5}
\begin{array}[t]{@{}r@{}l@{}} {\displaystyle \frac{1}{2 i(i+1)(j-3)(j-2)} \, }
\bigg[ & \begin{array}[t]{@{}l@{}l@{}l@{}} {\displaystyle \,\,
\sum_{l,k=1}^N \bigg( } & {\displaystyle \hphantom{+} \Bigl(
\sum_{p=1}^i \delta_{pk} - i\delta_{i+1,\,k} \Bigr)
\Bigl( \sum_{p'=1}^{j-1} \delta_{p'l} - (j-3)\delta_{jl} \Bigr) } \\
& {\displaystyle + \Bigl( \sum_{p=1}^i \delta_{pl} - i
\delta_{i+1,\,l} \Bigr) \Bigl( \sum_{p'=1}^{j-1} \delta_{p'k} -
(j-3)\delta_{jk} \Bigr) \bigg) } \\
\multicolumn{3}{l}{ {\displaystyle -2 \sum_{k=1}^N
 \bigg( \Bigl( \sum_{p=1}^i \delta_{pk} -
i\delta_{i+1,\,k} \Bigr) \Bigl( \sum_{p'=1}^{j-1} \delta_{p'k} -
(j-3)\delta_{jk} \Bigr) \bigg) \bigg] } } \end{array} \\
\times \, \tilde{\iota} \, \bigg[ & \begin{array}[t]{@{}l@{}l@{}}
{\displaystyle \,\, \sum_{n=2}^N \sum_{m=1}^{n-1} } &
{\displaystyle \hphantom{+} \Bigl( \sum_{p=1}^i \delta_{pm} -
i\delta_{i+1,\,m} \Bigr) \Bigl( \sum_{p'=1}^{j-1}
\delta_{p'n} - (j-3)\delta_{jn} \Bigr) } \\
& {\displaystyle + \Bigl( \sum_{p=1}^i \delta_{pn} - i
\delta_{i+1,\,n} \Bigr) \Bigl( \sum_{p'=1}^{j-1} \delta_{p'm} -
(j-3)\delta_{jm} \Bigr) \bigg] \,. } \end{array} \end{array}
\renewcommand{\arraystretch}{1}
\label{sigmaFG22N-2D}
\end{eqnarray}
Using $\sum_{k=1}^N ( \sum_{p=1}^i \delta_{pk} - i
\delta_{i+1,k} ) = 0$  yields:
\begin{eqnarray}
\lefteqn{ [W_{\overline{\bm{\gamma}}'}^{[N-2, \hspace{1ex} 2]}] \,
\tilde{\iota} \, J_{M} \, [W_{\overline{\bm{\gamma}}'}^{[N-2,
\hspace{1ex} 2]}]^{T}} \nonumber
\\ & = &
\renewcommand{\arraystretch}{1.5}
\begin{array}[t]{@{}r@{}l@{}} {\displaystyle \frac{-1}{i(i+1)(j-3)(j-2)}
\, \bigg[ } & {\displaystyle \,\, \sum_{k=1}^N
 \Bigl( \sum_{p=1}^i \delta_{pk} - i\delta_{i+1,\,k} \Bigr)
\Bigl( \sum_{p'=1}^{j-1} \delta_{p'k} - (j-3)\delta_{jk} \Bigr)
\bigg] } \\
{\displaystyle \times \, \tilde{\iota} \, } \bigg[ &
\begin{array}[t]{@{}l@{}l@{}l@{}} {\displaystyle \,\,
\sum_{n,m=1}^N \frac{1}{2} \bigg( } & {\displaystyle \hphantom{+}
\Bigl( \sum_{p=1}^i \delta_{pm} - i\delta_{i+1,\,m} \Bigr)
\Bigl( \sum_{p'=1}^{j-1} \delta_{p'n} - (j-3)\delta_{jn} \Bigr) } \\
& {\displaystyle + \Bigl( \sum_{p=1}^i \delta_{pn} - i
\delta_{i+1,\,n} \Bigr) \Bigl( \sum_{p'=1}^{j-1} \delta_{p'm} -
(j-3)\delta_{jm} \Bigr) \bigg) } \\
 \multicolumn{3}{l}{ {\displaystyle - \sum_{m=1}^N \bigg( \Bigl(
\sum_{p=1}^i \delta_{pm} - i\delta_{i+1,\,m} \Bigr) \Bigl(
\sum_{p'=1}^{j-1} \delta_{p'm} - (j-3)\delta_{jm} \Bigr) \bigg)
\bigg] \,. } } \end{array} \end{array}
\renewcommand{\arraystretch}{1}
\label{sigmaFG22N-2Da}
\end{eqnarray}
%
Since $i \leq j-2$\,, $ \Bigl( \sum_{p=1}^i \delta_{pk} -
i\delta_{i+1,\,k} \Bigr) \Bigl( \sum_{p'=1}^{j-1} \delta_{p'k} -
(j-3)\delta_{jk} \Bigr) = \,   \sum_{p=1}^i \delta_{pk} -
i\delta_{i+1,\,k}$ holds we obtain
\begin{eqnarray}
[W_{\overline{\bm{\gamma}}'}^{[N-2, \hspace{1ex} 2]}] \,
\tilde{\iota} \, J_{M} \, [W_{\overline{\bm{\gamma}}'}^{[N-2,
\hspace{1ex} 2]}]^{T} & = &
\renewcommand{\arraystretch}{1.5}
\begin{array}[t]{@{}r@{}l@{}} {\displaystyle \frac{-1}{i(i+1)(j-3)(j-2)}
\, \bigg[ } & {\displaystyle \,\, \sum_{k=1}^N
 \Bigl( \sum_{p=1}^i \delta_{pk} - i\delta_{i+1,\,k} \Bigr) \bigg] } \\
\times \, \tilde{\iota} \, \bigg[ & {\displaystyle \hphantom{-}
\sum_{m=1}^N \Bigl( \sum_{p=1}^i \delta_{pm} - i\delta_{i+1,\,m}
\Bigr) } \\ & {\displaystyle - \sum_{m=1}^i \Bigl( \sum_{p=1}^i
\delta_{pm} - i \delta_{i+1,\,m} \Bigr) \bigg] }
\end{array} \renewcommand{\arraystretch}{1} \nonumber \\
& = & 0 \label{sigmaFG22N-2Db}
\end{eqnarray}

Thus the reduced $\bm{FG}$ matrix element in the
symmetry-coordinate basis in the $[N-2, \hspace{1ex} 2]$ sector
is:

\begin{eqnarray}
[\bm{\sigma_{[N-2, \hspace{1ex}
2]}^{FG}}]_{\overline{\bm{\gamma}}',\, \overline{\bm{\gamma}}'} &
= & \, \tilde{g} \label{sigmaG22N-2}
\end{eqnarray}


\begin{thebibliography}{99}
\bibitem{loeser} J.G.\ Loeser, J.\ Chem.\ Phys.\ \textbf{86}, 5635 (1987).
\bibitem{FGpaper} B.A.\ McKinney, M.\ Dunn, D.K.\ Watson, and J.G.\ Loeser,
  Ann.\ Phys.\ \textbf{310}, 56 (2003).
\bibitem{energy} B.A.\ McKinney, M.\ Dunn, D.K.\ Watson,
  Phys.\ Rev.\ A \textbf{69}, 053611 (2004).
\bibitem{copen92} {\em Dimensional Scaling in Chemical Physics},
  edited by D.R.\ Herschbach, J.\ Avery, and O.\ Goscinski (Kluwer,
  Dordrecht, 1992).
\bibitem{wieman} S.L.\ Cornish, N.R.\ Claussen, J.L.\ Roberts, E.A.\ Cornell,
and C.E.\ Wieman,  Phys.\ Rev.\ Lett.\ \textbf{85}, 1795 (2000).
\bibitem{blume} D.\ Blume and C.H.\ Greene, Phys.\ Rev.\ A {\bf
63}, 63061 (2001).
\bibitem{paperI} M.\ Dunn, D.K.\ Watson, and J.G.\ Loeser,
submitted to Ann.\ Phys.\
\bibitem{avery}J.\ Avery, D.Z.\ Goodson, D.R.\ Herschbach,
Theor.\ Chim.\ Acta \textbf{81}, 1 (1991).
\bibitem{hamermesh} See for example M.\ Hamermesh, {\it Group theory and its
application to physical problems}, (Addison-Wesley, Reading, MA,
1962); and Appendix~C of Ref.~\cite{paperI}.
\bibitem{WDC} See for example E.B.\ Wilson, J.C.\ Decius and P.C.\ Cross, {\it
Molecular Vibrations: The Theory of Infrared and Raman Vibrational
Spectra}, (Dover, New York, 1980); Appendix XII, p.\ 347.
\bibitem{chat} A.\ Chatterjee, J. Phys. A: Math. Gen. {\bf 18}, 735
(1985).
\bibitem{strang} G.\ Strang, {\it Linear algebra and its
applications, Third ed.}. Harcourt Brace Jovanovich College
Publishers, Orlando, FL, 1988.
\bibitem{dcw} E.B.\ Wilson, Jr., J.C.\ Decius, P.C.\ Cross,
\textit{Molecular vibrations: The theory of infrared and raman
vibrational spectra}. McGraw- Hill, New York, 1955.
\bibitem{different} In Ref.~\cite{loeser}, Eq.~(\ref{eq:E1}) would read
\begin{eqnarray}
\overline{E} &=& \overline{E}_{\infty} + \delta \overline{E}_o + O(\delta^2)
\nonumber \\
&=&V_{\mathtt{eff}}(\bar{r}_{\infty},\overline{\gamma}_{\infty}) +
\hspace{0.50em} \delta \left\{  \hspace{-0.5em}
\sum_{\renewcommand{\arraystretch}{0}
\begin{array}[t]{r@{}l@{}c@{}l@{}l} \scriptstyle \mu = \{
  & \scriptstyle \bm{0}^\pm,\hspace{0.5ex}
  & \scriptstyle \bm{1}^\pm & , & \\
  & & \scriptstyle \bm{2} & & \scriptstyle  \}
            \end{array}
            \renewcommand{\arraystretch}{1} }
\hspace{-1.5em} ( n_{\mu}+\frac{d_\mu}{2} ) \, \bar{\omega}_{\mu}
+v_o \right\} + O(\delta^2) \,,
\end{eqnarray}

where $n_{\mu}$ is the total number of quanta in all the normal
modes with the same frequency $\bar{\omega}_{\mu}$, i.e.\

\begin{equation}
n_\mu = \sum_{\mathsf{n}_{\mu}=0}^\infty
           {\mathsf{n}}_{\mu} \, d_{\mu,\mathsf{n}_{\mu}} \,.
\end{equation}
%
\bibitem{Hamermesh_p._27} Ref.~\cite{hamermesh}, p.~27.

%



\end{thebibliography}
\end{document}